\documentclass [11pt, twoside] {thesis}

\usepackage{graphicx}
 
\include{SETEPS}

\begin{document}

%
%
\prelimpages

%
%
\Title{Event-by-event Analysis Methods and Applications to Relativistic Heavy-ion Collision Data}
\Author{Jeffrey Gordon Reid}
\Year{2002}
\Program{Physics}
\titlepage  

\Chair{Thomas A. Trainor}{Professor}{Physics}

%
%
\setcounter{page}{-1}
\abstract{Event-scale analysis techniques using correlation and fluctuation measures are applied to heavy-ion collision data with the goal of discovering, characterizing, and understanding deconfined quark matter.  In service of these goals development of a scale-local thermodynamics is discussed and applied to a broad range of toy models, simulations, and data.  Scale-local partitioning is discussed at length and an ideal theoretical reference is derived for comparison to data.  The dimension of the chaotic attractor of the H\'enon map is calculated as a function of scale, and the implications of scale-local dimension are explored.  Event spaces with discriminatory power are developed and applied to the face recognition problem, detector triggering, and STAR and NA49 data.  Event-wise mean transverse momentum fluctuations in STAR data are analyzed with a minimally biased central-limit based measure using numerical and graphical methods.  Cut efficiencies, corrections, and systematic error sources are all addressed.  A connection between non-statistical fluctuations and two-particle correlations is found and exploited.  Two-particle correlation space formation is discussed and methods for minimizing error from event mixing are discovered.  Central data from STAR and NA49 are analyzed, confirming the results of the fluctuation analysis and providing additional insight into the physics sources contributing to correlations in heavy-ion collisions.}

%
%
\tableofcontents
\listoffigures

%
%
\chapter*{Glossary}      
\addcontentsline{toc}{chapter}{Glossary}
\thispagestyle{plain}

\begin{glossary}

\item[bin occupancy] The data volume contained within a bin.  The bin occupancy of the $i$th bin is written as $n_i$.  For discrete point set data this is the number of points occupying the $i$th bin.  

\item[bin probability] The normalized bin occupancy.  The bin probability of the $i$th bin is written as $p_i$.  For discrete point set data this is the probability of a randomly chosen data point occupying the $i$th bin.

\item[central limit theorem] An important theorem from statistics that states that sampling from a parent distribution with mean $\mu$ and variance $\sigma^2$ results in a distribution approaching normal with mean $\mu$ and variance $\frac{\sigma^2}{N}$.

\item[dimension transport] The scale derivative of the information.  It can also be written as the difference between the scale-local object and reference dimension distributions.  The integral of the dimension transport is fixed by the relative data volume of the object and reference distributions.  Changing the correlations in the data only serves to transport dimension on scale, this is a measure of that transport.

\item[dithering] A binning procedure in which the partition is applied to the data many times, changing the relative position each time.  By averaging over these applications of the partition any bias from a single application is removed.

\item[effective width] The rank-$q$ width of a 1d distribution as seen by a correlation analysis.  

\item[ensemble average] A simulated event that is defined by the average of the entropy of all of the events within an ensemble.  The difference between the entropy of a single event in an ensemble and its ensemble average is a measure of the uniqueness of the event.

\item[event space] A space in which event characteristics are used to define the axes.  In an event space each event appears as a single point and the distance between two points is a measurement of the similarity of the events represented by those points.

\item[global variable] An event characteristic that is represented by a single number.   

\item[mixed pair] An ordinal pair made from the measured properties of two particles taken from different events.

\item[model-explicit] Containing model assumptions explicitly.  Scaled correlation analysis, for example, is model-explicit since one must explicitly choose a model to generate a reference for comparison to the object of the analysis.  The $\Phi_{p_t}$ measure is not model-explicit.  Its definition ($\sqrt{\frac{\overline{Z^2}}{\overline{N}}}-\sigma_{p_t}$) contains the reference ($\sigma_{p_t}$) and its model assumptions implicitly.

\item[nearest-neighbor events] The events closest to a given event within an event space.  These are the events that are most similar in terms of the event-space measures.

\item[object distribution] The distribution that is the object of an analysis.  To test a hypothesis the object distribution must be compared to a reference distribution consistent with the hypothesis.

\item[occupied bin] A bin in a partition that contains some fraction of the data.

\item[pile-up event] An event that contains multiple beam-target or beam-gas interactions.  These events are a source of error since they contain multiple beam interaction vertices.

\item[pre-binned data] Any data that takes the form of a list of bin occupancies.  All real data is pre-binned since we cannot in reality know the position of individual data points with perfect accuracy.  

\item[primary track] A track that originates from the primary vertex.

\item[primary vertex] The point of contact between colliding nuclei. 

\item[pseudoinformation] The apparent information contained in a randomly generated uniform distribution when compared to an ideal uniform distribution with a cutoff.  This is an effect of an approximation to the correlation integral used to simplify the topological measure calculations.

\item[$q$-tuple] A grouping of $q$ points.  A collection of N points contains $C^N_q=\frac{N!}{q!(N-q)!}$ unique $q$-tuples.   

\item[quark-gluon plasma] A theoretical state of quark matter in which quarks and gluons are asymptotically free.

\item[reference distribution] The distribution that serves as a baseline for comparison to the object of an analysis.  The reference is determined by the hypothesis that the analysis is testing.

\item[scale-local] As a function of scale.

\item[sibling pair] An ordinal pair made from the measured properties of two distinct particles taken from the same event.

\item[split track] A track that has been improperly reconstructed so that it appears as two or more disconnected segments instead of a single continuous track.

\item[support] The collection of occupied bins for a distribution.

\item[track] A reconstruction of the path taken by a particle through a detector.  

\item[vertex] A crossing of two or more tracks.

\item[void bin] A bin in a partition that does not contain any part of the data.
 
\end{glossary}

%
%
\acknowledgments{
\vskip2pc
{\narrower\noindent

I have had the privilege of working with the most interesting and dynamic people I have ever met while in graduate school.  I must thank all of the members of the NA49 and STAR collaborations for allowing me to be part of their work.  In addition, there are some specific people who deserve individual recognition and thanks:

Tom Trainor for continuing to believe in me, even when I don't.
Lanny Ray for too many things to mention, but mostly for listening.
Lee Barnby and Tim Yates for countless rounds at the co-op.
Yota Foka for her limitless kindness and hospitality.  
Gunther Roland for being a patient and expert fluctuation analysis guide.
Peter Seyboth for being intelligent and caring, the ideal physicist.
Peter Jacobs for his unvarnished insights on everything.
Peter Jones for being one of the lads.  
John Gans and Mike Miller for being {\em amigos}.
Aya Ishihara for her correlation analysis efforts.
Matt Lamont for being a great friend. UP THE BORO.
Ron Longacre for showing me what real genius looks like.
Jack Sandweiss for his encouragement of my error analysis in correlations work. 
Kathy Turner for being a woman and a single mother in physics; she is an inspiration.
Duncan Prindle for being inscrutable.
Dick Seymour for being the IT zen master.
Dhammika Weerasundara for believing in SCA and working hard to make it a reality.
Michael Jahn for letting me teach and being a friend.
Barb, Karin, and Kate for running the place.
Duba, Chapman, Brugalette, Beck, Henry, Hoyle and Mumm for being pals.
Sue Tollefson, Geoffrey Boers, John Cramer, Hamish Robertson, and David Thouless for their insights.
Derek and Simon for being.
Tammie, Brad, Janet, Judy, Tanya \& Pete, for all their love and support; and especially George and Sonja Reid for everything, I really appreciate it even if it doesn't show.

\par}
}

%
%

%
%
\textpages


%
%
%
\chapter{Motivation}
 
\section{Introduction and Overview}

\begin{figure}[ht]
\centering
\includegraphics[width=4in]{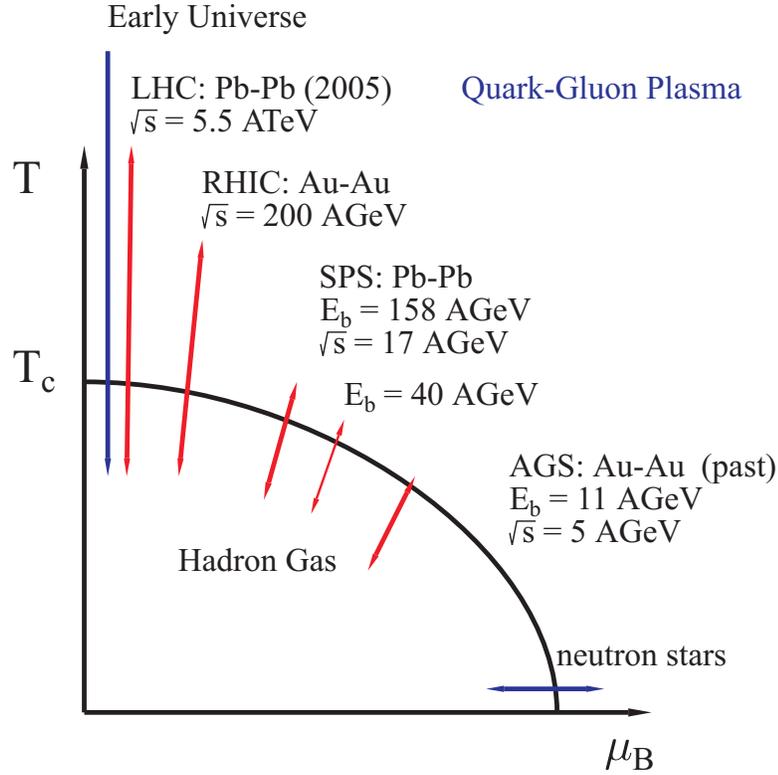}
\caption[A cartoon of the QCD phase diagram showing temperature vs. baryon chemical potential.]{A cartoon of the QCD phase diagram showing temperature vs. baryon chemical potential. (taken from \cite{FluctReview})}
\label{PhaseDiagram}
\end{figure}

The constituents of nuclear matter behave according to the rules of quantum chromodynamics (QCD).  These rules guide our expectations for the interaction of quarks via the exchange of gluons.  The most striking feature of QCD is the energy dependence of the strong coupling constant, $\alpha_s(q^2)$.  At short distances (large momentum transfer, high temperature) the coupling is small; at large distances (small momentum transfer, low temperature) the coupling increases to its eponymous strength \cite{PDG}.  The major consequence of the strong coupling at low energy is the hadronic confinement of quarks and gluons which is responsible for nucleonic structure.  At very high energies the coupling constant nearly vanishes and the quarks and gluons are asymptotically free.  This deconfined state of quark matter is known as the {\bf quark-gluon plasma} (QGP) \cite{HM}.  By creating a QGP and observing its behavior we hope to obtain a better understanding of strong interaction physics in the high temperature regime.  The work in this dissertation presents one approach to finding and understanding this new form of matter.  This is a small piece of a larger effort within the nuclear physics community to map out the nuclear matter phase diagram (see figure~\ref{PhaseDiagram}).

%
%

QGP physics is difficult because we cannot access the deconfined quark matter directly.  We can prepare high energy-density nuclear matter with the accelerators (CERN's Super Proton Synchrotron) and colliders (BNL's Relativistic Heavy-ion Collider) at our disposal, but the collision products dissipate very quickly.  Thus, our detectors only see the decay products of the excited matter after dissipation has occured.  However, this achievement is itself no small feat.  The fact that we are able to record a collision event with the necessary precision is a testament to modern experimental techniques.  We have gone from analyzing photographic bubble chamber data by eye to the modern time projection chamber, which provides an accurate digital snapshot of the charged collision products.  The tools of our trade have been revolutionized along with the rest of the world by the availability of fast, affordable, networked computing power.  As a consequence of this revolution of detectors and experimental methods we are now experiencing a painful period of rapid change in the development of analysis methods used to extract physics results from the data.  With precise, digital event reconstruction and high-energy collider-accelerators we can create and reconstruct individual events with thousands of produced charged particles.  This has given us, for the first time, the ability to do statistically meaningful event-by-event analysis of our data.  This type of analysis is essential if we are to achieve the goal of creating deconfined quark matter and understanding the nuclear matter phase diagram.

Consider the data that comes from a single collision event.  After the high energy-density matter created in the collision has dissipated, we are left with only a detector image of the produced particles.  Along with the trajectory information we record as much as we can about the properties of the particles we detect (charge, momentum, energy-loss, etc).  Because this matter cannot be kept in the interesting region of the phase diagram, a wide variety of possible decay-product signals of the formation of a QGP have been proposed \cite{Harris}.  Ultimately the data available to us are so far removed from the physics of interest (the state of the collision participants at the instant of the collision) that the final state particle distributions are nearly correlation free.  From this barely-correlated distribution of particles we are trying to piece together the convoluted journey that they take during the evolution of the collision.  This requires an approach to data analysis that is sensitive to very small non-statistical effects so that we can fully exploit the correlations present in the data.  Until recently the application of such methods was in its infancy because the small number of particles detected per event at lower energies made it difficult to reach statistically meaningful conclusions.  With the recent increases in collider energy and the resulting increases in event multiplicity a new frontier of event-by-event data analysis has opened.  Event-by-event approaches are gaining popularity where inclusive measures once dominated, and the physics possibilities are very exciting.  There are a variety of ways in which the QGP phase transition may present itself on an event-by-event level, two distinct approaches to QGP detection adopted here utilize measures of fluctuations and correlations.  

Event-by-event fluctuation analysis relies on measurements of fluctuations present in event variables \cite{GeneralizedPhi} like multiplicity \cite{SPH} and particle species ratios \cite{QGPFluct}.  All of the event-by-event fluctuation measures have some degree of utility \cite{Compress} \cite{FluctReview} \cite{APPFluct}, but for this analysis effort we have chosen to focus on fluctuations in mean transverse momentum.  $<p_t>$ fluctuations have the most theoretical support of all event-by-event approaches and there are models that both contradict \cite{LUCIAE} and agree with \cite{QGSM} \cite{FirstOrder} the experimental fluctuation results.  There are a variety of proposed momentum fluctuation analysis measures \cite{PhiPt} \cite{SubEvent} \cite{CLT}.  All are closely related and come out of comparisons of measured fluctuations to statistical expectations.  Of these the $\Delta\sigma_{p_t}$ measure is minimally biased \cite{TATMeasureBias} and thus the most useful for our purposes.  

From a theoretical perspective fluctuation analysis is an exploitation of canonical thermodynamics which tells us that the non-statistical fluctuations of collective observables are related to physically interesting measures.  For example, fluctuations in the slope of the transverse momentum distribution can provide information about the temperature of the collision at freeze-out \cite{TFluct}.  We need event-by-event approaches because they have access to information unavailable to inclusive measure analysis \cite{TptFluct}.  Event-by-event signatures can be used to map the phase diagram with information like the order of the transition \cite{FirstOrder} and the presence of a critical point \cite{CriticalPoint}.

Event-by-event correlation analysis has been used in various forms since the inception of experimental QCD physics.  Parton jet finding \cite{jets}, event-plane determination for flow analysis \cite{flow} \cite{STARFlow}, and Hanbury-Brown Twiss interferometery for source size and lifetime calculations \cite{HBT} all depend on correlation analysis.  All of these methods have been (and continue to be) applied on an event-by-event basis.  We have chosen to take a different and completely general approach to correlation analysis.  Using tools derived from the entropy \cite{EbyEEntropy}, factorial moments \cite{FacMo}, and autocorrelation \cite{AutoCorr}, we study QCD through the arbitrary correlation content of the final state measured with model-independent correlation analysis systems.  

Here we present novel methods that are suitable for analyzing data from single relativistic heavy-ion collision events and map out the analysis tasks that lie before us.  We will always have an eye on the greater goal of discovering and understanding deconfined quark matter, but the initial task is to construct a robust analysis system for event-by-event analysis, which has become increasingly important in experimental QCD physics.

\section{Approaches to Event-by-event Analysis}

At relatively low energies ($\sqrt{s_{NN}}\sim$1-5 GeV) the low multiplicity of a single event makes analyzing single heavy-ion events of limited utility.  With the larger energies seen at the CERN SPS ($\sqrt{s_{NN}}=17$ GeV) and recently at BNL with the RHIC ($\sqrt{s_{NN}}=$130-200 GeV) have come larger event multiplicities, and thus the possibility of making statistically meaningful physics measurements with event-by-event analyses.  Of course, accelerator/collider experiments have often focused on a kind of event-by-event physics.  When searching for a new particle one often hopes to find an elusive single event that contains clear evidence that the particle exists and has been created successfully in the lab.  Parton searches were finally successful because single events with dramatic jet structures were observed \cite{jets}; event-by-event searches for specific decay toplogies in bubble chamber images are responsible for the discovery of a plethora of particles ({\em e.g.,} the discovery of the $\Omega^-$ \cite{omegadiscovery}).  In this sense event-by-event physics has had a long history in nuclear and particle physics.  

However, now that our experiments have matured we can do more.  We can now move beyond specific, model-dependent event-by-event physics searches and make statistically significant measurements of the general properties of individual events.  This  experimental evolution has been accompanied by five years of rapid growth in the number of publications aimed at the quantification of event-wise physical quantities \cite{FluctReview}.  The work reported here has been aimed at developing a set of tools for the analysis and characterization of event-by-event RHI data.  The interpretation of results from these new tools is often difficult, and a large portion of this work will be aimed at addressing the issue of making our results meaningful.  

By maintaining a connection to the existing canon of analysis tools the job of interpreting our results is made much easier.
This is an important consideration.  The more care that is taken in the crafting of the analysis system, the easier it will be to interpret the results and generate meaningful conclusions.  One could easily become lost in a sea of questionable results when dealing with poorly designed, misunderstood measures.  Avoiding this is a large part of the motivation for this work.  In addition to using the connection to established measures as a touchstone for interpretation of our results, we wish to provide a new context for understanding existing techniques.  We hope to address and resolve the shortcomings of existing measures by developing a holistic approach to event-by-event analysis.

Towards this end, we consider three distinct but interconnected types of correlation analysis, each based (in part) on a different piece of the existing analysis canon.  First, we will take on the task of creating a generalized correlation analysis system appropriate for event-by-event analysis.  The available QCD theory of RHI collisions is not very predictive.  We need an analysis system that is able to find something interesting even though we may not know how to define what ``interesting" is.  By carefully examining this problem we can identify the characteristics of an appropriate analysis system to help with the actual implementation (see chapters 2, 3, and 4 for details).

Secondly, we will formulate an event-by-event analysis system that uses fluctuation analysis to learn about the collision.
The most straightforward method used for characterizing events is by the calculation and analysis of event-wise properties.  We will refer to these as {\bf global variables}.  By analogy to thermodynamics there are two types of global variables, extensive (multiplicity $N$, total momentum $P$, etc.) and intensive (temperature T, event-wise mean momentum $<p>$, etc.).  We will concentrate on intensive variables since we want to compare events with different centrality, and the increase in multiplicity with centrality makes extensive quantities more difficult to deal with.  By looking at the fluctuations present in global variables we can already understand a great deal about the evolution of the collision.  The $\Phi_{p_t}$ variable of Ga\'zdzicki and Mr\'owczy\'nski \cite{PhiPt} provides a starting point for the development and refinement of fluctuation analysis in RHI collision data.  The details of our global variable analysis approach and its relationship to $\Phi_{p_t}$ are discussed in chapter 5.

Finally, we will show that two-particle correlations are uniquely important in understanding the early state of the collision.  We want to understand the hadronization process that created the detected particles.  Thus, we are naturally drawn to the correlations between particles, which can contain significant amounts of information about the early collision.  While higher-order correlations are also interesting, two-particle correlations are the easiest to calculate and interpret.  The exponential growth of the combinatorial background with correlation order makes higher-order correlation calculations too computationally expensive to be useful for this analysis.  Not only do the two-particle correlation measures provide valuable insight into the correlation structure of the data, but there is also a deep algebraic connection between these measures and the fluctuations present in certain global variables \cite{CLT}.  This is explored in detail in chapter 6 where we examine the relationship between the $\Phi_{p_t}$ variable and two-particle correlation measures.

\section{Generalized Model-independent Correlation Analysis}

While it is important to understand and incorporate existing event-by-event analysis methods into this work, the main purpose is to develop new techniques applicable to (but not constrained by) event-by-event physics.  The most significant part of this work is in the development of a system for quantifying the information contained in an arbitrary data set and interpreting its results to better understand possible correlation mechanisms.  Statistics, probability, topology, information theory, statistical mechanics, and thermodynamics have all been applied, with varying degrees of effectiveness, as probes of correlation structure.  Using these canonical tools as a starting point, we will formulate a robust analysis system appropriate to the problems of event-by-event physics. 

We want to study a collision after it has heated up to the point that quark deconfinement is possible, but before it has cooled again and made the detectable hadrons we observe.  Because the detected particles are the only connection we have to the collision we need an analysis system that will allow us to see through the veil of hadronization back to the early collision.  This alone is a daunting task, but the real difficulty comes when we consider the state of QCD theory.  Since the QCD Lagrangian has no closed-form solution we are dependent upon a wide variety models and assumptions to help us know what to expect from our data \cite{Intro}.  These expectations are often contradictory and rife with theoretical oversimplification and experimentally unsupportable assumptions.  Because of this cacophony of available models we have chosen an explicit approach when incorporating any model into the analysis system.  One simple way to do this is to design a system that is model-independent.  This decouples our results from any current theory, but still allows us to make comparisons with theory expectations when appropriate.  While all of our techniques are not model-independent, it is important to stress that when a model is incorporated implicitly into an analysis we will address the model choice directly.  

An example of {\bf model-explicit} analysis can be found in detector triggering.  When colliding nuclei as we do, it is impossible to precisely control the time at which the collision occurs ($t_0$).  This necessitates a triggering system to determine $t_0$ and set in motion the chain of events that record the data present in the detector at that instant.  A simple trigger might look for significant energy deposition in the detector and/or a lack of significant energy deposition along the beam path.  This selects events where a significant amount of momentum transfer has occured.  Modern triggers have become more sophisticated.  We can now choose to record data from a collision based on how interesting that collision appears to be according to a variety of criteria.  The trigger system in the STAR experiment can operate several triggers in parallel to generate a variety of data streams from a single set of collisions.  This trigger system is analogous to a model-explicit analysis because it incorporates a variety of different triggers based on explicit criteria.  When analyzing a data set one must explicitly define the trigger that was used in creating it.  Similarly, when analyzing data with a model-explicit analysis system one must explicity specify the reference model used, otherwise the results are meaningless.

The motivation for the development of a general model-independent analysis was to develop a smart trigger for STAR similar to the one described above.  We approached the problem by trying to answer the universal question: ``How can we quantify the amount and type of information contained in some general distribution?"  This problem arises in many different contexts, from data compression to image analysis.  Our approach was to characterize the information contained in the data using a model-independent analysis, then incorporate a model explicitly by performing an identical analysis on the model.  A comparison between these two results gives a measure of their similarity, and allows us to quantify how interesting the data might be.  Building such an analysis system was more subtle than we first imagined, so we chose to abandon the smart trigger and move to offline analysis.

\section{Quantifying Information}

We want to measure the low-level structure present in a {\em nearly} statistical distribution of particles produced from a RHI collision.  While there are many analyses designed to address similar problems, there is no existing analysis system that can fulfill all of our needs.  The problem calls for an abstraction of various techniques so we can create a generalized, universal language for model-independent correlation analysis, applicable and useful in a variety of contexts.

Since we approach the problem from a physicist's perspective we start by considering statistical mechanics and thermodynamics.  The simplest mathematical measure of information content or correlation is entropy.  The Boltzmann entropy, $S=k_b\:log(g)$ is a measure of the number of states that are accessible to a given system \cite{KandK}.  Clearly, the more states accessible to a system, the more information the system can contain.  The computing problems that arose at the turn of the millennium are a good example of this.  Most computers had been programmed to represent the date using two digits in base 10.  This allowed for only 100 unique states in the system.  At the turn of the millennium suddenly we needed to avoid the redundancy of 1900 and 2000  being represented by the same state (00).  This was solved by adopting a system that could contain more information.  Many systems went to a four digit year, insuring no redundancy for 10,000 years because of the extra information content possible in a system with a factor of 100 more states.  The {\em relative} entropy between two systems tells us how much information one system can contain as compared to another.  This is why information is defined as the relative entropy \cite{Info}.

This example suggests a straightforward {\em information-based} solution to the triggering problem.  By comparing the entropy of a known reference distribution ({\em e.g.,} an ``uninteresting" event from an appropriately chosen statistical model) to the entropy of the data in a particular event, we can measure the mutual information.  This will tell us if the data contain any information beyond what we expect to see in the reference distribution.  As long as the reference we have chosen is appropriate, this will tell us which events are ``interesting" and worth further study, and which events look so similar to the ``uninteresting" model prediction that they can be ignored.

\section{Reference Comparisons}

Reference comparison is an important aspect of any information or correlation analysis.  The scientific method itself is based on the idea of comparisons between data and reference.  We compare data to a model to see if the information in the data matches the information predicted by the model.  The model's information content serves as the reference, with the comparison to data telling us if the data and model are in agreement.  

Describing a set of data with a fitting function is an illustrative example of a model-explicit reference-based analysis.  Consider a goodness-of-fit measure such as $\chi^2$.  By minimizing $\chi^2$ we are minimizing the difference between the data and a reference defined by the function we are fitting to the data.  In effect, we are changing the fitting parameters using the assumption that the fitting function being considered is a good reference to compare to the data.  If it is a good reference, then we can determine the parameters of the fitting function so the data deviate minimally from this reference.  The final minimized $\chi^2$ calculated gives us an idea of the quality of fit, if the fit is bad we must reexamine the assumption that the reference chosen for the fitting function is appropriate for the data \cite{Bevvy}.  

In this example the operating assumption is that the reference (fitting function) is robust and its parameters can be tuned to match the data so that meaningful conclusions can be drawn from the parameters.  However, if there is no good reference available, this approach is not useful.  The goodness-of-fit procedure cannot make suggestions as to what reference is appropriate; it can only make a comparison using an explicitly specified reference.    

An example that illustrates a model-dependent reference-based analysis is the linear correlation coefficient.  The linear correlation coefficient is obtained by performing a least-squares fit on a two-dimensional data set with straight lines of the form $y=a+bx$ and $x=a'+b'y$ \cite{Bevvy}. The coefficient is a measure of the correlation covariance of the data, $r=\sqrt{bb'}$.  Because of the integrated reference, the data can contain significant correlation to which this measure is totally insensitive.  Namely, if the assumption about the linear relationship between $x$ and $y$ is incorrect the LCC may be useless.  Any circularly-symmetric data distributions will have a linear correlation coefficient of zero.  This makes it impossible to distinguish between a totally uncorrelated distribution and a tightly correlated, circularly-symmetric distribution when using the LCC as the correlation measure.  The LCC is a useful measure, but one must be aware of the correlations to which it is sensitive and use it only in situations where it is well matched to the problem being addressed.  

These examples make it clear that the choice of reference should be made explicit in our correlation analysis system.  Since we aren't sure exactly what correlations will present themselves it is especially important that we use the most flexible approach possible.  This is also consistent with our desire to create a system that is inherently model-independent.  Using this model-explicit approach the analysis can be done without any model.  This would yield an absolute measure of the information contained in the data.  If a model comparison is appropriate the absolute results from the data and a reference (from an explicitly chosen model) can be used to calculate the mutual information between the two distributions.

\section{Scale Generalization}

As we know from thermodynamics, the choice of partition is an essential part of calculating thermodynamic quantities \cite{Huang}.  Thus, one of the first problems that we encounter in designing an entropy-based correlation analysis system is choosing the proper partition.  

As the number of degrees of freedom available to a system undergoing a phase transition changes self-similar scaling behavior often arises in the ordering of the system's constituents ({\em e.g.,} crystal formation, cluster agglomeration) \cite{Fractal}.  This is particularly true in first order phase transitions.  Because some theorists suggest that the QGP transition should be first order \cite{FirstOrder}, we have chosen to insure that our analysis be sensitive to correlations at a variety of scales.  Thus, we can easily identify events with self-similar scaling behavior if the phase transition presents itself in such a direct way.  

Unfortunately, standard thermodynamic entropy calculations are relevant only in the zero-scale limit.  This restricts us to analyzing distributions in the region where the partition size $e$ is taken to be very near zero.  Since we want to be sensitive to correlations present at any scale, we cannot depend on the standard methods for calculating thermodynamic quantities.  We need a generalization of entropy calculation methods that can be extended to instances where the number of bins is small.

The solution to our problems with the standard entropy lie in a generalization of the partition scale.  By abandoning the restrictive zero-scale limit we gain sensitivity to information content at all scales.  A scale-generalized (or {\bf scale-local}) entropy also allows direct comparison of results at different scales, making self-similar (fractal) behavior easily detectable.  Furthermore, this generalization incorporates the thermodynamic limit (a zero-scale partition) seamlessly with the continuum of other binning scales.  This creates a larger context that is conceptually satisfying; the zero-scale limit of the scale-local entropy is the standard entropy.

With such a seemingly simple generalization as a solution to our analysis problem one might wonder why this approach has not been applied to entropy calculations before.  As we will see in the coming sections the computational task of calculating the entropy of a data distribution as a function of scale is substantial.  It requires significant numerical computation to characterize even the simplest distribution.  With cutting-edge computing technology these calculations are quite time consuming, and it is clear that they would not have been technically feasible even 10 years ago.  It is possible to calculate the scaled entropy of the simplest distributions analytically, but the practical application of such a calculation requires the large and reliable computing power afforded by modern computers.  The development of this analysis method has been intimately tied to recent developments in computing power.

\section{Conclusions}

The physics of high temperature quantum chromodynamics is itself undergoing an exciting phase transition.  The advent of fast and affordable computing coupled with new experimental techniques has given us high-energy accelerators and exquisitely sensitive detectors that are revolutionizing experimental QCD.  No subfield feels this more acutely than event-by-event physics.  For the first time statistically meaningful event-scale analysis has become possible.  These developments have motivated us to tackle the problem of developing new analysis techniques to exploit the exciting new possibilities of meaningful event-by-event physics.  

This work presents three distinct, but interconnected, approaches to event-by-event analysis.  First is the most general, scale-local model-independent correlation analysis.  This approach eschews any connection to the specifics of QCD and event-by-event physics.  By designing a completely general analysis system we hope to be sensitive to any possible QGP signal.
Second is an analysis of fluctuations in the event-wise transverse momentum distribution.  This analysis is based on the work of Ga\'zdzicki and Mr\'owczy\'nski \cite{PhiPt} and addresses the expected effect of a QGP transition on temperature fluctuations.  Our goal with this analysis has been to clarify the meaning of existing fluctuation measures and design a framework for fluctuation analysis that is applicable in a broader context.  Finally, is the analysis of two-particle correlations.  This effort grew out of the fluctuation measure work and has since taken on its own life.  Here we to look at the full two-particle correlation map instead of a specific region or measure of correlation strength.  This gives us a broad view of the correlation content of the data and gives us access to nearly all of the physics of the collision.

While these approaches are diverse in scope and method, they are bound together by the common thread of event-scale analysis.  The results born of these techniques represent the realization of the promise of event-by-event physics in heavy-ion collisions.

%
%
%

\chapter{Scaled Correlation Analysis}

\section{Introduction}

Our goal is to detect anomalous QGP events.  However, the scientific method works best when there are clear, testable hypotheses and direct measurement methods for testing them.  As we discussed in the previous chapter, that is not the case for QCD theory and experiment.  This is not the fault of the theory community.  It simply indicates the depth and complexity of the issues that we must address.  To accommodate this situation we have built an analysis system independent of theory expectations.  In this system we characterize the information contained in the data without interpretation.  As much as we value this theoretical independence, we will eventually want to make a variety of model comparisons.  Thus, we have built a general analysis that 1) does not rely on any specific model or theory and 2) can nevertheless be used to make model comparisons.  We will start by focusing on model-independence, then move to developing tools for making explicit comparisons with theory predictions.

\section{Simplifying the Analysis Task}

Using thermodynamics and information theory as a starting point we have found that each collision event can be specified uniquely and independently.  The entropy of a single event can be calculated by treating the data as a set of points in a $d$-dimensional space (where $d$ is the number of unique variables needed to completely specify the particle data), then partitioning that space and applying the methods of statistical mechanics to calculate properties of the single-event data.  For example, we can calculate an entropy $S_0=log[M_0]$, where $M_0$ is the total number of bins in the partition that contain {\em at least one data point}.  We refer to these as {\bf occupied bins}.  The set of occupied bins is the {\bf support} of the distribution.  Of course, it is not necessary to use all of the data variables from an event in our analysis.  Calculating the entropy of an arbitrary point set in a $d$-dimensional space could be prohibitively difficult and time-consuming.  Thus, we will ignore most of the event variables and focus initially on a one-dimensional analysis.  

We select our initial analysis variable to be the transverse momentum ($p_t$) of the detected collision products.  The slope of the $p_t$ distribution is a measure of the temperature of the collision \cite{TFluct}, and so $p_t$ is a logical place to look first for signs of interesting physics.  By focusing on $p_t$ we keep the problem tractable computationally, and also make the physics interpretation of our results simpler.  This does not rule out later, more complex analyses including other variables.  Rather, it serves as a starting point for our analysis development.

To calculate the entropy of a point set we must select a scale for the binning that will be applied to the data.  This decision may bias our results.  For example, we might choose a binning scale at which we expect to find interesting features in the data.  This provides sensitivity in that scale region at the cost of sensitivity elsewhere in scale.  If we instead consider the entropy of each event as a function of the scale of the partition we remove any scale bias present in the standard entropy formulation.  This allows us to apply the results of information theory in a general way, independent of assumptions about the scale of anticipated correlation features.  Thus, we write the entropy $S_0$ as a function of scale, $S_0(e)=log[M_0(e)]$ where $e$ is the scale of the binning system.  This novel approach to the entropy is outlined in great detail in \cite{SLTM}.

\section{Partitioning the Embedding Space}

Before we discuss the details involved in calculating the scaled entropy we must first decide what type of partition to use.  There are a couple of requirements that the chosen partition must obey.  First, we need a partition that minimizes scale-mixing; at each point in scale we would like the partition to be sensitive to correlations at that specific scale and no others.  A careful choice of partition can maximize our resolution by minimizing the sensitivity window.  Secondly, it is important that the partition completely cover the embedding space with no holes or overlapping bins.  If there were holes then we would miss some correlations in the data dependent upon where the correlated points presented themselves in the embedding space.  If the bins overlapped then we would weight the correlations in the region of space where the overlap occured more heavily then the correlations in other regions of the embedding space.  

With a one-dimensional embedding space the choice of bins for a scale $e$ is obvious, each bin is a line segment of length $e$.  Points that fall on the $i$th line segment are considered to be in the $i$th bin.  Partitioning the space with a lattice of line segments insures minimal scale mixing.

It is worth noting that a variety of other binning methods could be used.  For example, one could choose the binning system to be data-dependent.  In this system we might define the partition such that each bin contains the same number of data points (this method has useful applications in astrophysics \cite{AstroBin}). In this way the features of the data are reflected in the binning itself.  In this system areas with a high density of points would be covered with many small bins and low-density areas with a few large bins.  Such data-dependent approaches are difficult to apply in a scale-local way, so we avoid them in favor of simpler partitions.

Moving to higher dimensional embedding spaces adds further complexity to the binning problem.  Consider data in a two-dimensional embedding space.  A logical choice for a partition of scale $e$ would be a grid of squares with side length $e$.  However, because the diagonal of the each square bin is naturally $\sqrt{2}e$ this will cause some undesirable mixing of correlations from different scales.  This could be avoided by using circular bins of diameter $e$, but making a compact tiling of circles to cover a 2d embedding space isn't possible.  The best we could do would be to use a tiling of hexagons with side length $\frac{e}{2}$.  This would give us a compact covering of the embedding space while minimizing the sensitivity of the binning to correlations at scales other than $e$.  This would be the ideal solution if computational time and effort was irrelevant.  Unfortunately, it is not.

With a square binning, determining if a data point is in a particular bin is trivial: if the position of the data point is greater than that of the lower bin edge and less than that of the upper bin edge (on every axis) the data point can be said to be contained by the bin in question.  Using a hexagonal binning, or even a rotated rectangular binning, the problem can no longer be trivially factored so that each axis is considered separately.  This would make analysis of data in two dimensions difficult, and higher-dimensional analysis effectively impossible given the restrictions of currently available computing technology.  Thus, we are restricted to a square binning system by the limitations of the computers available to us.  It is not significantly more difficult to extend a square binning system to a rectangular binning system, but for the sake of clarity at this point we consider only square bins.  An example of the application of a rectangular binning can be found in section 2.11.

\section{Relative Binning Placement}

Now that we have selected a square binning to partition the space and calculate the entropy we address the issue of the spatial relationship between the binning and the embedding space in which the data resides.  For a generic distribution there is no way of determining a preferred relationship between the binning and the embedding space.  One might argue that the edges of the bins ought to coincide with the edges of the embedding space.  However, this would place restrictions on the scale points used to calculate the entropy.  We would be forced to use only bin scales that are an integer fraction of the side length $L$ of the embedding space.  

In general there are an infinite number of different ways to lay down the binning relative to the embedding space.  If none of these bin orientations is preferable to the others then we must show that the relative position of the binning to the embedding space is irrelevant, or find a way of incorporating all possible bin positions to avoid biasing our results by making an arbitrary choice.

\begin{figure}[ht]
\centering
\includegraphics[width=4in]{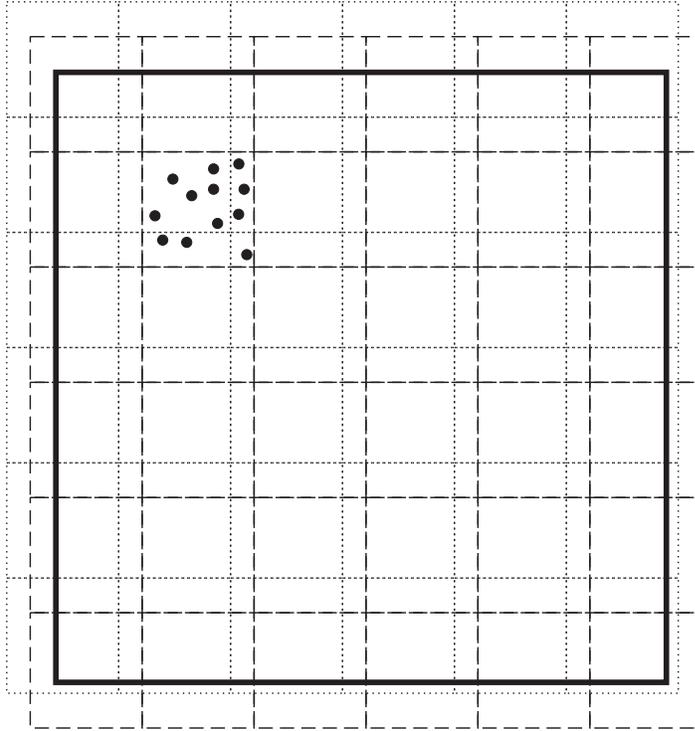}
\caption[An illustration of the importance of bin placement.]{An illustration of the importance of bin placement.  The solid dark line marks the border of the embedding space in which the data resides.  This space is partitioned using a square binning of scale $e$.  Two different relative placements of this same partition system are shown.  The first is placed in such a way that all of the data points are contained in a single bin (dashed lines), this bin placement would yield an entropy $S_0=log [1]=0$.  The second is placed so that four different bins are occupied (dotted lines), this bin placement calculates the entropy of the data set to be $S_0=log [4]=0.60206$.}
\label{DitheringCartoon}
\end{figure}

Consider the example of a data distribution that is much smaller than the embedding space.  If the data distribution contains a bin edge then it will be split among multiple bins.  However, if the binning is placed in precisely the right place then the data will fit entirely in a single bin.  This can make a large difference in the resulting entropy calculation (see figure~\ref{DitheringCartoon}).  Clearly, the relative position of the binning and the embedding space is important to the entropy calculation.    

%
%

This example also illustrates the utility of averaging to incorporate all possible relative binning positions.  The contribution from the rare bin positioning that contains all of the data points is made insignificant by the inclusion of many other binning positions in which the data points are contained in multiple bins.  This gives us a better picture of the correlation content of the data than the calculation of the entropy from a single, arbitrary bin position.

This averaging over different binnings is a very powerful tool.  Consider the example of data in two dimensions being partitioned with circular bins.  We cannot cover the entire embedding space with a single lattice of circles.  If we instead chose to partition the space by placing a very large number of circles randomly we will cover the space many times over.  We can then average over the redundant binnings (divide our results by the ratio of the total area of circles to the area of the embedding space) and we have effectively applied a compact binning using circular bins.  One could envision using a similar approach averaging over fractal bins to try and minimize any bin-edge effects.  With exotic binning approaches such as these we could try to solve many of our analysis problems, but it isn't that simple.  The science of partitioning spaces is non-trivial and a seemingly well-designed partition can have unintended consequences.  Thus, we will avoid exotic solutions in favor of simplicity and focus on the details of applying a simple square partition.

\section{Bin Dithering}

\begin{figure}[ht]
\centering
\includegraphics[width=4.5in]{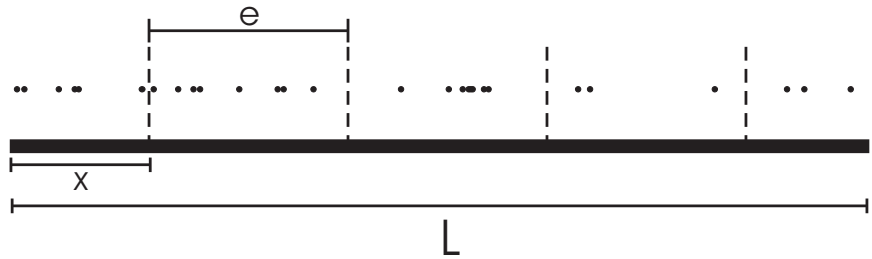}
\caption[A cartoon of a data set in a one-dimensional embedding space that illustrates the relationships among the relevant binning, event, and dithering variables.]{A cartoon of a data set (black dots) in a one-dimensional embedding space (thick solid line) that illustrates the relationships among the relevant binning, event, and dithering variables.}
\label{DitherVariables1d}
\end{figure}

In the previous section we showed that to maintain the generality of the binning we must average over all possible translations of the bin edges with respect to the embedding space.  We do this with a process we call {\bf dithering}.  First, we define a dithering phase $\alpha$ to be $\frac{x}{e}$ where $x$ is the distance from the lower edge of the embedding space to the nearest bin edge. $e$ is the size of a bin (distance between bin edges) and $L$ is the size of the embedding space (see figure~\ref{DitherVariables1d}).  

%
%

All of the unique relationships between the partition and embedding space (along this axis) are represented by taking $\alpha$ from $0$ to $1$.  In general, we will need $d$ such dithering phases for data existing in a $d$-dimensional embedding space.  It is important to note that $\alpha \in [0,1)$ and behaves like a phase.  This is a consequence of the definition $x \equiv 0$ when the bin edge coincides with the edge of the embedding space.  With the goal of simplifying the dithering algebra we define $\phi=1-\alpha$ ($\phi \in (0,1]$) and use $\phi$ as the relevant dithering phase variable.  The utility of this substitution can be seen in section 2.11 where we derive an analytical form for the entropy of an ideal uniform distribution.  

To properly incorporate bin dithering into entropy calculations we must take the dithering phase average over the number of occupied bins.  According to this prescription we calculate $M_{\phi}(e)$ of an event $J$ times at each scale, where each time we increment $\phi$ by $\frac{1}{J}$, and average over these results to get the entropy for this event at scale $e$.  Thus, we must now write the entropy $S_0$ as the log of the $\phi$-averaged number of occupied bins: $S_0(e)=log[\frac{1}{J}\sum_{\phi}M_{\phi}(e)]=log[<M(e)>_{\phi}]$.  

This treatment only applies to translations of the binning lattice.  Even though we want to carefully consider and include all possible binning positions and orientations, we will not consider rotations of the binning system relative to the embedding space.  Because of the rotational asymmetry of the square bins this would lead to a more complicated scale mixing than we already have, and would prohibitively increase the computational task.

\section{Measuring Multiparticle Correlations}

Near a phase boundary in a classical bulk-matter system we expect to see clusters and droplets form as the number of degrees of freedom available to the system changes.  These clusters and droplets can vary significantly in size and density, depending on the physics of the relevant phase transition.  Thus, we must insure that our analysis is sensitive to arbitrary multiparticle correlations.  To incorporate sensitivity to correlated clusters involving $q$ particles (known as {\bf $q$-tuples}) we look to existing analysis tools to avoid needlessly creating new language.  The analysis and quantification of information is common to many disciplines and we must build a necessary and sufficient language that does not replace existing methods unless absolutely necessary.  All of the tools and language we are developing must be consistent with existing knowledge, as with the generalization of entropy to a scale-local form.

To calculate the entropy of a $q$-tuple ensemble we refer to the work of Hungarian mathematician Alfred R\'enyi.  R\'enyi formulated a general approach to the calculation of entropy \cite{Renyi} that is directly applicable to our problem.  Incorporating the scale generalization and bin dithering we can write the rank-$q$ R\'enyi entropy as:  
\begin{eqnarray}
S_q(e)&=&\frac{1}{1-q}log[<\sum_{i=1}^{M_{\phi}(e)}p_{i,\phi}^q(e)>_{\phi}]. 
\end{eqnarray}

Where $p_{i,\phi}(e)$ is the {\bf bin probability} of the $i$th bin at dithering phase $\phi$, which varies as a function of the scale $e$ of the binning.  This bin probability is simply the normalized {\bf bin occupancy}, or number of data points that occupy the $i$th bin ($n_i$) divided by the total number of data points present in the embedding space ($N$).  $p_i$ is defined so that $\sum_ip_i=1$ at a given scale $e$ and dithering angle $\phi$, with the sum taken over all bins in the partition.  Of course, for bins that contain no points ({\bf void bins}) $p_i=0$ and these bins do not contribute to the sum.  We can represent this explicitly by taking the sum to be over only the $M_{\phi}(e)$ bins that are occupied.  

The argument of the logarithm in the R\'enyi entropy can be more generally written as a normalized correlation integral $C_q(e)$.  In this case we can use the approximation $C_q(e) \approx <\sum_{i=1}^{M(e)}p_i^q(e)>_{\phi}$.  The reader will find the more general correlation integral approach completely described in \cite{SLTM}.  For this treatment it is sufficient to adopt the scale-generalized R\'enyi entropy given above.  

The rank-0 R\'enyi entropy is exactly the entropy we have used in previous examples:
\begin{eqnarray}
S_0(e)&=&\frac{1}{1-0}log[<\sum_{i=1}^{M_{\phi}(e)}p_{i,\phi}^0(e)>_{\phi}] \\ \nonumber
&=&log[<\sum_{i=1}^{M_{\phi}(e)}1>_{\phi}] \\ \nonumber
&=&log[<M_{\phi}(e)>_{\phi}].
\end{eqnarray}
Expressing $S_0$ in this form instead of in the R\'enyi form makes its meaning much clearer.  It is a special entropy because it measures only the number of occupied bins as a function of scale and is insensitive to interparticle correlations within each individual bin.  

The other case that must be considered separately is $q=1$:
\begin{eqnarray}
S_1(e)&=&\frac{1}{1-1}log[<\sum_{i=1}^{M_{\phi}(e)}p_{i,\phi}^1(e)>_{\phi}] \\ \nonumber
&=&\frac{log[1]}{0}.
\end{eqnarray}

This calls for an application of L'H\^opital's rule:
\begin{eqnarray}
\lim_{q\rightarrow 1}S_q(e)&=&\lim_{q\rightarrow 1}\frac{log[<\sum_{i}p_{i,\phi}^q(e)>_{\phi}]}{1-q} \\ \nonumber
&=&\lim_{q\rightarrow 1}\frac{\frac{\delta}{\delta q}\{log[<\sum_{i}p_{i,\phi}^q(e)>_{\phi}]\}}{\frac{\delta}{\delta q}\{1-q\}} \\ \nonumber
&=&\lim_{q\rightarrow 1}\frac{-1}{ln[10]}\frac{\delta}{\delta q}\{ln[<\sum_{i}p_{i,\phi}^q(e)>_{\phi}]\} \\ \nonumber
&=&\lim_{q\rightarrow 1}\frac{-1}{ln[10]}\frac{1}{<\sum_{i}p_{i,\phi}^q(e)>_{\phi}}<\sum_{i}\frac{\delta}{\delta q}\{p_{i,\phi}^q(e)\}>_{\phi} \\ \nonumber
&=&\lim_{q\rightarrow 1}\frac{-1}{ln[10]}\frac{<\sum_{i}p_{i,\phi}^q(e)\:ln[p_{i,\phi}(e)]>_{\phi}}{<\sum_{i}p_{i,\phi}^q(e)>_{\phi}} \\ \nonumber
&=&-\frac{<\sum_{i}p_{i,\phi}^1(e)\:log[p_{i,\phi}(e)]>_{\phi}}{<\sum_{i}p_{i,\phi}^1(e)>_{\phi}} \\ \nonumber
&=&-<\sum_{i}p_{i,\phi}(e)\:log[p_{i,\phi}(e)]>_{\phi}. \\ \nonumber
\end{eqnarray}

Thus, we can write $S_1$ in a form that is equivalent to the Shannon entropy \cite{Info}:
\begin{eqnarray}
S_1(e)&=&-<\sum_{i=1}^{M_{\phi}(e)}p_{i,\phi}(e)\:log[p_{i,\phi}(e)]>_{\phi}. \\ \nonumber
\end{eqnarray}

For all $q > 1$ the R\'enyi entropy is well behaved, and most clearly expressed in the form written in equation (2.1).

\section{Scale-local Information}

Now that we have found an expression for the scale-local entropy we can formulate other scale-local measures.  We start with the information.  It is defined as the difference between two entropies, so given the scale-local R\'enyi entropy the scale-local information must be:
\begin{eqnarray}
I_q(e)&=&S_{q,obj}(e)-S_{q,ref}(e).
\end{eqnarray}

We have chosen to write the entropies as $S_{q,obj}$ and $S_{q,ref}$ to reflect the interpretation of the information as we will be applying it.  When we calculate the information we are making a comparison between the data or {\bf object} distribution and the results we would expect from some model that we adopt as a {\bf reference}.  This allows us to be model-explicit, whenever we calculate the relative information contained in some data distribution we must explicitly choose the model used to generate the reference.  The selection of a reference is a key issue in this analysis, so we will explore it in detail in later sections.

\section{Scale-local Dimension}

In addition to writing entropy, volume, and information as functions of scale we can also express the dimension of a distribution as a function of scale.  We start with a standard definition of the dimension from \cite{Gollub}: 
\begin{eqnarray}
d_q&=&\frac{1}{q-1}\lim_{e\rightarrow 0} \Biggl[ \frac{log [\sum_{i=1}^Mp_i^q]}{log [e]} \Biggr] \\ \nonumber
&=&\lim_{e\rightarrow 0} \Biggl[ -\frac{S_q(e)}{log [e]} \Biggr].
\end{eqnarray}
In this definition $M$ is the number of phase space elements required to cover the distribution (the number of bins in the support), $p_i$ is the normalized bin probability measure, and $e$ is the characteristic scale of the binning used to partition the phase space.  In this standard approach, the dimension of a set is measured only in the limit of a zero-scale partition.  As we did with the entropy, we can loosen this unnecessary restriction.  After making an obvious substitution for the scale-local entropy we apply the definition of the derivative and rewrite the dimension as a scale-local function in terms of the scale derivative of the entropy:
\begin{eqnarray}
d_q(e)&=&-\frac{\partial S_q(e)}{\partial log[e]}.
\end{eqnarray}

Treating dimension as a scale-local function is a novel consequence of a scale-local entropy.  To understand how the scale of a measurement can effect the perceived dimension of an object, consider the dimension of the planet Mercury.  From the perspective of an observer on Earth, Mercury appears to be a single point (an object with zero dimension) to the naked eye.  If we instead observe Mercury with a telescope that is orbiting the Earth we will increase the resolution of our observation (decrease the binning size) significantly.  Mercury will now appear to be a small disc, betraying the fact that it has a rich internal structure that would not be present in a single point.  In this regime the planet appears to be a two-dimensional object.  Now consider the observations made by a probe orbiting Mercury itself.  Since the orbiting probe moves around the planet it is clear that Mercury is a three-dimensional object in this regime.  If the probe lands on the planet and moves around a small region of the surface, the planet will look essentially flat (two-dimensional again).  This is the region of scale occupied by the ``flat Earth" society.  Finally, if the probe does an analysis of the surface material on an atomic scale, it will find that the planet is made of atoms and molecules that appear to be individual points.  Thus, at this scale the planet is seen to be a collection of disconnected points and its dimension returns to zero.

\section{A Simple Example}

Now that we have done the work of deriving an appropriate form for the scaled entropy and related topological measures, we should look at a simple application of this analysis to gain insight into the meaning of its results.  We start by calculating the scaled entropy and dimension (for a few different $q$ values) of a randomly generated 1d uniform distribution of $N$ points. 
%
\begin{figure}[ht]
\centering
\includegraphics[width=4in]{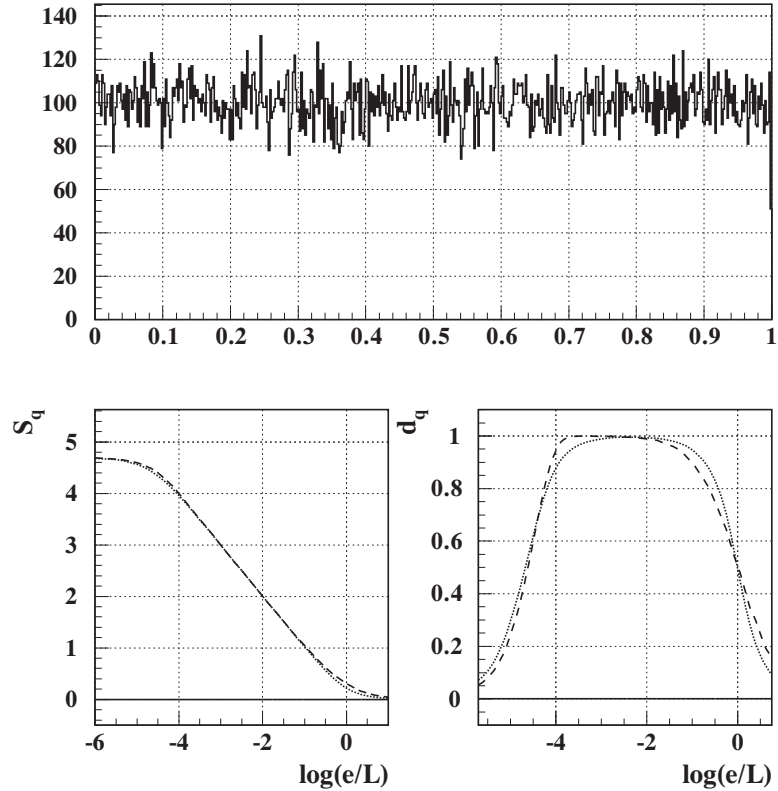}
\caption[Results of scaled correlation analysis applied to a distribution of 50,000 uniformly thrown random points in 1d.]{Results of scaled correlation analysis applied to a distribution of 50,000 uniformly thrown random points in 1d.  The distribution itself is shown in the top panel.  The bottom panels show the entropy and dimension for $q=0$ (dashed lines) and $q=1$ (dotted lines).}
\label{1dRGUD}
\end{figure}
%

To understand these results, consider the small, intermediate, and large scale regimes.  In the small scale limit ($e << \frac{L}{N}$) each of the $N$ points in the data distribution occupies it's own bin, so there are N occupied bins ($M=N$) each with bin probability $p_i=\frac{1}{N}$.  Thus, the entropy approaches $log[N]$ as the bin size becomes vanishingly small:
\begin{eqnarray}
\lim_{e\rightarrow 0}S_q(e)&=&\frac{1}{1-q}log[\sum_{i=1}^{N}(\frac{1}{N})^q] \\ \nonumber
&=&\frac{1}{1-q}log[N^{1-q}] \\ \nonumber
&=&log[N]. 
\end{eqnarray}
This result is extremely general, being independent of the details of the correlations present in the data.  This makes it independent of the rank ($q$) of the entropy being calculated and the dimension ($d$) of the embedding space in which the data resides.

At intermediate scales where the details of the correlations present in the data are probed by the binning we cannot make a convenient general statement about the expected results.  However, if we idealize our uniform $N$ point 1d data distribution to one that is perfectly dense ($N\rightarrow \infty$) and of infinite extent ($L\rightarrow \infty$) then we know that at all scales the dimension of the distribution will be one.  This is the same as saying that the dimension of a line is one.  This generalization is also independent of the rank of the entropy calculated, but depends totally on the dimension of the embedding space.  Thus, at intermediate scale regions where the point-like nature of this data is invisible to the binning (there are no void bins), the slope of the entropy distribution must be one.

At large scales ($e >> 1$) the entire distribution fits in a single bin, so $M=1$ and $p_i=1$.  This yields the general result that the rank-$q$ entropy vanishes (for a bounded distribution) as the bin size approaches infinity.

To recap these results, the entropy and dimension are zero in the large-scale regime where the data looks like a single point.  As the scale decreases and we pass through the scale of the embedding space (the unit interval; $e = 1$) we begin to see structure in the data.  In this mid-scale regime, we find the entropy has a slope of one because the data behaves like a dense 1d distribution at these scales.  This continues until the scale of the bins gets near the mean interparticle spacing of the distribution ($L/N$).  At this point we begin to see the space between the points and find an increasing number of void bins as the scale continues to decrease.  Finally, we reach the point where each of the $N$ points is in its own bin and the entropy levels off at $log[N]$.  Thus, the results of figure~\ref{1dRGUD} make sense algebraically and intuitively.

\section{Selecting a Reference}

To calculate the relative information between a data set and a reference, we must first select the reference distribution we wish to use as a model and calculate its entropy.  There are two distinct approaches that one can take in selecting a reference.  

%
\begin{figure}[ht]
\centering
\includegraphics[width=4in]{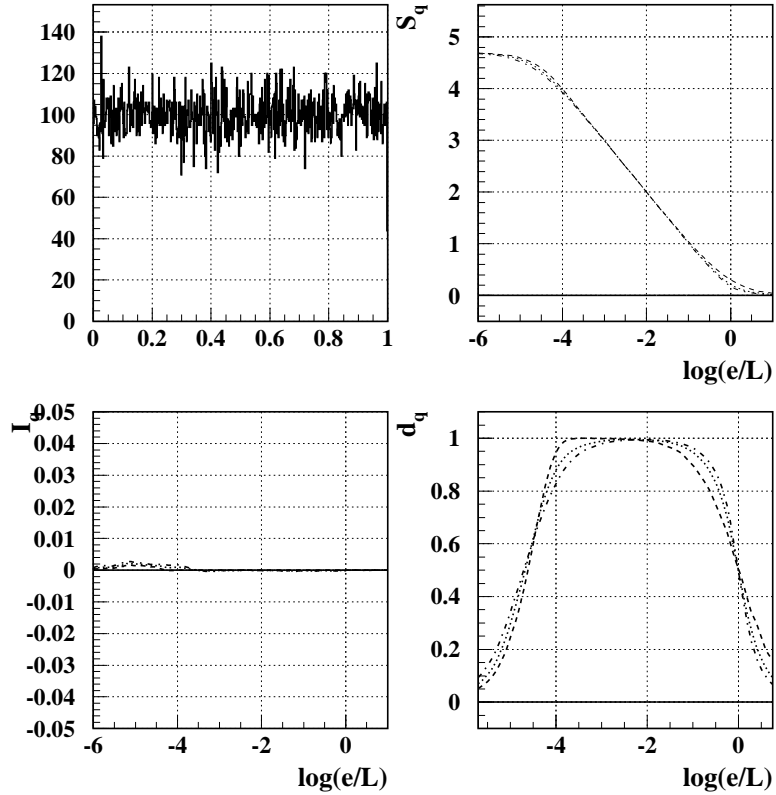}
\caption[SCA results for a 1d randomly generated uniform distribution of 50k points using the data analyzed in figure~\ref{1dRGUD} as a reference.]{SCA results for a 1d randomly generated uniform distribution of 50k points using the data analyzed in figure~\ref{1dRGUD} as a reference.  The distributions are identical in the large- and mid-scale regimes, only differing at scales in the vicinity of $-4.7$ where the differences in the distributions arising from statistical fluctuations are visible.  The curves are $q=0$ (dashed), $q=1$ (dotted), and $q=2$ (dot-dashed).}
\label{1d50k}
\end{figure}
%

Our first approach to generating a reference is to use an event as the reference model.  In this simple example, we will use the data from the analysis of the previous section (see figure~\ref{1dRGUD}) as our reference event.  We will create a data event by randomly generating a uniform distribution using the same process used to generate the reference.  Because they contain the same correlation content, the only difference between these two distributions will be from statistical fluctuations, which appear only at small scales as in figure~\ref{1d50k}.

\begin{figure}[ht]
\centering
\includegraphics[width=4in]{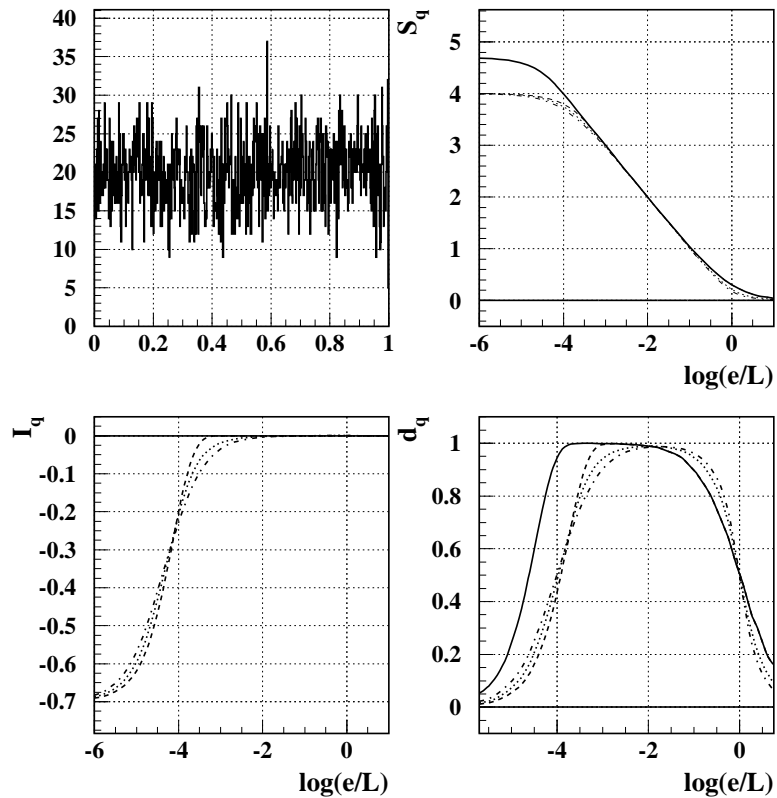}
\caption[SCA results for a 1d randomly generated uniform distribution of 10k points using the data analyzed in figure~\ref{1dRGUD} as a reference.]{SCA results for a 1d randomly generated uniform distribution of 10k points using the data analyzed in figure~\ref{1dRGUD} as a reference.  The distributions are identical at large scales, but diverge rapidly when the difference in event multiplicities becomes visible (near $log[e/L]=-4$).  In the small-scale limit the relative information is $I_q=log[10,000]-log[50,000] \approx -0.7$.  The solid lines show the $q=0$ results for the reference distribution.}
\label{1d10k50r}
\end{figure}

In the previous example we have chosen the multiplicity of the data and reference events to be the same.  If we instead choose events with differing multiplicities, we get very different results (see figure~\ref{1d10k50r}).
%
%
As one might expect, the information calculated for these two different distributions is essentially the same in the large- and mid-scale regimes.  It is only in the small-scale region, where the difference in the multiplicities becomes visible to the binning, that the information becomes significantly non-zero.  This is good to know because we have no control over the exact multiplicities of the events that come out of an actual experiment, and we would like to have the ability to make viable event comparisons.  As long as we keep in mind that the difference in event multiplicities effects the information in the small-scale region increasingly with the magnitude of the difference, we can make meaningful comparisons between events of differing multiplicity.

The second approach to generating a reference is to use a model that allows us to calculate $S_{ref_q}(e)$ analytically.  The most basic (and most useful) distribution for which the entropy can be solved analytically is the ideal uniform distribution (IUD).  This is the distribution that will have the maximum entropy for a given multiplicity because it uniformly fills the embedding space.  Because the IUD represents the maximum entropy hypothesis, using it as a reference yields an absolute measure for the information contained in a given event.

\section{Derivation of $S_q(e)$ for an Ideal Uniform Distribution}

The scaled entropy of an ideal uniform distribution is special because it represents a maximum entropy hypothesis.
This is the $N \rightarrow \infty$ ideal to which a uniform, random distribution aspires; the maximal filling of the embedding space.  In some sense, this is the distribution that contains the least information.  This makes it the logical choice for a benchmark from which any correlated distribution will vary.  

\begin{figure}[ht]
\centering
\includegraphics[width=4in]{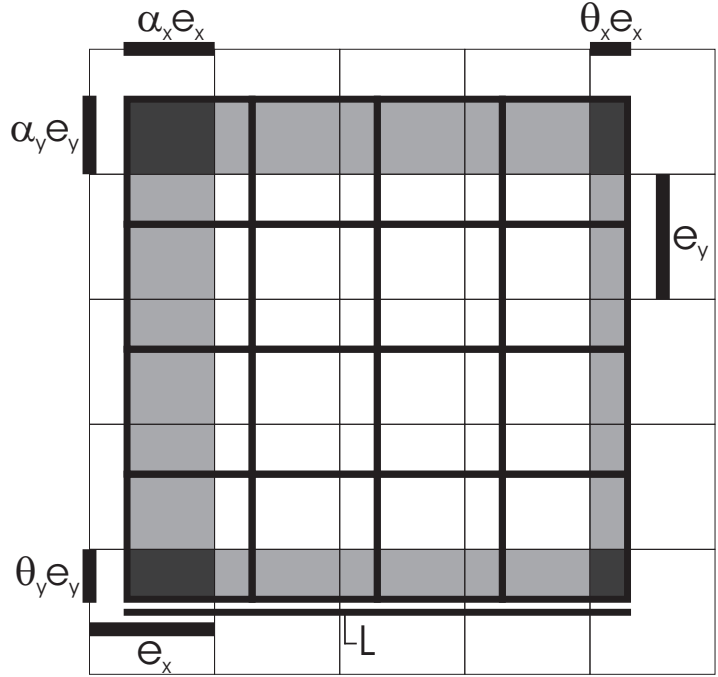}
\caption{A binning cartoon that illustrates the relationships among the relevant binning, event, and dithering variables in two dimensions.}
\label{DitherVariables2d}
\end{figure}

To derive an analytic form for the scaled entropy ($S_q(e)$) of an IUD we will consider a two-dimensional distribution (and then extrapolate to $d$ dimensions).  This is an ideal uniform distribution that exists in a square embedding space of side length L.  The unique feature of the IUD is that it is distributed totally evenly throughout the embedding space.  Thus, the probability of finding a portion of data from the distribution in any given bin is just a function of the bin size.  
The cartoon in figure~\ref{DitherVariables2d} shows the IUD binned with a general rectangular binning (thin dark lines).  We need to calculate the bin contents for each bin (as a function of scale) and integrate over all possible dithering configurations to determine the analytic form for the entropy.  
%
%
For bins that are entirely contained within in the embedding space (interior bins), calculating the normalized bin contents (bin probability) is trivial.  For these bins, $p_i$ is just the area of the bin divided by the total area of the data, $\frac{e_xe_y}{L^2}$.  This is independent of any bin dithering, although the dithering must be taken into account in determining the total number of interior bins present.  For edge and corner bins the problem is more difficult, so we will consider each type of bin, interior (white), edge (grey) and corner (dark grey) separately.  

First consider the corner bins.  These are bins that contain both an x- and y-axis embedding space edge.  For all dithering configurations there are always four corner bins.  The effective area of these corner bins is $e_xe_y$ scaled down by $f_xf_y$ where $f_i$ is the fraction of the bin along the $i$th axis that overlaps the data.  By definition, the dithering phase along an axis is $\phi_i=(1-\alpha_i)$.  Furthermore, we define $\Delta_i\equiv \frac{L}{e_i}-int(\frac{L}{e_i})$ for convenience.  Thus, we can now express the amount of overlap between the last bin on axis i and the edge of the data along that axis as $\theta_i=\phi_i+\Delta_i-int(\phi_i+\Delta_i)$.  With these definitions, calculating the contribution to the correlation integral of the corner bins is a matter of integrating over all $\phi$ values using the relevant bin probabilities.  Labeling the corner bins from right to left starting with the upper left bin, we can write down the $\phi$-dependent corner bin probabilities:
\begin{eqnarray}
p_{A}^q&=&(1-\phi_x)^q(1-\phi_y)^q(\frac{e_xe_y}{L^2})^q \\ \nonumber
p_{B}^q&=&(1-\phi_x)^q[\phi_y+\Delta_y-I(\phi_y+\Delta_y)]^q(\frac{e_xe_y}{L^2})^q \\ \nonumber
p_{C}^q&=&[\phi_x+\Delta_x-I(\phi_x+\Delta_x)]^q(1-\phi_y)^q(\frac{e_xe_y}{L^2})^q \\ \nonumber
p_{D}^q&=&[\phi_x+\Delta_x-I(\phi_x+\Delta_x)]^q[\phi_y+\Delta_y-I(\phi_y+\Delta_y)]^q(\frac{e_xe_y}{L^2})^q.
\end{eqnarray}
We can calculate the dithering-averaged correlation integral $C_q=<\sum_i p_i^q>_{\phi}$ by integrating over the different dithering configurations for each bin and summing ($<\sum_i p_i^q>_{\phi}=\sum_i <p_i^q>_{\phi}$).  Following this approach we simply need to integrate the $p_i^q$ expressions over the two dithering variables and sum their results to calculate the correlation integral.  The $\phi$ integral over the first corner bin yields:
\begin{eqnarray}
<p_{A}^q>_{\phi}&=&\int_0^1\int_0^1(1-\phi_x)^q(1-\phi_y)^q(\frac{e_xe_y}{L^2})^q d\phi_x d\phi_y \\ \nonumber
&=&(\frac{e_xe_y}{L^2})^q(\frac{1}{1+q})^2.
\end{eqnarray}

The second term is a bit trickier:
\begin{eqnarray}
<p_{B}^q>_{\phi}&=&[\int_0^1\int_0^1 (1-\phi_x)^q[\phi_y+\Delta_y-I(\phi_y+\Delta_y)]^q d\phi_x d\phi_y][(\frac{e_xe_y}{L^2})^q] \\ \nonumber
&=&(\frac{e_xe_y}{L^2})^q(\frac{1}{1+q})[\int_{\Delta_y}^1 v^q dv+\int_0^{\Delta_y} v^q dv] \\ \nonumber
&=&(\frac{e_xe_y}{L^2})^q(\frac{1}{1+q})^2.
\end{eqnarray}
The third and fourth terms are similar to the first and second, so we see that each of the four corner bins contributes a term of $(\frac{e_xe_y}{L^2})^q(\frac{1}{1+q})^2$ to the total dithering-averaged correlation integral.  

The contribution from edge bins is simpler to calculate than the corner bins.  The overlap fraction $f_i=1$ in the direction parallel to the edge, so along that axis we simply need to count the number of edge bins.  The integral along the other axis is the same as those from the corner bin calculations.  Again there are four terms, but again there is a symmetry that simplifies the problem:
\begin{eqnarray}
<p_{edge}^q>_{\phi}&=&\int_0^1\int_0^1 (\frac{e_xe_y}{L^2})^q[I(\frac{L}{e_x}+1+\phi_x)-2](1-\phi_y)^q d\phi_x d\phi_y \\ \nonumber
&=&(\frac{e_xe_y}{L^2})^q(\frac{L}{e_x}-1)(\frac{1}{1+q}) \\ \nonumber
&=&\int_0^1\int_0^1 (\frac{e_xe_y}{L^2})^q[I(\frac{L}{e_x}+1+\phi_x)-2][\phi_y+\Delta_y-I(\phi_y+\Delta_y)]^q d\phi_x d\phi_y \\ \nonumber
&=&(\frac{e_xe_y}{L^2})^q(\frac{L}{e_x}-1)(\frac{1}{1+q}).
\end{eqnarray}
This is for the x-axis border bins, of course calculating the contributions from the y-axis bins is a simple matter of switching the indices.  Thus, the contribution from border bins is:
\begin{equation}
2(\frac{e_xe_y}{L^2})^q[(\frac{L}{e_x}-1)(\frac{1}{1+q})+(\frac{L}{e_y}-1)(\frac{1}{1+q})].
\end{equation}

The only piece of the correlation integral calculation remaining is the integral over interior bins:
\begin{equation}
\int_0^1\int_0^1 (\frac{e_xe_y}{L^2})^q[I(\frac{L}{e_x}+1+\phi_x)-2][I(\frac{L}{e_y}+1+\phi_y)-2] d\phi_x d\phi_y = (\frac{e_xe_y}{L^2})^q(\frac{L}{e_x}-1)(\frac{L}{e_y}-1).
\end{equation}
Now that we have calculated the contributions from all bins the full correlation integral can be calculated:
\begin{eqnarray}
C_q(e)&=&<\sum_i p_i^q>_{\phi}=\sum_i <p_i^q>_{\phi} \\ \nonumber
&=&[(\frac{e_xe_y}{L^2})^q][4(\frac{1}{1+q})^2+2(\frac{1}{1+q})(\frac{L}{e_x}-1)+2(\frac{1}{1+q})(\frac{L}{e_y}-1)+(\frac{L}{e_x}-1)(\frac{L}{e_y}-1)] \\ \nonumber
&=&[(\frac{e_xe_y}{L^2})^q][2(\frac{1}{1+q})+(\frac{L}{e_x}-1)][2(\frac{1}{1+q})+(\frac{L}{e_y}-1)] \\ \nonumber
&=&[(\frac{e_xe_y}{L^2})^{q-1}][1+(\frac{1-q}{1+q})(\frac{e_x}{L})][1+(\frac{1-q}{1+q})(\frac{e_y}{L})].
\end{eqnarray}
Plugging this result back into the definition of the rank-$q$ entropy, we find that:
\begin{eqnarray}
S_q(e)&=&\frac{1}{1-q} log[C_q(e)] \\ \nonumber
&=&\frac{q-1}{1-q} log(\frac{e_xe_y}{L^2})+\frac{1}{1-q} log[1+(\frac{1-q}{1+q})(\frac{e_x}{L})]+\frac{1}{1-q} log[1+(\frac{1-q}{1+q})(\frac{e_y}{L})] \\ \nonumber
&=&log(\frac{L^2}{e_xe_y})+\frac{1}{1-q}(log[1+(\frac{1-q}{1+q})(\frac{e_x}{L})]+log[1+(\frac{1-q}{1+q})(\frac{e_y}{L})]) \\ \nonumber
&=&\Biggl[log(\frac{L}{e_x})+\frac{1}{1-q}log[1+(\frac{1-q}{1+q})\frac{e_x}{L}]\Biggr]+\Biggl[log(\frac{L}{e_y})+\frac{1}{1-q}log[1+(\frac{1-q}{1+q})\frac{e_y}{L}]\Biggr].
\end{eqnarray}

Because of the additivity of the log of a product we are able to break the entropy up in this case into two terms, one for each axis.  This is suggestive of the $d$-dimensional generalization, namely simply adding another term for each axis that is relevant to the problem.  Thus, we can generate the scaled entropy for a $d$-dimensional ideal uniform distribution based on an extrapolation of the 2d result.

\section{Derivation Limitations} 

We have found the rank-$q$ scaled entropy of an IUD (along a single axis) to be:
\begin{equation}
S_q(e)=log(\frac{L}{e})+\frac{1}{1-q}log[1+(\frac{1-q}{1+q})\frac{e}{L}]
\end{equation}
in the scale regions where interior bins are relevant.  In the derivation of equation 2.18 we considered contributions from three types of bins: corner, border, and interior.  In the 2d example used in the derivation this is only appropriate when both $e_x \le L$ and $e_y \le L$.  If either of the bin scales is larger than the embedding space then there are no interior bins along that axis and 2.18 does not apply.

To derive the entropy for the IUD at large scales consider a single axis (we have shown the entropy to be factorizable so we will now take advantage of that fact) with $e \ge L$.  In this case, when there is a bin edge in the embedding space we have exactly two corner bins, so we can express the bin probability of the second bin in terms of the first:
\begin{eqnarray}
p_1^q&=&[(\alpha)(\frac{e}{L})]^q \\ \nonumber
p_2^q&=&[1-p_A]^q.
\end{eqnarray}
When there is no bin edge in the embedding space then the entire space fits inside a single bin, and so $p_0=1$.  This occurs when the distance from the edge of the embedding space to the nearest bin edge is larger than the size of the embedding space ($\alpha e \ge L$), thus we can write down and solve the relevant integrals:
\begin{eqnarray}
<p_{0}^q>_{\alpha}&=&\int_{\frac{L}{e}}^1 [1]^q d\alpha \\ \nonumber
&=&1-\frac{L}{e},
\end{eqnarray}
\begin{eqnarray}
<p_{1}^q>_{\alpha}&=&\int_0^{\frac{L}{e}} \alpha^q(\frac{e}{L})^q d\alpha \\ \nonumber
&=&(\frac{e}{L})^q\frac{1}{1+q}(\frac{L}{e})^{q+1}=\frac{L}{e}(\frac{1}{1+q}),
\end{eqnarray}
\begin{eqnarray}
<p_{2}^q>_{\alpha}&=&\int_0^{\frac{L}{e}} [1-\alpha(\frac{e}{L})]^q d\alpha \\ \nonumber
&=&\frac{L}{e}\int_0^1 u^q du = \frac{L}{e}(\frac{1}{1+q}).
\end{eqnarray}

Since the definition of the corner bins is arbitrary, we could relabel them and get the same results.  This symmetry requires that $<p_{1}^q>_{\alpha}=<p_{2}^q>_{\alpha}$, as we have confirmed analytically.  These results can now be put together to calculate the scaled entropy:
\begin{eqnarray}
S_q(e)&=&\frac{1}{1-q} log[C_q(e)] \\ \nonumber
&=&\frac{1}{1-q} log[1-\frac{L}{e}+\frac{L}{e}(\frac{2}{1+q})] \\ \nonumber
&=&\frac{1}{1-q} log[1+\frac{L}{e}(\frac{2}{1+q}-1)] \\ \nonumber
&=&\frac{1}{1-q} log[1+(\frac{1-q}{1+q})\frac{L}{e}].
\end{eqnarray}
This result allows us to write down the rank-$q$ scaled entropy for a 1d IUD in all scale regions:
\begin{equation}
S_q(e)=\cases{log(\frac{L}{e})+\frac{1}{1-q}log[1+(\frac{1-q}{1+q})\frac{e}{L}],&if $e \le L$\cr
\frac{1}{1-q} log[1+(\frac{1-q}{1+q})\frac{L}{e}],&if $e \ge L$.\cr}		
\end{equation}
Of course, this result can be trivially extrapolated to the $d$-dimensional case.  Notice that $S_q(e)$ is continuous at $e=L$.  This is useful since it allows us to apply equation 2.8 and take the scale derivative of the entropy to calculate the scaled dimension (details are left as an exercise for the motivated reader).  This yields:
\begin{equation}
d_q(e)=\cases{\frac{1+q(1-\frac{e}{L})}{1+q(1-\frac{e}{L})+\frac{e}{L}},&if $e \le L$\cr
\frac{1}{1-q(1-\frac{e}{L})+\frac{e}{L}},&if $e \ge L$.\cr}		
\end{equation}

The scaled dimension provides a very nice consistency check.  At all scales significantly smaller than the embedding space ($e \ll L$) the scaled dimension of the IUD should reflect the dimension of the embedding space that it fills independent of $q$.  Taking the limit of the scaled dimension as $e \rightarrow 0$ yields:
\begin{equation}
\lim_{e\rightarrow 0}d_q(e)=\frac{1+q}{1+q}=1,
\end{equation}
as it must.  Thus, we have derived believable analytic forms for the relevant topological measures for a 1d IUD, which can be trivially extrapolated to the $d$-dimensional case because of the extensivity of the scaled R\'enyi entropy.

\section{Special Cases}

As we saw earlier the $q=0$ and $q=1$ cases are somewhat special, so we will write them out explicitly.  First, for $q=0$:
\begin{equation}
S_0(e)=log[1+\frac{L}{e}].		
\end{equation}
At the beginning of this chapter we saw that the rank-0 entropy is the log of the number of occupied bins $M_0$.  Because the IUD is perfectly dense (by definition) there are never any void bins.  Thus, in this case $M_0=1+\frac{L}{e}$, the total number of (dithered) bins in the partition at scale $e$.  

As we saw in section 2.6, an application of L'H\^opital's rule was required to derive a well-behaved form for the $q=1$ R\'enyi entropy.  This problem is also present in the analytical form for the entropy of the IUD.  After applying L'H\^opital's rule (again the details are left as an exercise for the motivated reader) one finds:
\begin{equation}
S_1(e)=\cases{log(\frac{L}{e})+\frac{1}{2}\frac{e}{L},&if $e \le L$\cr
\frac{1}{2}\frac{L}{e},&if $e \ge L$.\cr}		
\end{equation}

\section{Adding a Cutoff to the IUD Reference}

To test the results of the previous section we can use the entropy of an IUD as a reference in the SCA analysis of a discrete, randomly generated uniform distribution.  Because the discrete nature of a distribution is invisible at scales much larger than the mean interparticle spacing, we expect to see exact agreement between the IUD reference and the discrete data distribution at medium and large scales.  Only at small scales should there be any disagreement. 

%
\begin{figure}[ht]
\centering
\includegraphics[width=4in]{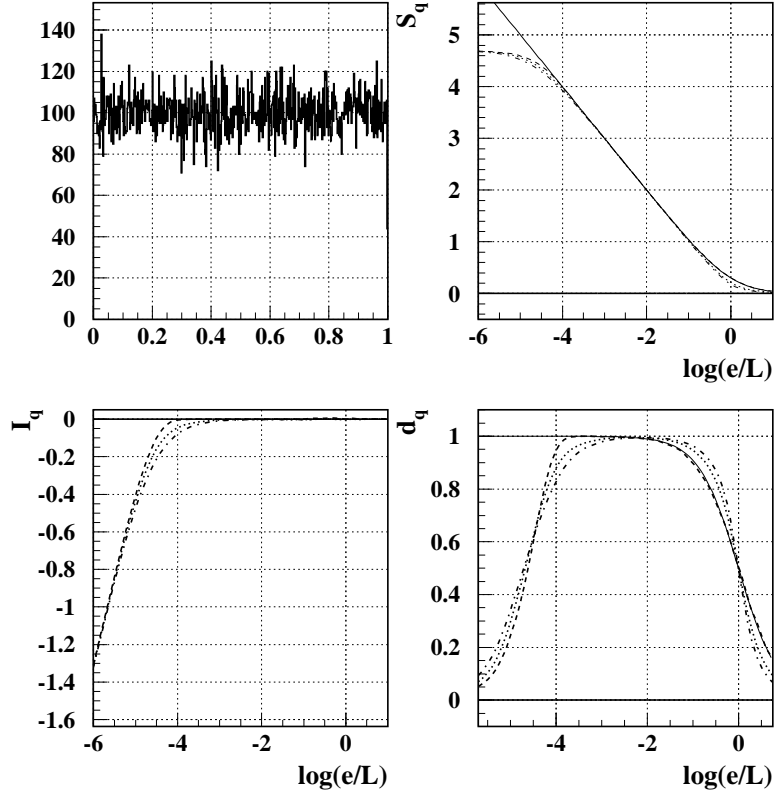}
\caption[SCA results for a 1d randomly generated uniform distribution of 50k points using the IUD entropy as a reference.]{SCA results for a 1d randomly generated uniform distribution of 50k points using the IUD entropy as a reference.  The solid line is the IUD reference for $q=0$, comparable to the dashed line ($q=0$ for the data).  Agreement is perfect at medium and large scales.  At small scales the results diverge as the discrete nature of the data becomes visible.}
\label{fig2.7}
\end{figure}
%

There indeed is significant disagreement between the IUD reference and the randomly generated data results at small scale.  The distribution of ``points" in the IUD is continuous and infinite, whereas in any real distribution of points it is finite and discrete.  Thus, there arises a difference in the entropy at scales where the finite, discrete nature of the non-ideal distribution becomes visible to the binning system.  To successfully make the IUD reference look more like the reference from a finite, discrete distribution we need to incorporate an appropriate cutoff factor into equations 2.24 and 2.25.  

The data distribution in fig 2.7 rises with the IUD reference curve, but then smoothly goes to log N.  We can make the IUD reference mimic this behavior by including a multiplicative factor in the reference entropy which is 1 at large scale and $\frac{N}{M_{q,eff}(e)}$ at small scale, where $M_{q,eff}(e)$ is the effective number of occupied bins in the IUD and is defined by $S_q(e)\equiv log[M_{q,eff}(e)]$.  One simple function that fulfills this criteria is $f(e)=1-e^{-\frac{N}{M_{q,eff}(e)}}$.  At large scale $M_{q,eff}(e) \rightarrow 1$, so $f(e) \rightarrow 1$ for large $N$.  More importantly, at small scale $M_{q,eff}(e) \rightarrow \infty$, so if we expand the exponential term to first order in $M_{q,eff}(e)^{-1}$ then we are left with exactly the term we need:
\begin{eqnarray}
f(e)&=&1-e^{-\frac{N}{M_{q,eff}(e)}} \\ \nonumber
&=&1-[1-\frac{N}{M_{q,eff}(e)}+\frac{1}{2}(-\frac{N}{M_{q,eff}(e)})^2+...] \\ \nonumber\
&\approx&\frac{N}{M_{q,eff}(e)}.
\end{eqnarray}
Thus, incorporating the cutoff factor $f(e)$ into the IUD entropy yields the result
\begin{eqnarray}
S_q(e)&=&log ([M_{q,eff}(e)][1-e^{-\frac{N}{M_{q,eff}(e)}}]),
\end{eqnarray}
where $M_{q,eff}(e)$ is:
\begin{eqnarray}
M_{q,eff}(e)&=&\cases{[\frac{L}{e}][1+(\frac{1-q}{1+q})\frac{e}{L}]^{\frac{1}{1-q}},&if $e \le L$\cr
[1+(\frac{1-q}{1+q})\frac{L}{e}]^{\frac{1}{1-q}},&if $e \ge L$.\cr} 		
\end{eqnarray}
Naturally, we can also calculate the scaled dimension for an IUD with a cutoff:
\begin{equation}
d_q(e)=\cases{\frac{1+q(1-\frac{e}{L})}{1+q(1-\frac{e}{L})+\frac{e}{L}}[1-(\frac{N}{M_{q,eff}(e)})(e^{\frac{N}{M_{q,eff}(e)}}-1)^{-1}],&if $e \le L$\cr
\frac{1}{1-q(1-\frac{e}{L})+\frac{e}{L}}[1-(\frac{N}{M_{q,eff}(e)})(e^{\frac{N}{M_{q,eff}(e)}}-1)^{-1}],&if $e \ge L$.\cr}		
\end{equation}
To test the quality of this cutoff function a comparison of these results to a discrete uniform distribution is in order.

\section{Understanding the IUD Reference}

\begin{figure}[ht]
\centering
\includegraphics[width=4in]{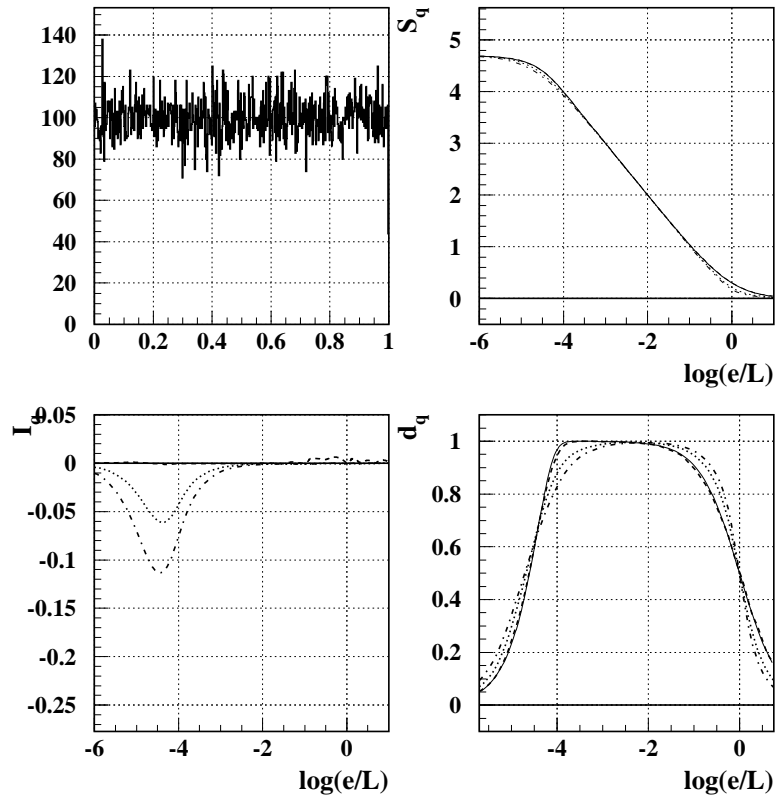}
\caption[SCA results for a 1d randomly generated uniform distribution of 50k points using an IUD with a cutoff as the reference.]{SCA results for a 1d randomly generated uniform distribution of 50k points using an IUD with a cutoff as the reference.  The distributions are again identical at large scales, but diverge near the mean interparticle spacing scale ($\sim$-4.7) where the reference does not model the discrete nature of the data distribution well.}
\label{50kRUD}
\end{figure}

Comparison of the 1d randomly generated uniform distributions analyzed in the previous sections to the analytically derived IUD reference with a cutoff yields a somewhat disappointing result (see figure~\ref{50kRUD}).

%
%

The IUD entropy that includes the cutoff factor $f(e)=1-e^{-\frac{N}{M_{q,eff}(e)}}$ gives a fair approximation to the entropy of a discrete, randomly generated uniform distribution for $q=0$.  However, as $q$ increases this reference increasingly underestimates the presence of void bins.  This leads to an apparent information or {\bf pseudoinformation}.  

For a quantitative understanding of the pseudoinformation effect one must look at the approximation to the correlation integral made by calculating the entropy from a partitioned space.  Ideally, the normalized correlation integral of a data set would be calculated by counting all $q$-tuples of scale $e$ or smaller and dividing by the total number of $q$-tuples in the set.  For computational simplicity we are approximating the correlation integral as $C_q(e) \approx <\sum_{i=1}^{M(e)}p_i^q(e)>_{\phi}$.  The pseudoinformation is a byproduct of this approximation and with a more complicated cutoff function one can get rid of it.  I direct interested readers to an excellent and rigorous description of the mathematics of the pseudoinformation in \cite{SLTM}.

\section{Conclusions}

We have succeeded in our goal of creating a model-independent correlation analysis system that can be made to explicitly incorporate model predictions when appropriate.  By making a scale generalization of the R\'enyi entropy we have found a way to calculate the entropy, volume, information, and dimension of a data set all as a function of scale.  This gives us the power to characterize the scale dependent correlation content in a data distribution and explicitly compare those results to various models and expectations.  

With a little algebra we have derived an analytical form for the scaled topological measures of an idealized (continuous and infinite) uniform distribution.  This reference is important because it represents a maximum entropy hypothesis.  We have also seen the utility of using other data distributions as references.  These analysis methods clearly illuminate the scale regions where the distributions are different, and can give unique insight into the differences in the correlation structure of the events analyzed.  Now that we have built the analytical framework for SCA and analyzed some simple data to further our understanding of the results, we can begin to analyze a variety of more complicated distributions to gain deeper insight into the interpretation of SCA results.

%
%
%

\chapter{SCA Applications: Analysis of Toy Models}

\section{Introduction}

The randomly generated uniform distributions analyzed in the previous chapter provided useful insights into the interpretation of scaled correlation analysis results.  In this chapter we will discuss a variety of computer generated model distributions to further our understanding of the results of the SCA system.  As we will find, a great deal can be learned about the meaning of SCA results by generating simple distributions and analyzing them.  It is essential that we learn how to interpret SCA results if we are to reach meaningful conclusions based on this work.

\section{Scaling the Uniform Distribution}

In the previous chapter we exhaustively analyzed the correlation structure of the one-dimensional randomly generated uniform distribution (RGUD).  In those studies we generated data to fill the embedding space as completely as possible.  Now we will consider an example in which the scale of the data is decreased, but the scale of the embedding space remains fixed.  If our analysis is truly scale-local we should find that increasing the relative size of the space by an order of magnitude will only shift the analysis results down a decade in scale.  The shape of the scaled topological measure curves should remain the same.
%
\begin{figure}[ht]
\centering
\includegraphics[width=4in]{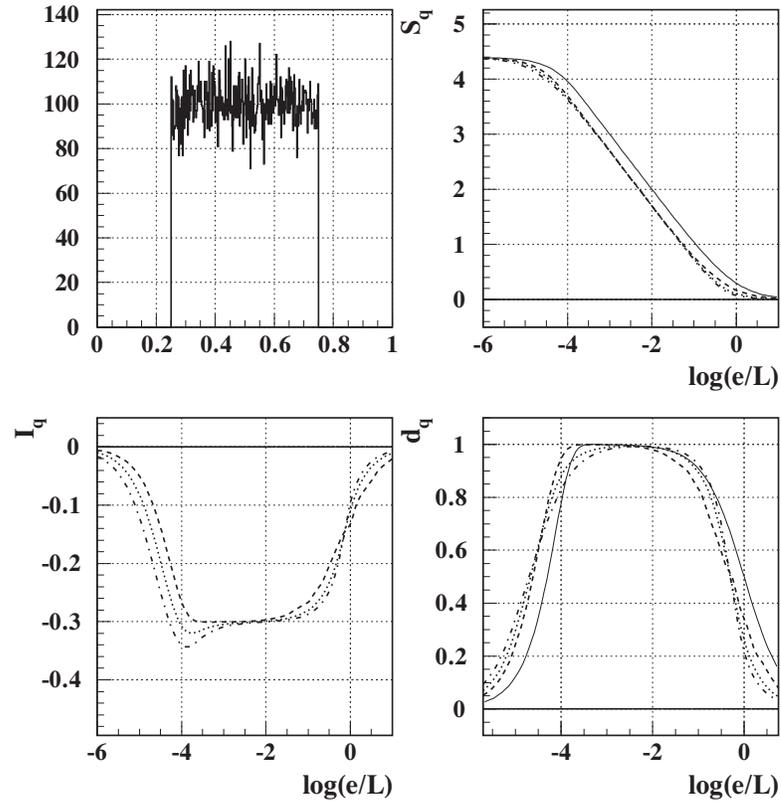}
\caption[A 1d randomly generated uniform distribution scaled down by a factor of 2 relative to the embedding space.]{A 1d randomly generated uniform distribution scaled down by a factor of 2 relative to the embedding space.}
\label{1dRGUD/2}
\end{figure}
%

In the example shown in figure~\ref{1dRGUD/2} a one-dimensional randomly generated uniform distribution is scaled down by a factor of two relative to the embedding space.  Comparing these results to those of figure~\ref{1dRGUD}, it is clear that the form of the topological measures is the same.  The only difference is a scale shift of the entropy by -0.3 corresponding to the base 10 logarithm of the distribution width scaling factor (0.5).  The relative information between the scaled and unscaled uniform distributions shows us exactly what their relative widths are.  Thus, the information can be interpreted as a measure of the {\bf effective width} of the object distribution relative to the reference.  In this example the scale shift is independent of the rank ($q$) of the information calculation, but this is not true in general.  As we will see in the next example the rank-$q$ information is really a measure of the rank-$q$ effective width.

\section{A Simple Correlation Example}

To begin to understand how the SCA system works on data with interesting (non-trivial) correlations we begin by analyzing a data distribution with a single correlation feature constructed at a specific scale.  Consider two uniform one-dimensional point distributions on the unit interval, one with $N_1$ points, and the second containing $N_2$ points.  We can create a third distribution from these two by removing the center half of the second distribution and replacing the center of the first distribution with it.  This distribution has a single correlation feature at a scale half the size of the embedding space.  

\begin{figure}[ht]
\centering
\includegraphics[width=4in]{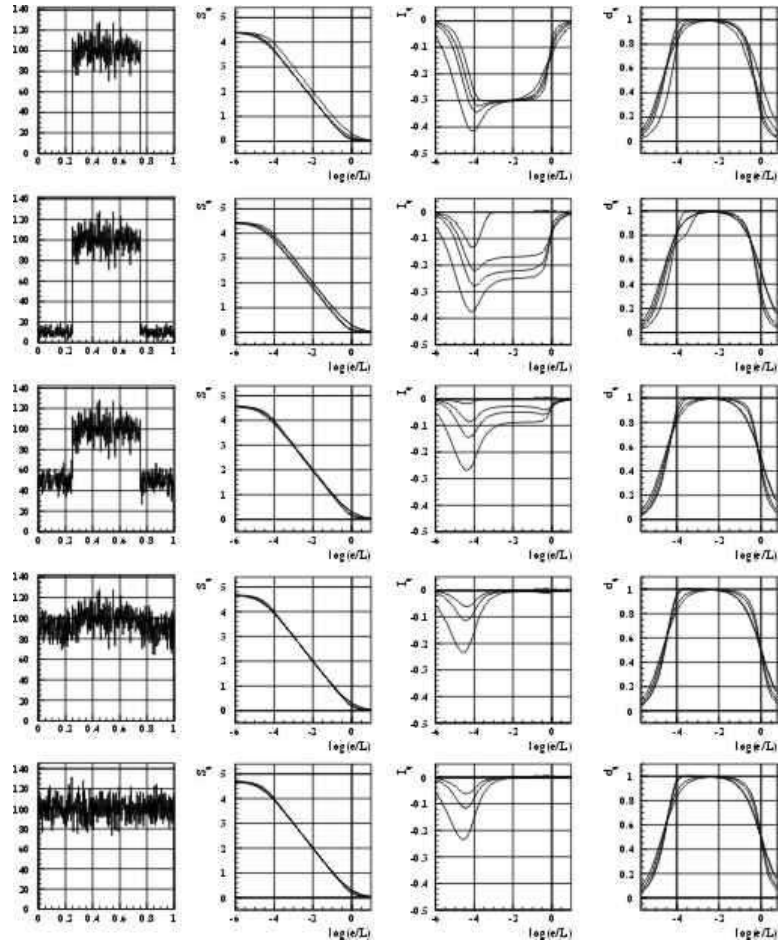}
\caption[A study of a signal on an increasing background.]{A study of a signal on an increasing background.  Results for $q=0,1,2,5$ are shown.}
\label{1dBossBkgd}
\end{figure}

This distribution resembles the scaled RGUD of the previous example.  This resemblance is no accident, in fact, if we choose $N_1=0$ and $N_2=50,000$ this simple correlation example reduces to the scaled RGUD.  Note that the uniform distribution can also be characterized in this way ($N_1=N_2$).  We can take the same philosophical approach to this correlation example as we would to a signal on a background.  For the uniform distribution the signal and background are at the same level, and thus indistinguishable.  For the scaled RGUD the background vanishes and we have only the signal.  This provides significant guidance for our expectations.  As the background rises from nothing to the level of the signal the SCA results should smoothly go from those of the scaled RGUD to the RGUD that fills the embedding space (see figure~\ref{1dBossBkgd}).
%
%

It is important to note that because the background is increasing while the signal stays fixed the multiplicity increases from 25,000 to 50,000 in this example.  However, the differing multiplicities are only relevant at small scales.  This can be seen in the relative information between the data distribution and the ideal uniform distribution of the same multiplicity.  The peak of the pseudoinformation moves down in scale as the multiplicity increases.    

Focusing on a mid-scale point ($log (e/L)=-2$) the information results show the utility of performing an analysis for a variety of ranks.  In the mid-scale region for $q=0$ we get no information about the structure of the data, we only see its support.  This explains the dramatic difference in $I_0$ for the signal with no background when compared to the signal with a very slight background.  The difference in the support of the data is dramatic, so the difference in $I_0$ is also dramatic.  For $q>0$ the $I_q$ curves change smoothly as the background increases.

This example confirms that we can interpret the $I_q$ value at a given scale as the effective $q$-width of the distribution at that scale.  For the case of no background the effective width at mid-scale is independent of $q$, and tells us that the distribution has a width of half of the embedding space.  When a slight background is added the effective width of the distribution increases, so $I_q$ must increase to reflect the change.  As $q$ increases the analysis system becomes increasingly sensitive to large clusters.  Thus, when a background is added the results are no longer independent of $q$.  The higher rank information measures are more sensitive to the large clusters of points in the signal than the smaller clusters of points in the background.  This is why the effective width seen by the information decreases as $q$ increases.  Of course, with even a slight background and large $q$ the effective width is larger than the actual width of the signal and with a significant background level the effective width increases to nearly the width of the embedding space.

\section{Beyond the First Dimension}

\begin{figure}[ht]
\centering
\includegraphics[width=4in]{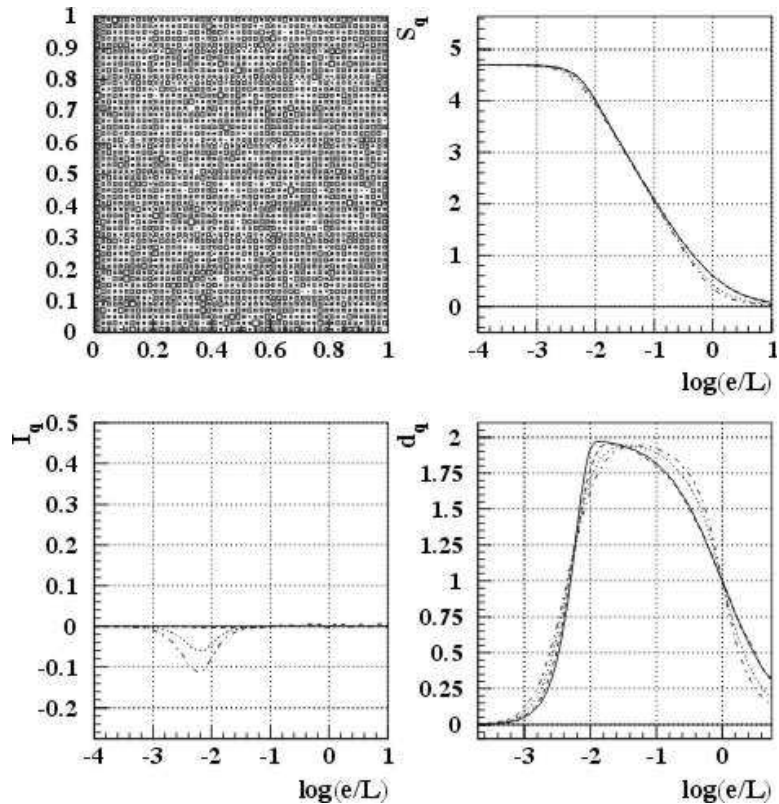}
\caption[SCA results for a 2d randomly generated uniform distribution with a 2d IUD reference.]{SCA results for a 2d randomly generated uniform distribution ($N=50,000$) with a 2d IUD reference.}
\label{2dRGUD}
\end{figure}

Thus far we have only considered one-dimensional data distributions in the interest of simplicity.  Now that we have laid the groundwork for a basic understanding of the SCA results it will be useful to extend the scope of our data analysis to a two-dimensional embedding space.  Since we derived the entropy of the IUD in two dimensions in the previous chapter, the logical starting point for moving beyond the first dimension is the two-dimensional RGUD with a two-dimensional IUD reference (see figure~\ref{2dRGUD}).

%
%

In this case we are distributing $N=50,000$ points randomly on the unit square instead of the unit interval, but otherwise the 1d and 2d distributions are in principle equivalent.  Thus, for the 2d case we expect to see behavior similar to the 1d entropy result.  Indeed, $S_q(e)$ behaves exactly as it does in 1d except the number of occupied bins increases much faster with decreasing scale in the 2d case.  We know this must happen because the slope of the entropy curve of an RGUD is determined by the dimension of the embedding space.  In terms of bin counting, this difference comes from the number of bins needed to cover the embedding space at a given scale.  This is $M^1$ for the one-dimensional distribution, and $M^2$ for the two-dimensional case.  

The difference between the 1d and 2d cases also has consequences for the illumination.  Since the slope is larger in the midscale region we run out of illumination much sooner.  Thus, the increase in the number of bins needed to cover the 2d space also accounts for the poor illumination of the scaled dimension curve afforded by 50,000 points.  In the 1d case (see figure~\ref{1dRGUD}) a 50,000 point RGUD fully illuminates the embedding space; in the $-3.5 < log(e/L) < -2.5$ region the embedding space is fully illuminated as one-dimensional (for $q=0$).  This is possible because the mean interparticle spacing is $\frac{L}{N}$ and so the granularity of the data is not fully apparent until we are in the neighborhood of $log(e/L)=log(1/N)\approx -4.7$.  In the 2d case the mean interparticle spacing occurs at $\sqrt{\frac{L^2}{N}}$ so its granularity becomes apparent at larger scales, in the scale region near $\frac{1}{2}log(1/N)=-2.35$.  To get the same quality of illumination as the 50,000 point 1d RGUD we would need 2.5 billion points for a 2d RGUD!  One should note here that the scale axis in the 2d case refers to the side length of a square bin, so at scale $e/L$ we need $(L/e+1)^2$ bins to cover the whole embedding space.

Other than the difference of illumination and magnitude, the scaled dimension behaves exactly as one might expect from our previous experience.  At large scales, the distribution looks like a point so the dimension is zero.  As we move down in scale across the scale of the embedding space the dimension rapidly moves up toward the dimension of the embedding space (two in this case).  As the scale gets very small and we resolve the individual points that make up the distribution the dimension again falls to zero.  Now that we have confirmed that the intuition we have gained in analyzing 1d distributions is applicable to 2d distributions we will consider 2d data with some interesting correlation structures.

\section{Analysis of Hierarchically Organized Point Distributions}

One approach to generating a distribution with correlations at a characteristic scale is to simulate a system of clusters.  We can model cluster formation via condensation by generating a hierarchically organized point distribution.  This will yield a distribution that has a multitude of correlation features rather than a single correlation feature as in the signal-on-a-background example.  This type of model is particularly relevant to phase transition searches.    

To create a two-dimensional, two-tier cluster hierarchy we start by generating a uniform distribution of $N_0$ cluster sites.  This gives the distribution correlations at the characteristic length scale of $\frac{1}{\sqrt{N_0}}$, the mean separation between sites.  At each cluster site we can now place a different randomly generated uniform distribution of $N_1$ points with width $\delta_1$.  This gives the distribution a second characteristic length scale of $\frac{\delta_1}{\sqrt{N_1}}$.  

\begin{figure}[ht]
\centering
\includegraphics[width=4in]{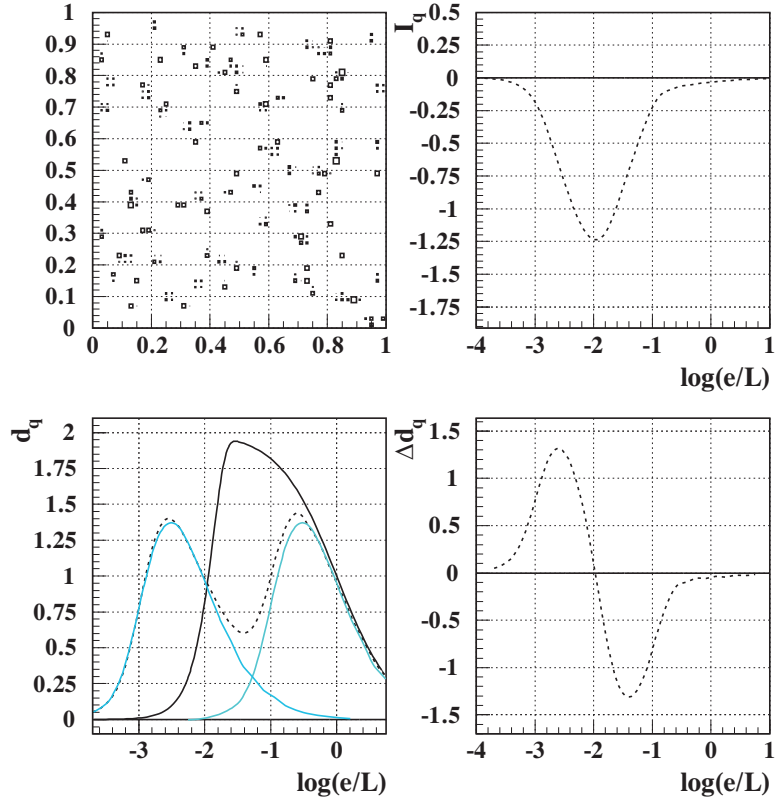}
\caption[SCA results for a two-tier hierarchy.]{SCA results for a two-tier hierarchy with $N_0=100$, $N_1=100$ \& $\delta_1=0.001$.  The scaled dimension of the data (dashed line) is compared to the IUD reference with the corresponding number of points ($N=N_0*N_1=10,000$) at scale $e/L=1$ (black solid line) as well as the IUD reference for $N=N_0=N_1=100$ at scales $1$ and $0.001$ (blue solid lines).}
\label{2d2stepH}
\end{figure}

Assuming the two tiers of the hierarchy are separated by a significant scale interval the sub-structure of the clusters will be invisible to the analysis at large scales ($e >> \delta_1$) .  In that regime we expect the hierarchy to look exactly like an RGUD of $N_0$ points.  Conversely, at small scales ($e \sim \delta_1$) the only apparent structure is the cluster structure that behaves like an RGUD of $N_1$ points.  Thus, this two-tiered hierarchically organized distribution has a unique self-similarity that allows it to be an RGUD at scales $L$ and $\delta_1$ simultaneously (see figure~\ref{2d2stepH}).  This is exactly the same type of self-similarity that one observes in fractal point distributions, only in this case it is restricted to a small scale range.  

%
%

The lower right panel of figure~\ref{2d2stepH} shows the {\bf dimension transport} for the two-tier hierarchy.  The dimension transport is the scale derivative of the information.  It can be equivalently expressed as the difference between the scaled dimension of the data distribution and the scaled dimension of the reference distribution.  As long as the two distributions compared in the calculation of the information have the same number of points the scale integral of the dimension transport is zero.  The dimension transport shows how (relative to the reference distribution) the correlation of the data distribution has been transported on scale.  In the example of the two-tier hierarchy there is an anti-correlation (relative to the RGUD of the same multiplicity) at large scale that causes the points to condense onto the cluster sites.  The cost of creating this anti-correlation at large scale is an increase in the amount of correlation at small scale.  

%
\begin{figure}[ht]
\centering
\includegraphics[width=4in]{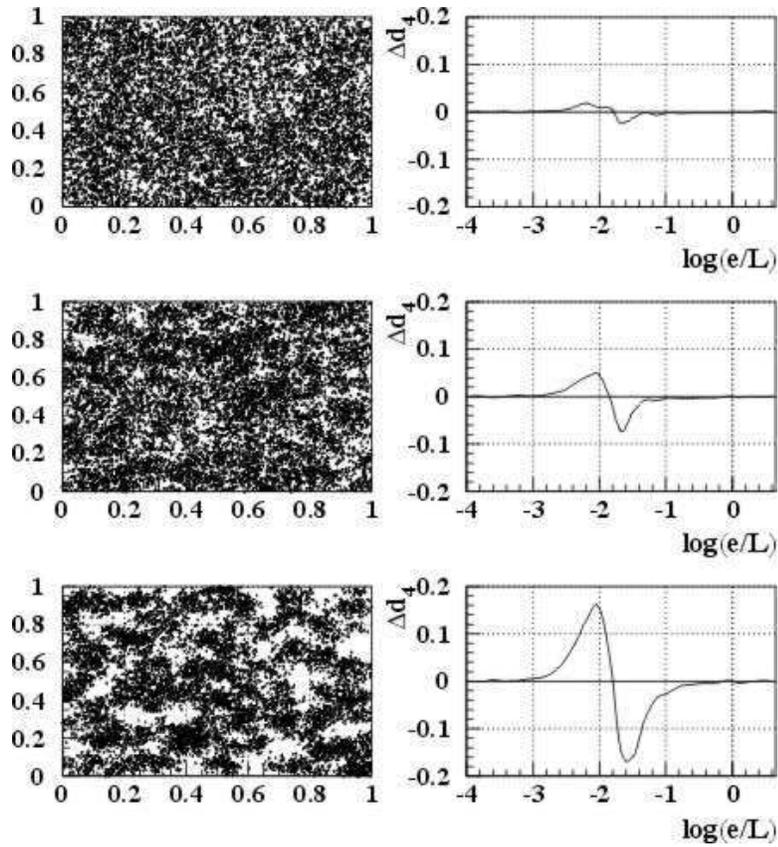}
\caption[A picture of the gradual onset of cluster formation in a system with a fixed multiplicity.]{A picture of the gradual onset of cluster formation ($\delta_1=0.03$) in a system with a fixed multiplicity ($N=10,000$).}
\label{6panelH}
\end{figure}
%

An extended two-tier condensation example shows how small scale correlations are generated by condensing points of an RGUD onto cluster sites (see figure~\ref{6panelH}).  At the onset of cluster formation ($\sim$3000 cluster sites, $\sim$3 points per cluster) the transport of dimension to smaller scale is barely visible (but still non-statistical).  Once the size of the clusters becomes significant ($\sim$1000 cluster sites, $\sim$10 points per cluster) it becomes clear to the analysis (and the eye) that the distribution of points is in some way correlated.  By the time the cluster size is 10\% of the number of clusters ($\sim$300 cluster sites, $\sim$30 points per cluster) the dimension transport shows dramatically the movement of correlation from large to small scales.

\begin{figure}[ht]
\centering
\includegraphics[width=4in]{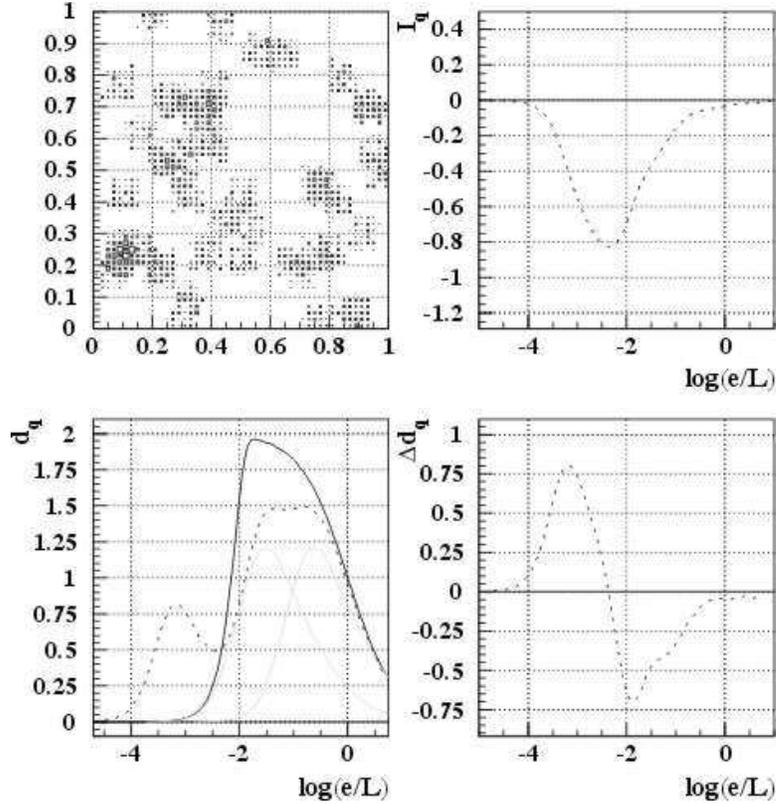}
\caption[SCA results for a three-tier hierarchy.]{SCA results for a three-tier hierarchy.  The scaled dimension is compared to the IUD reference with the corresponding number of points ($N=N_0*N_1*N_2=25,000$) at scale $e/L=1$ (black solid line) as well as the IUD reference for $N=N_0=N_1=50$ at scales $1$ and $0.1$ (green solid lines).  The scaled dimension for an IUD reference with $N=N_2=10$ at scale $0.001$ is not shown explicitly.}
\label{3tierH}
\end{figure}

The two-tier hierarchy can be extended by treating the points of the second tier as cluster sites for a third tier.  At each of these $N_0*N_1$ sites uniform distributions of $N_2$ points and width $\delta_2$ would be placed giving the hierarchy a third characteristic scale of $\frac{\delta_2}{\sqrt{N_2}}$. Continuing in this manner it is in principle possible to generate a hierarchy with an arbitrarily large number of levels.  The results of the application of the SCA system to a three-tier hierarchy with $N_0=50$, $N_1=50$ \& $\delta_1=0.1$, $N_2=10$ \& $\delta_2=0.001$ can be seen in figure~\ref{3tierH}.  Notice how the small scale separation between the first two tiers ($\delta_1=0.1$) presents itself as a bimodal peak at large scale in the scaled dimension curve.

%
%

\section{A 1d Distribution in a 2d Embedding Space}

We have already seen in the example of the scaled RGUD how the relationship between the data and the embedding space is irrelevant to the {\em shape} of the scaled topological measure curves.  The data analysis results must be independent of the topology of the embedding space since it only serves as a platform for the data.  The extrapolation of an embedding space from one to two dimensions creates an opportunity to test this idea.  If the embedding space behaves as we expect it must then the results of an analysis of a 1d RGUD in a 2d embedding space should not substantively differ from the previous result of the same analysis in a 1d embedding space.
%
\begin{figure}[ht]
\centering
\includegraphics[width=4in]{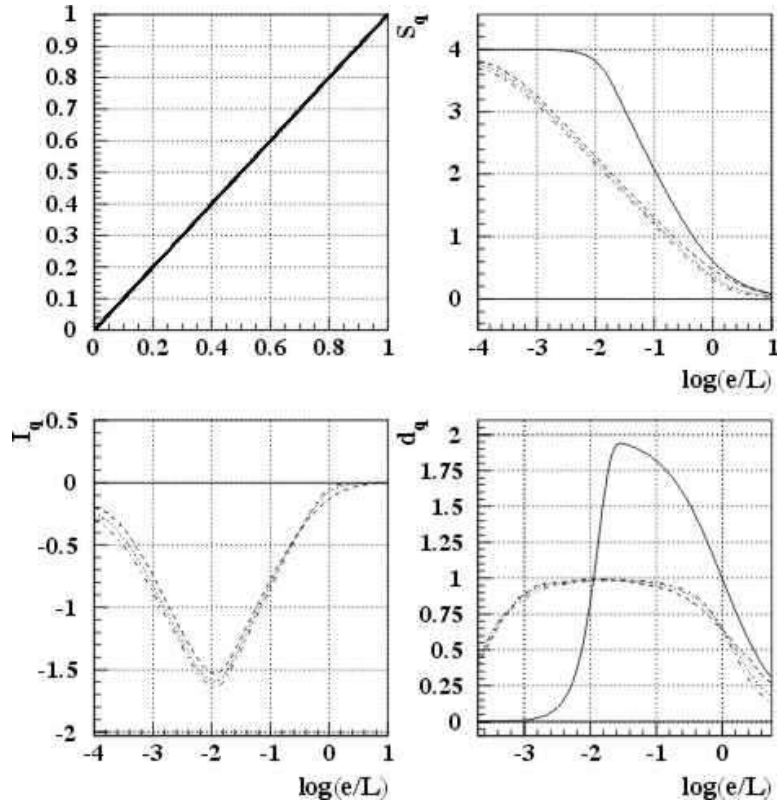}
\caption{Analysis results for a 1d RGUD in a 2d embedding space relative to a 2d IUD reference.}
\label{sfc4panel}
\end{figure}
%
Indeed, as figure~\ref{sfc4panel} shows, the presence of the 2d embedding space does not substantively effect the analysis of a 1d distribution.  The scaled entropy has exactly the slope that we expect (leading to a significant difference between the 1d RGUD data and a 2d IUD reference) and never rises above one.

\section{The Space Filling Curve}

Taking the previous example of a 1d distribution in a 2d embedding space further we will now focus our attention on a space filling curve \cite{Peano}.  Like all previous examples, the distribution will be a collection of some number of points (in this case $N=10,000$) and not actually a proper curve.  However, for the scale region in which we are most interested the particulate nature of the data will not be relevant.  

To create a space filling curve we systematically fold a line ($x=y$) back on itself in 2d while holding fixed the number of component points.  For an arbitrarily large number of iterations the data will be indistinguishable from a 2d RGUD with the same multiplicity.  In this study we will examine the first 5 folding iterations to see how the space filling curve moves from a purely 1d distribution to nearly filling a 2d embedding space.  This speaks to the greater issue of a system expanding to take advantage of new degrees of freedom, which is relevant to phase transition analysis.

\begin{figure}[ht]
\centering
\includegraphics[width=3.5in]{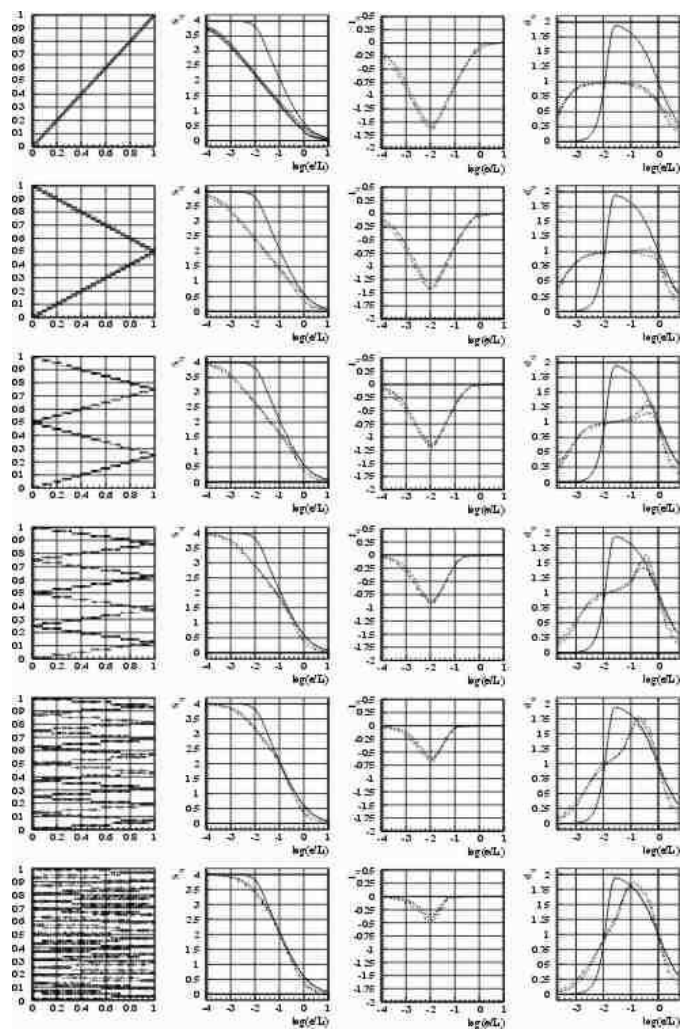}
\caption[SCA analysis of 6 iterations of a line-folding generated space filling curve.]{SCA analysis of 6 iterations of a line-folding generated space filling curve.  Data are shown in box plot format with a 50x50 grid resolution.}
\label{sfc24panel}
\end{figure}
%

The change in the scaled dimension as the space filling curve moves from one to two dimensions is quite revealing (see figure~\ref{sfc24panel}).  Even as the large scale dimension increases toward that of a 2d RGUD, the small scale dimension retains knowledge of the inherent 1d nature of the space filling curve.  However, the cost of meeting the demands of the 2d nature of the distribution at large scale is paid by the loss of illumination of its 1d nature at small scale.  This interplay between the 1d and 2d nature of the data suggests that the scaled dimension can be a powerful tool for characterizing the data distribution; it has access to all scales which allows it to see the distribution as 2d, 1d, and something in between (in the relevant scale regimes).  This is essential for understanding data that have complex and varying correlation structures over a broad range of scale.

\section{A Fractal Example}

The space filling curve example suggests that our scaled dimension results will be most interesting for data that are exquisitely correlated over a large scale range.  With this in mind, we will now calculate the scaled dimension of a well-known strange attractor and compare our results to the results in the literature.  This will provide insight into the significance of a scale-local approach to dimension.  For this analysis we choose to use a strange attractor of the H\'enon map because of the calculational simplicity involved in such a straightforward mapping:
\begin{eqnarray}
x&\mapsto&a+by-x^2 \\ \nonumber
y&\mapsto&x.
\end{eqnarray}
The chaotic attractor we will analyze occurs for parameter values $a=1.4$ and $b=0.3$, and has been thoroughly investigated in the existing literature.  Following the example of \cite{BRHunt} we analyze the dimension of whatever attractors exist in the square $-1.8 \le x,y \le 1.8$ for the aforementioned parameter values.  It has been determined that all orbits are either trapped in this region or become unbounded.  The reported $q=0$ dimension (box dimension) of this attractor is $\sim$1.28 \cite{PGrass} and the $q=1$ dimension (information dimension) is $\sim$1.258 \cite{BRHunt} using a standard $e\rightarrow0$ limit approach.  Using scaled correlation analysis this can be extended to calculate the rank-$q$ dimension of the strange attractor as a function of the bin size $e$.  

%
\begin{figure}[ht]
\centering
\includegraphics[width=4in]{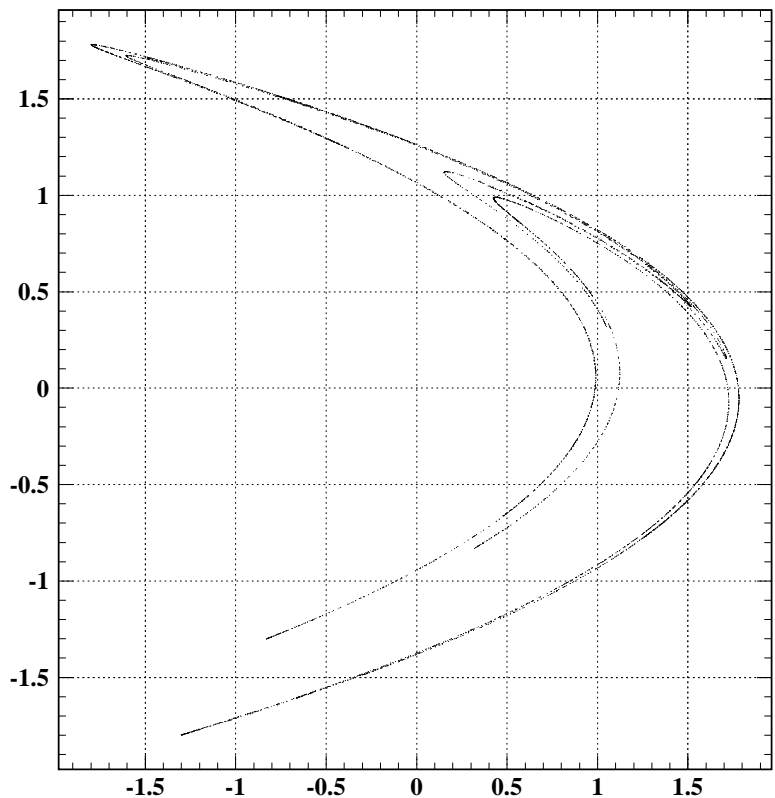}
\caption{The strange attractor of the H\'enon map for parameter values $a=1.4$ and $b=0.3$.}
\label{Henon}
\end{figure}
%

To generate the attractor that we wish to analyze it is a simple matter of choosing some initial values and plugging them into the mapping.  Of course, we must go through several iterations of the mapping until the initial state transients have vanished and we are left with only points on the attractor.  By continuing to run iterations of the mapping we can generate this attractor with an arbitrarily large number of points.  This is necessary to push down the small scale limit of the analysis; we need a high density of points on the attractor to allow us to analyze its small scale (non-limit) behavior.  We can also generate the attractor using different initial conditions to test if the attractor's scaled dimension is independent of the specific points used for illumination.  In analyzing the attractor we first rescale the data points to fit within the unit square ($[0,1]\otimes[0,1]$).  This makes no difference to the analysis as long as we are careful to maintain the relative position of points in the data.  

%
\begin{figure}[ht]
\centering
\includegraphics[width=4in]{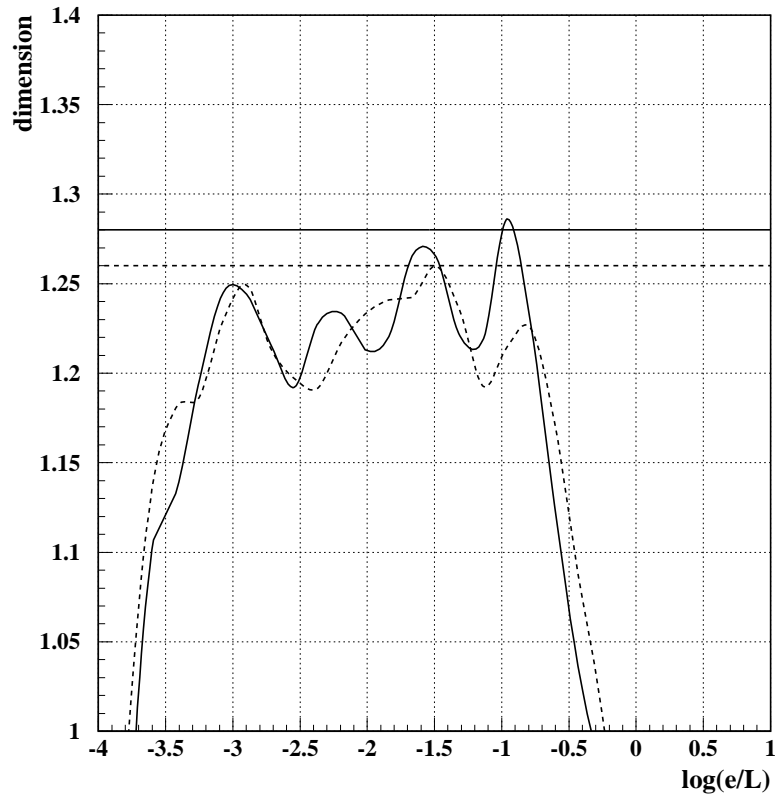}
\caption[Dimension of the H\'enon attractor as a function of scale.]{Dimension of the H\'enon attractor as a function of scale.  Curves for $q=0$ (solid line) and $q=1$ (dotted line) are shown.  Reference lines are shown marking the zero scale limit values for $d_0$ and $d_1$.  Below $log(e/L)=-3$ the analysis runs out of illumination ($N=10,000$), causing the dimension of the attractor to fall sharply.}
\label{HenonD}
\end{figure}
%

\begin{figure}[ht]
\centering
\includegraphics[width=4in]{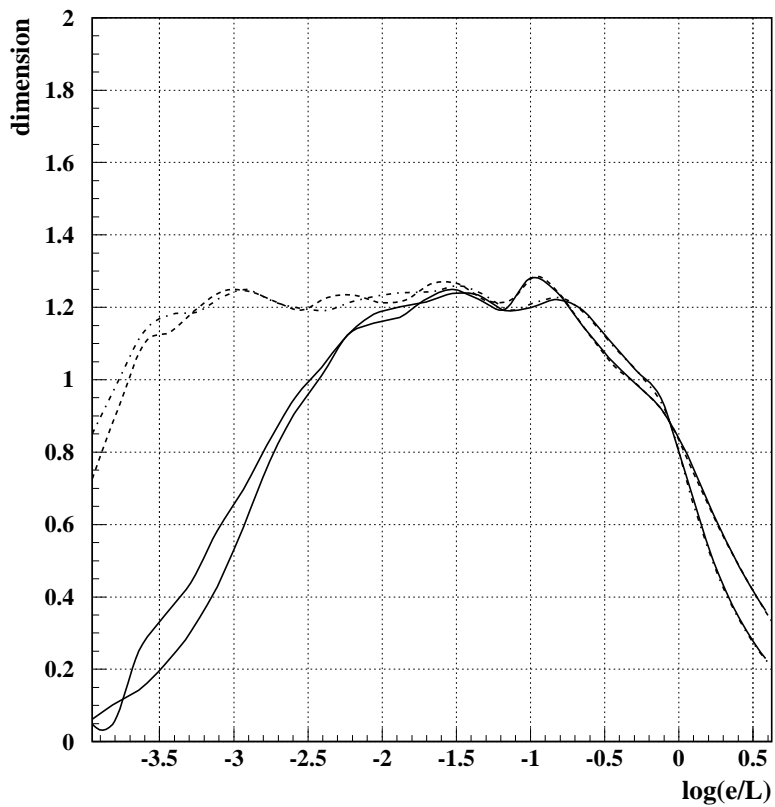}
\caption[Dimension of the H\'enon attractor as a function of scale with differing levels of illumination.]{Dimension of the H\'enon attractor as a function of scale with differing levels of illumination ($N=10,000$ \& $N=300,000$).}
\label{HenonDillum}
\end{figure}

The scaled dimension curves shown in figures~\ref{HenonD} and~\ref{HenonDillum} are calculated from two different point distributions, one with $N=10,000$ and another with $N=300,000$.  These distributions were generated using different initial conditions, so the actual points illuminating the H\'enon attractor are different in both data sets.  Figure~\ref{HenonD}
shows a close-up of the rich scale structure of the ($q=0$ and $q=1$) dimensions of the attractor over a limited scale range.  Figure~\ref{HenonDillum} compares the dimension results of the two data sets and shows that the intricate structure of the scaled dimension extends as far as the available illumination will allow us to see.  More importantly, the comparison between the different point distributions shows that in the large scale region where the illumination is good for both distributions the results are the same.  Thus, we conclude that the scaled dimension results are in general independent of the specific points chosen to illuminate the attractor.  This suggests an exciting result, that any strange attractor has a unique dimension profile that varies as a function of scale.

%
%

The rich, characteristic structure of the scaled dimension is another representation of the attractor's unique correlation structure.  It is only logical that the scaled dimension of an attractor will have meaningful features over the same scale range as the correlation content of the attractor.  This is a significant improvement upon the conventional approach that summarizes the dimension of a complex fractal object in a single number at the asymptotic small-scale limit.  While the zero-scale dimension is undeniably useful and easy to calculate, it cannot provide the full picture of the complex correlations present in a fractal system.  A scaled dimension approach provides a broader view of the fractal and provides the tools to create meaningful alternatives to the simplistic formulation of canonical dimension definitions.

\section{Conclusions}

With a solid understanding of this menagerie of model distributions we now have the tools to analyze and interpret real data.  We have seen how the scaling of the data effects the scale-local topological measures, highlighting the versatility of this analysis method.  By varying the signal strength at various scale points in the hierarchical point distributions we have seen how clustering phenomena at a phase transition might look to a scaled correlation analysis.  We saw in the space filling curve example how the opening of degrees of freedom effects the topological measures of a distribution.  The analysis of the attractor of the H\'enon map showcases the power of the scale-local dimension and gives us insight into exploring self-similar distributions with scaled correlation analysis.  All of these toy models have given us a broad view of SCA results, and so we are now ready to move beyond simulation to real data analysis.

%
%
%

\chapter{SCA Applications: Analysis of Physical Models and Data}

\section{Event Spaces}

The whole purpose of the development of scaled correlation analysis has been to understand heavy-ion collision data.  Thus, we must find a way to reach meaningful physics conclusions based on SCA results.  Towards this end, we create an {\bf event space} in which we can measure the similarity of two events by their spatial proximity.  This allows us to easily identify unique and rare events within our experimental data.  

We have seen how the correlation content of a data distribution can be fully described by its scaled dimension and so we will use this measure to characterize the data in the event-space system.  The simplest way to create such an event space is to treat each point on the scaled dimension curve as a coordinate in this space.  This gives us a $k$-dimensional space (where $k$ is the number of points on each scaled dimension curve) in which each event appears as a point.  The distance between these event-space points is then a quantitative measure of the similarity of the analyzed events.  To see how this system can work effectively consider an example where the system is applied to the well-known problem of face recognition.

The human detector system is spectacularly good at face recognition.  Even when a person is disguised our brain's software can often detect the nuances of individuality and correctly identify the subject.  This problem is not so easy for computers and has only recently been approached seriously.  To solve the computational face recognition problem, we need to be able to analyze a set of images of faces and find an event space in which these faces are reliably and quantitatively distinguishable.  This is quite similar to the problem of event discrimination in heavy-ion collisions.  In the heavy-ion collision analysis problem we want to differentiate between events in which a QGP has been made and events that have stayed in the hadronic regime (according to a class of theoretical predictions) \cite{hadronVqgp}.  Thus, we will tackle the problems of face recognition and QGP finding in heavy-ion collision data with the same event-space approach.

\section{Face Recognition Analysis}

The data used in this face recognition analysis consists of 23 grey-scale images (29x38 pixels) of five different individuals.  The images were composed uniformly with the subjects wearing shower caps to minimize the effect of differing hair styles on the final results.  Subjects were asked to present a variety of differing facial expressions to do their best to fool the analysis system.  

%
\begin{figure}[ht]
\centering
\includegraphics[width=5in]{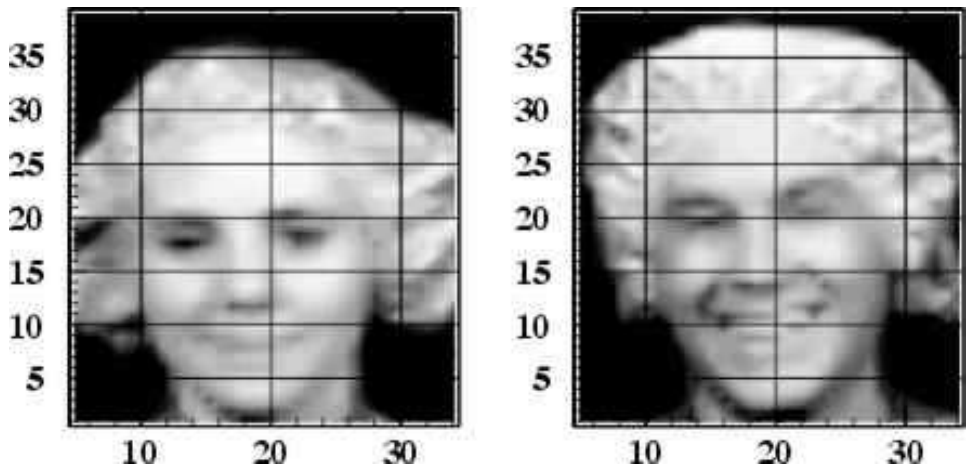}
\caption{Sample data images from the face recognition analysis study.}
\label{Faces}
\end{figure}
%

Before discussing the face recognition results, we must point out that the details of this analysis differ slightly from the previous analyses we have performed.  Because this is real data it cannot be treated as a collection of individual points.  The detector (in this case the CCD in a digital camera) has already performed a binning and so the data we have to work with is a list of bin occupancies $n_i$.  We call data of this form {\bf pre-binned data}.  To calculate the scale-local topological measures we need a scheme for the scale-local rebinning of this data in a way that will not insert bias into the results.  Because the only information we have is the list of bin occupancies at scale $e_{detector}$, the only fair assumption that can be made is that the data is uniformly distributed within each detector bin (a maximum entropy hypothesis).  This serves to make the task of rebinning easy since the amount of data in a fraction of a bin will be the geometrical fraction of the bin considered times the occupancy of that bin ($f*n_i$).  Of course, rebinning at scales smaller than $e_{detector}$ is irrelevant since that would only serve to illuminate the assumptions we have made about the distribution of data within a detector bin; no analysis can increase the physical resolution of the detector.  Other than the procedural difference between rebinning a pre-binned data distribution and binning a simulated point distribution, the analyses of these two types of data are the same.  

%
\begin{figure}[ht]
\centering
\includegraphics[width=3.8in]{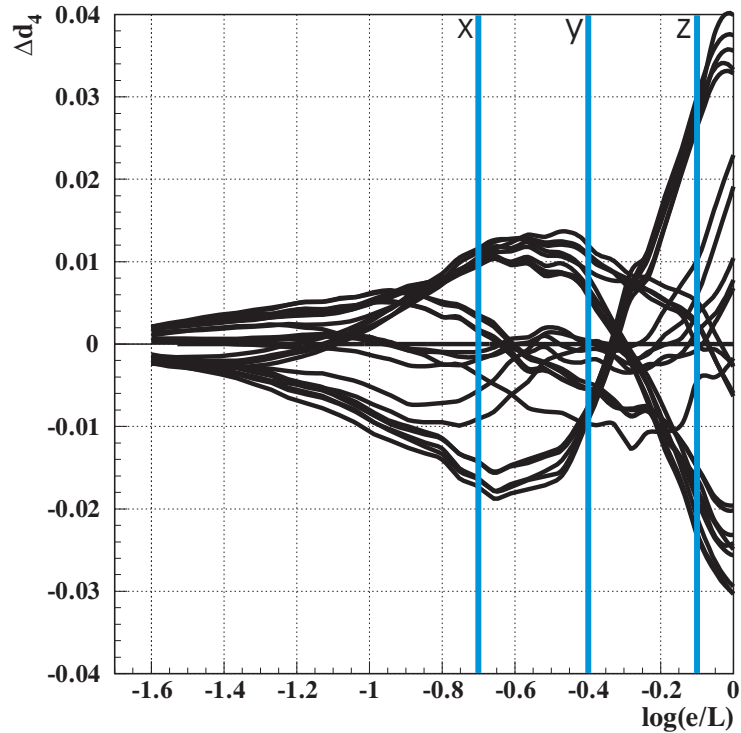}
\caption[Dimension transport results for the face recognition example using an ensemble average reference.]{Dimension transport results ($q=4$) for the face recognition example using an ensemble average reference.  Blue bars mark the points in scale used to build the Cartesian event space.}
\label{FaceDdq}
\end{figure}
%

To generate the event space we must first apply the SCA system to the data and calculate the dimension transport for each image.  However, to calculate the dimension transport we must select a reference.  It doesn't make sense to use an IUD or RGUD reference because the difference between an image of a subject's face and a uniform reference is vastly greater than the difference between any two images in the data set.  Since we want to discriminate between the slight differences in the data it makes sense to use an {\bf ensemble average} reference.  This reference allows us look for correlation content that deviates from the mean of the data ensemble and is particularly useful when looking for slight deviations between events.  In this case the ensemble average is the most appropriate reference because we need to be sensitive to the details of an individual face, as they deviate from the average face shape.

Once the analysis has been done and we have the dimension transport results in hand we can use those results to build an event space that will allow us to discriminate between the subject's faces.  There are 30 points on the dimension transport curves in figure~\ref{FaceDdq} so we could build a 30-dimensional event space, but for the purposes of this example it will be sufficient to use a much simpler 3d space.  In practice, a 30-dimensional space would contain a great deal of redundant information anyway, and it is much easier to understand the event-space results when they are easily visualizable, as in the case of a 3d Cartesian space.  We choose scale points at $-0.7$, $-0.4$, and $-0.1$ for event-space coordinates $x$, $y$, and $z$, respectively.  These scale points were selected specifically to maximize the descriminatory power of the event space (minimize the redundancy of scale points).  This is a crude version of a principal component analysis in which the significant degrees of freedom in the data are obtained by a formal procedure \cite{PCA}.

%
\begin{figure}[ht]
\centering
\includegraphics[width=4in]{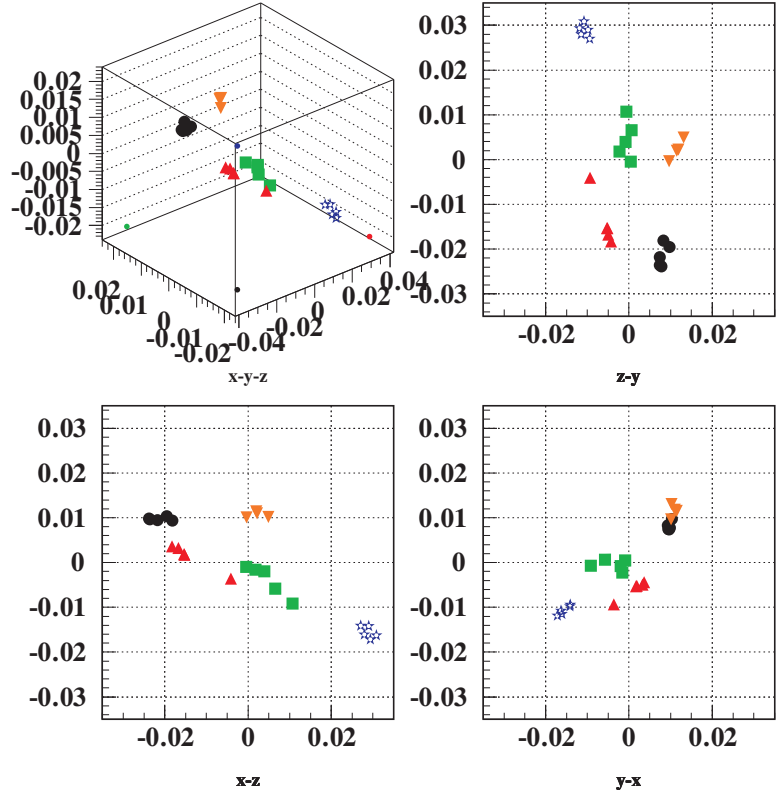}
\caption[The Cartesian event space formed from the face recognition analysis results.]{The Cartesian event space formed from the face recognition analysis results.  The five different marker types each represent a different subject.}
\label{FaceEvSpc}
\end{figure}
%

The 3d Cartesian event space does an excellent job of discriminating among the different face images in the data set.  Each marker type in figure~\ref{FaceEvSpc} represents a different person's face, and in spite of the subjects' attempts to fool the recognition system by varying their expressions the clustering of each individual's data points is quite dramatic.  By calibrating an event space such as this with some reference images one could easily apply this system to the face recognition problem in the field and successfully identify an individual based solely on an image of their face.  

\section{STAR Trigger Simulations}

\begin{figure}[ht]
\centering
\includegraphics[width=4in]{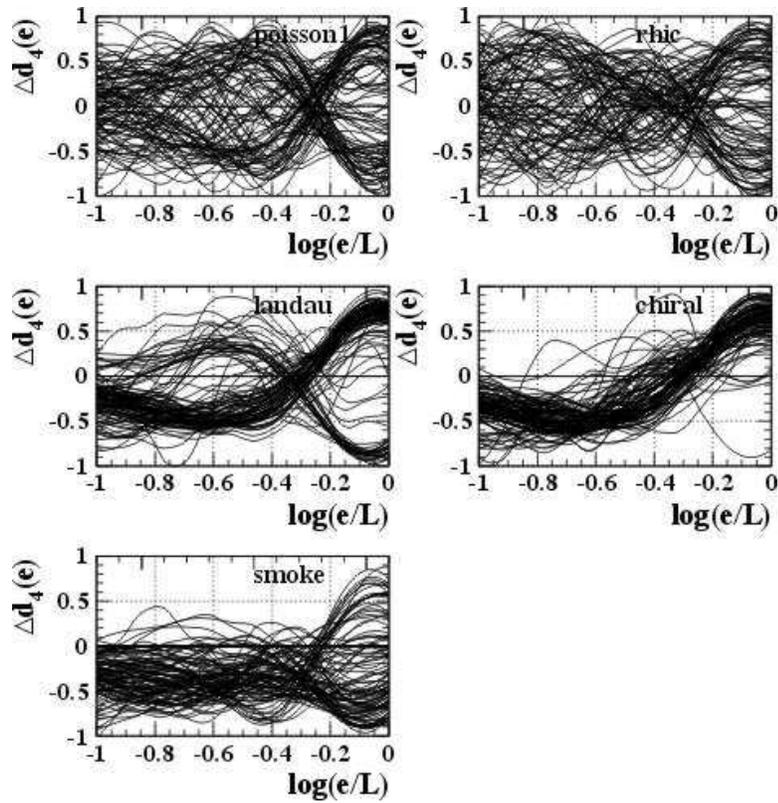}
\caption[Dimension transport results from simulated STAR trigger models.]{Dimension transport results ($q=4$) from simulated STAR trigger models.}
\label{STARDdq}
\end{figure}

The first application of SCA to heavy-ion collisions came in a trigger study for the STAR experiment in 1995.  Even before the detector was built we were trying to create a trigger that could distinguish between normal events and events in which a phase transition was made into the QGP regime.  By making this distinction at the trigger level we hoped to maximize the number of QGP events recorded.  To calibrate this SCA-based trigger, simulated data was generated and modified according to a variety of QGP models \cite{Ron}.  The events were designated as rhic, landau, chiral, and smoke, named according to the method used to modify the simulated data.  These event tags were hidden from the analysis system, which was given the whole data set blindly along with two sets of toy model data (poisson1 and poisson2).  Taking all of the data together the SCA system compared each event to the ensemble average reference and the rank-4 dimension transport was calculated.  Figure~\ref{STARDdq} shows the different event classes' dimension transport results.

%
%

\begin{figure}[ht]
\centering
\includegraphics[width=4.5in]{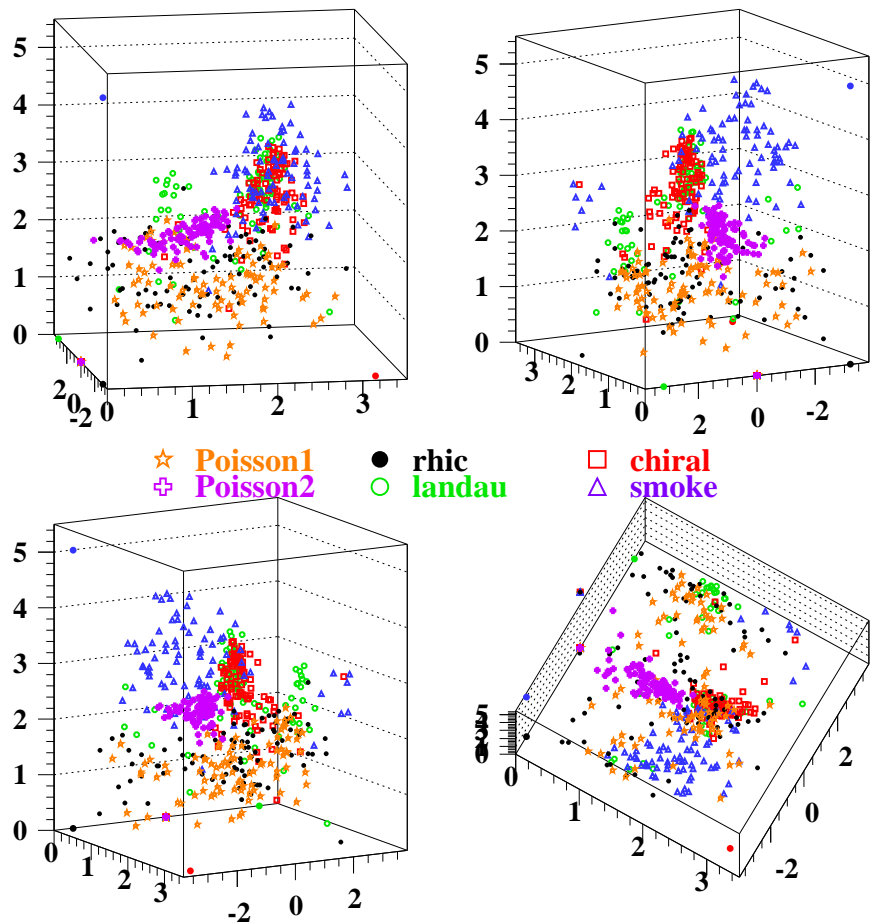}
\caption{Event-space results from the STAR trigger study.}
\label{STAREvSpc}
\end{figure}

The dimension transport results alone show how distinct the different event classes are, so it came as no surprise that we had good event class separation in the Cartesian event space (see figure~\ref{STAREvSpc}).  This was a great success, but one major obstacle remained.  How could we streamline the analysis to run on a time scale appropriate for a triggering analysis?  Running the analysis over the full scale window was prohibitively slow.  Most of our time was being lost in the binning of the data, so the solution was to calibrate the analysis on these simulations (our primitive approach to principal component analysis) and determine which scale points to use for the event space.  Once those determinations were made we were able to streamline the analysis.  By restricting the scale axis to three points (one for each axis of the Cartesian space) we were able to get away with applying only three different partitions to the space.  This, coupled with some clever code optimization, made it possible to take the event data available to the trigger and calculate an event's position in event space in 5 milliseconds.  This was necessary to process events at the design rate of the level-three trigger.  The funding for the high-level triggers in STAR never materialized, but this successful study lead us to take SCA seriously as an analysis tool.  Based on these results we moved beyond triggering into developing the full scaled correlation analysis system presented in chapters 2 and 3.

%
%

\section{Scaled Correlation Analysis of NA49 Data}

The scaled correlation analysis system was applied to main TPC central collision data from the NA49 experiment.  This analysis was done for each event on the 1d $\Delta m_t=m_t-m_{\pi}$ distribution.  The scaled-entropy of the transverse mass distribution was calculated for each event and compared to an ensemble average reference.  The rank-4 dimension transport was used to form an event space in which a class of anomalous events was identified.  

%
\begin{figure}[ht]
\centering
\includegraphics[width=4in]{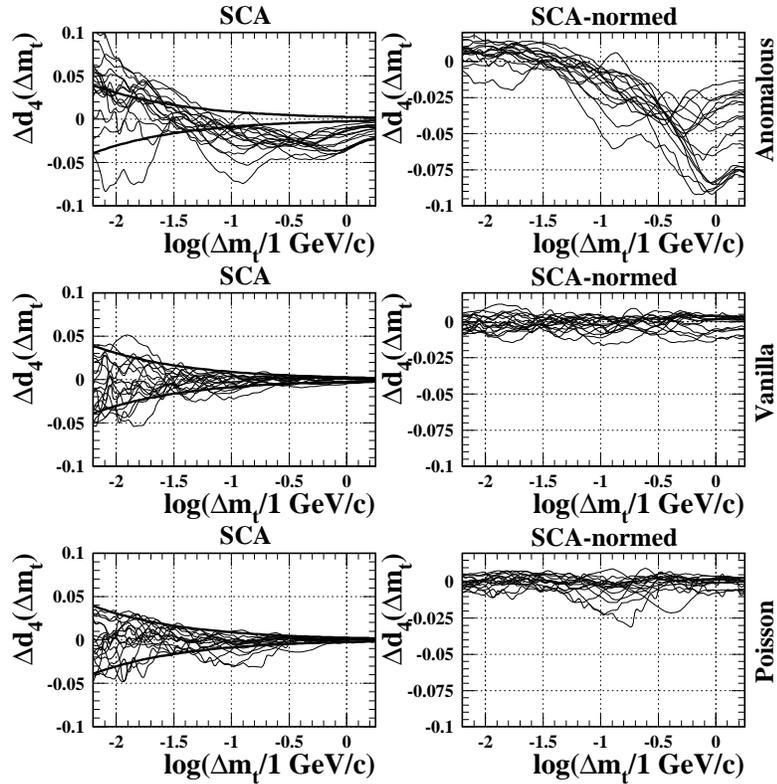}
\caption[Dimension transport results from NA49 central data.]{Dimension transport results ($q=4$) from NA49 central data.}
\label{NA49Ddq}
\end{figure}
%

\begin{figure}[ht]
\centering
\includegraphics[width=5.4in]{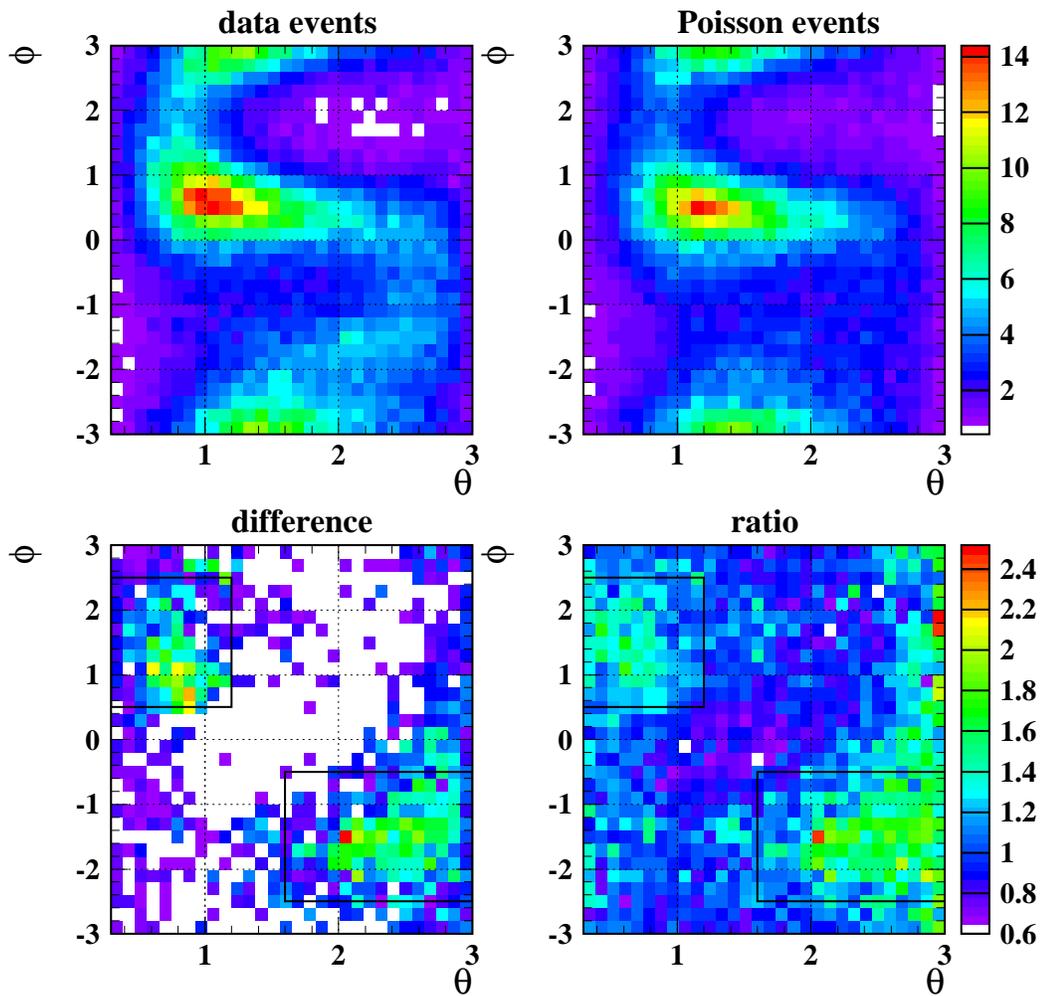}
\caption{Event-space results from NA49 data.}
\label{NA49EvSpc}
\end{figure}

Typical dimension transport results for three different classes of data are shown in figure~\ref{NA49Ddq}.  The vanilla event class is a random selection of a few of the events that were found to behave similar to the ensemble average.  This population serves to show the increasing statistical spread of the dimension transport with decreasing scale and defined the normalization procedure that was used to generate the SCA-normed results.  By definition these events must be uninteresting, and they are.  The Poisson event class is from a toy model generator that created events with no correlation features beyond Poisson statistics and with multiplicity matching exactly those of existing data events.  These results prove that there are no extra-statistical correlations in the vanilla event sample.  The anomalous event class was defined by forming an event space with ~300k events and identifying regions of the event space that differed from the vanilla/Poisson expectation (see figure~\ref{NA49EvSpc}).  

%
%

In this case three scale points were used to build the Cartesian event space that was then transformed into spherical coordinates.  The spherical space is preferable because the distance of an event from the origin of this space is a direct measure for the difference between an event and the ensemble average reference.  If less than 10\% of the central events make the phase transition into the QGP regime at the CERN SPS (as some theories suggest), then we expect these few QGP events to appear significantly distant from the origin of the event space.  Since we calculate the dimension transport with respect to an ensemble average reference we expect to see all ``normal" events clustered near the origin, and any rare anomalous events at a radius much larger than the average.  

Anomalous events (occuring at about the 10\% level) were indeed found.  Upon further investigation we discovered the anomalous event structure was consistent with data contamination by electron pairs from $\pi^0$ decay (and subsequent gamma conversion).  This was suggestive of beam-gas interactions contaminating the anomalous events, and led to the eventual discovery that these events with anomalous topology were in fact {\bf pile-up events} \cite{NA49SCA}. These were events that had two primary interaction vertices; one beam particle interacted with the target, and another interacted independently with the TPC gas.  This caused a unique correlation signature that was readily apparent to the SCA system.  

%
\begin{figure}[ht]
\centering
\includegraphics[width=5.4in]{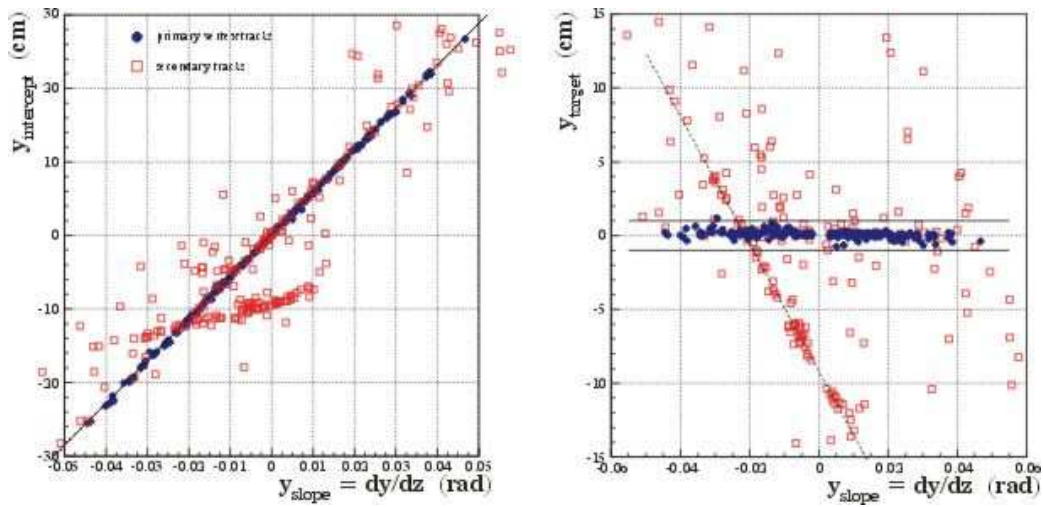}
\caption{Plots showing the presence of a secondary vertex in an anomalous event found in the NA49 data.}
\label{PileUp}
\end{figure}
%

Figure~\ref{PileUp} shows plots of the slope against the intercept in the non-bending plane for tracks from an anomalous event.  Particles originating from a common vertex will appear on the same line in this space.  In the left-panel plot particles originating from a primary vertex in the target will appear on the main diagonal, in the right-panel plot the space has been transformed so that particles originating from the target are clustered at $y=0$.  In both plots one can clearly see the presence of a secondary vertex displaced from the target position, in the right-panel plot a fit has been made to the secondary (pile-up) vertex.  Using the slope-intercept method shown in figure~\ref{PileUp} a pile-up rejection system was developed and added to the detector.  In future data taking runs these events were rejected at the trigger level.  Other than these pile-up events no significant class of anomalous events was identified by SCA in NA49 data.

\section{Conclusions}

The SCA system was successful in analyzing NA49 data in that it was able to identify rare events using a model-independent scale-local approach.  It was unsuccessful in that the analysis was unable to identify events that underwent a phase transition into the QGP regime.

To judge if this is a failure or not we need to compare the SCA non-result to the results of other analysis tools.  If the other analyses find QGP events where SCA finds none then it is clearly a failure as a QGP event finder.  There is also the possibility that the SPS is at too low an energy to make collisions that cross the phase transition into the QGP regime.  In spite of CERN's very public statements to the contrary \cite{CERNPressQGP} many in the field doubt that deconfined quark matter was made at the SPS.

At this point there is not enough data to make a definitive statement about QGP formation at SPS energies; no one really understands quark matter at those energies.  However, we can look to the other event-by-event measures used on NA49 data.  As we will see in the following chapters none of our other NA49 measurements supports a phase transition scenario in the observed collisions.  Thus, the scaled correlation analysis system seems to have performed as well as could be expected on NA49 data given the relevant physics at the SPS.

\section{SCA After NA49}

Shortly after the anomalous NA49 events were found and identified as pile-up events, we also discovered a significant non-zero charge-dependent signal in the $p_t$ fluctuations (and related two-particle correlations).  This provocative result served to draw our attention to fluctuation measure approaches at the same time that we discovered that the community was not ready to embrace a bold, mathematically complex approach like SCA.  We had succeeded in discovering and understanding an anomalous event population, but it taught us an important lesson: a model-independent correlation analysis is just a likely to discover systematic error effects as it is interesting physics.  This ruled out SCA as an effective tool for early STAR analysis since none of the instrumental effects were well understood at that point in the experiment's evolution.  Thus, motivated by the provocative NA49 fluctuation results, the state of STAR data at that point, and the pressures of the community, we turned our focus to global variable fluctuation analysis and two-particle correlations.  We brought the same mathematical rigor to this problem, but its grounding in the $\Phi_{p_t}$ measure and our suggestive preliminary results made this approach more palatable to the community.

%
%
%
\chapter{Fluctuation Analysis}

\section{Introduction}

In contrast to the exquisite complexity of scaled correlation analysis, we have also taken a more simplistic approach to event-by-event analysis of relativistic heavy-ion data.  Measuring fluctuations in global variables is calculationally simple, and in the case of event-wise mean transverse momentum ($<p_t>$), easily related to the physics of the collision.  Because transverse phase space is initially empty (before the collision occurs all of the energy is in axial degrees of freedom), $<p_t>$ is a good measure of the energy involved in the collision.  Thus, a measurement of the non-statistical fluctuations present in the $<p_t>$ distribution can be interpreted as a measurement of temperature fluctuations beyond statistical expectations.  

The analysis presented here will cover numerical and graphical results for $<p_t>$ fluctuations in STAR central and minbias triggered events.  This is an analysis of the unidentified charged hadrons produced in Au-Au collisions at $\sqrt{s_{NN}}=130$ GeV.  This represents the first analysis of fluctuations in a large acceptance detector at these energies.  While more recent data taken at higher energies (with significant detector improvements) exists, analysis of that data must be left to future work.  The newest data ($\sqrt{s_{NN}}=200$ GeV) is not yet thoroughly understood and the improved state of the detector requires a reassessment of systematics and detector effects before that data can be properly analyzed.

\section{$<p_t>$ Fluctuation Analysis Motivation}

We expect to see non-statistical fluctuations in the region of the QGP phase boundary.  This makes the measurement of $<p_t>$ fluctuations a simple probe that can tell us if we have indeed crossed into the QGP regime.  Furthermore, a substantial change in fluctuations with energy or centrality could serve to pinpoint the region of the phase diagram in which the QGP transition resides.  Additionally, fluctuation measures may be able to provide valuable insight into the properties of the QGP medium, early stage scattering, and the equilibration process \cite{FluctReview}.

The event-wise mean transverse momentum,
\begin{eqnarray}
<p_t>&=&\frac{1}{N}\sum_{i=1}^{N}p_{t_i},
\end{eqnarray}
is an estimator of the collision temperature for transverse phase space; hotter collisions involve more energy, which is then free to dissipate through the initially empty transverse degrees of freedom.  If we are to have any chance of understanding the formation and decay of QGP matter in RHI collisions we must be able to measure and understand its temperature.  $<p_t>$ is a good starting point for this work since it is trivial to calculate from the measured track parameters and it relates directly to the temperature of the particles produced in the collision, assuming particles produced in secondary interactions can be filtered out of the data sample.

\section{Data Filtering}

The sensitivity of event-scale fluctuation analysis makes it particularly prone to experimental and systematic errors.  To insure robust results an effort has been made to understand and minimize these effects.  The most important aspect of this error minimization effort is to create the cleanest possible data sample.  Any deviation in the reconstructed $p_t$ values from the parent $p_t$ distribution can appear as an excess fluctuation signal.  Thus, tracks that have been poorly reconstructed will have a significant negative impact on this analysis.  Particles that arise from secondary interactions
such as resonance decays would also contribute excess non-statistical fluctuations, so we must also be careful to include only particles produced by the primary interaction in this analysis.  To address these possible sources of error and insure that only highest possible quality data is used, a strict series of track and event cuts has been applied.

\subsection{Event Cuts}

First of all, at the event level, we remove all events that do not have a reconstructible {\bf primary vertex} within 75 cm of the center of the detector ($|v_z| < 75$ cm).  To properly identify tracks that originate from the primary interaction, it is essential to determine the precise position of the interaction within the detector.  It isn't always possible to definitively reconstruct the primary vertex from the tracks measured by the detector, particularly for events that have been corrupted in some way.  It is impossible to separate secondary from primary tracks in an event without a primary vertex, so any such events must be rejected for logistical as well as quality considerations.  In addition to finding the primary vertex, it is important to find a {\em centered} primary vertex.

The TPC gas drifts longitudinally along the beam axis away from a central membrane independently in the east and west halves of the TPC.  Thus, it is difficult to reconstruct particle tracks that travel across this membrane.  This problem is irrelevant for a properly centered vertex where nearly all of the tracks are heading radially away from the center, and only a few tracks will bend in such a way as to cross the membrane.  A significantly off-center primary vertex also changes the $\eta$ acceptance of the detector for that event. 

Even at $75$ cm the vertex displacement can cause problems, but the current limitations of the collider make it impossible to place the vertex position closer to the central membrane reliably.  To maintain a significant amount of data for this analysis we were forced to include events that have substantially displaced primary vertices.  Systematic errors arising from displaced vertices are further addressed in section 5.7.  The primary vertex cut is the only cut made at the event level.  At the track level however, a variety of cuts are applied, mostly to insure that tracks arising from secondary interactions are removed from the data volume.

\subsection{Track Cuts}

After the reconstruction software has processed all of the found clusters in the TPC volume and determined which clusters are to be associated with which tracks, each of the tracks that pass within 3 cm of the reconstructed primary vertex are considered to be primary track candidates.  The reconstruction then performs a track refit that requires the first point on every primary track candidate to coincide with the position of the primary vertex.  Tracks that can be satisfactorily refit in this way are considered to be primary tracks.  The first track cut we apply restricts the active data volume to tracks identified as {\bf primary tracks} according to this definition.  This population will certainly contain some tracks arising from secondary interactions, but this process has been designed to minimize such contamination.

Sometimes the track reconstruction will not properly associate sections of a single track.  This may cause a single normal track to appear as two or more short segments.  These are known as {\bf split tracks}.  This can be a significant source of error because it makes a single particle appear to be two or more particles.  The track splitting problem is easily dealt with by applying a cut on the ratio of the number of fit points for a track to the maximum number of points possible ($\frac{n_{fit}}{n_{max}} > 0.5$).  This insures that we are not counting any particles multiple times;  a track cannot be split into two segments in which both are more than half of the total combined length.

The next cut is applied purely to remove tracks that represent a failure in the reconstruction software.  When the reconstruction is done all tracks that it finds are saved, even tracks that the software itself knows are likely to be incorrectly reconstructed.  These are removed by cutting on a track quality parameter that is positive for good tracks and negative for bad ones ($iFlag > 0$).  

Additional quality information can be found in the $\chi^2$ parameter calculated during the reconstruction.  This parameter measures the quality of the track fit to the found clusters in the TPC.  Ideally, large $\chi^2$ values correspond to tracks that are poorly reconstructed and a cut would be made to remove them.  However, problems with the reconstructed $\chi^2$ prevent this (details are discussed in section 5.7.3).

The $p_t$ of accepted tracks is restricted ($0.1 < p_t < 2.0$ GeV) to remove abnormally high momentum tracks and cut out the low $p_t$ region where the reconstruction process is most vulnerable to error.  This also serves to limit the scope of our analysis to the soft-physics regime in which we are most interested.  High $p_t$ jets would likely add excess fluctuation to the $<p_t>$ distribution that might obfuscate the fluctuation contributions from the effects we are looking for.

We also apply a cut on pseudorapidity ($-1.0 < \eta < 1.0$).  The full pseudorapidity coverage of the STAR detector ranges from about -1.2 to 1.2, but at extreme pseudorapidity the track quality is rather low because of the limited detector volume covering that region.  For quality reasons it might be preferable to make an even tighter cut on $\eta$, but we have found that the results do not change substantially, and the uncertainty would be significantly larger if we rejected so much data.

Finally, after all of these track cuts have been applied we check to insure there is at least one track of each sign present.  If not, then that event is rejected since we need at least one particle of each charge species to calculate the charge-dependent fluctuation measures.

\section{Cut Efficiency, Acceptance Correction, and Measure Bias}

For the accepted events, the track quality cuts accept $\sim$60\% of the viable primary track candidates.  Because we have chosen to utilize the $\Delta\sigma_{p_t}$ fluctuation measure an extrapolation to full detector acceptance based on this cut efficiency is necessary.  One might naively assume that this problem could be effectively avoided by measuring the fluctuation per particle, or using a similar ``acceptance independent" approach.  This is not the case.  The available so-called acceptance independent fluctuation measures simply avoid acceptance correction issues by harboring inherent assumptions about the scalability of measured fluctuations.  The necessary work of studying the variation in fluctuations with acceptance must be done explicitly (see section 5.6).  It cannot be avoided by making simplistic and wrong assumptions about scalability.  A great effort has been made to utilize the least biased possible fluctuation measure for this analysis.  We have found this to be the $\Delta\sigma_{p_t}$ measure that is minimally biased, but requires the aforementioned acceptance correction.

\section{The $\Delta\sigma_{p_t}$ Numerical Fluctuation Measure}

The absolute fluctuation in $<p_t>$ is trivially calculable.  However, much like the calculation of information or dimension transport in the SCA system we are not interested in measuring only the absolute fluctuations.  We want to know if the measured fluctuations differ from our expectations derived from known physics and statistics.  To calculate the non-statistical fluctuations we need a reference to tell us what we should expect in the case of an uninteresting (purely statistical fluctuation) result.  The simplest theoretical reference one can use in this comparison is the distribution that would arise from a purely statistical sampling of the parent physics.  The {\bf central limit theorem} gives us a simple form for such a reference, assuming the parent distribution is fixed.  This means that the collisions made would all have the same physical properties (centrality, temperature, etc.); each collision would simply be a different glimpse of the same physics picture.

The central limit theorem (CLT) tells us that when sampling from a fixed parent distribution $x$, the deviation of the parent will be equal to the number of samples times the deviation of the mean of $x$, as calculated from the different sampling events \cite{CLT}.  For the purposes of this analysis the parent distribution is the inclusive transverse momentum distribution (the aforementioned $x$ is $p_t$ in this case) and each collision event provides a sampling of $N_e$ particle $p_t$ values from this parent.  Thus, the CLT requires that the width of the inclusive $p_t$ distribution ($\sigma_{p_t}^2$) be equivalent to the mean event multiplicity ($\overline{N}$) times the width of the event-wise mean $p_t$ distribution ($\sigma_{<p_t>}^2$).  A comparison of these widths yields a numerical measure of the fluctuations in excess of expectation:
\begin{equation}
\Delta\sigma_{p_t}^2=\overline{N(<p_t>-\hat{p_t})^2}-\sigma_{p_t}^2.
\end{equation}
Where we use $\overline{N(<p_t>-\hat{p_t})^2}$ instead of $\overline{N}\sigma_{<p_t>}^2=\overline{N}\:\overline{(<p_t>-\hat{p_t})^2}$ because of the varying event-by-event multiplicity of accepted particles.  Of the variety of possible formulations of a CLT-based width measure this one is the least biased \cite{TATMeasureBias}. 

By calculating the $\Delta \sigma_{p_t}$ values for the different charge species independently we can investigate the charge-dependence of the non-statistical $<p_t>$ fluctuations.  We combine $\Delta\sigma_{{p_t}_{+}}$ and $\Delta\sigma_{{p_t}_{-}}$ as follows: 
\begin{equation}
\Delta{\sigma^2}_\Sigma=\frac{1}{N}(N_+\Delta\sigma^2_{{p_t}_+}+N_-\Delta\sigma^2_{{p_t}_-}+2\sqrt{N_+N_-}\Delta\sigma^2_{{p_t}_+{p_t}_-})
\end{equation}
\begin{equation}
\Delta{\sigma^2}_\Delta=\frac{1}{N}(N_+\Delta\sigma^2_{{p_t}_+}+N_-\Delta\sigma^2_{{p_t}_-}-2\sqrt{N_+N_-}\Delta\sigma^2_{{p_t}_+{p_t}_-}),
\end{equation}
to highlight the charge-dependent and -independent contributions to the non-statistical fluctuations.  As defined above, $\Delta{\sigma^2}_\Sigma$ measures the charge-independent component and $\Delta{\sigma^2}_\Delta$ measures the charge-dependent component.  

If there are some events that cross the phase boundary into the QGP region then we expect to see substantial non-statistical fluctuations in the event-by-event temperature.  This means that each collision would not be sampling from the same parent because the parent $p_t$ distribution depends on temperature ($\frac{1}{T}$ is the slope).  Significant event-by-event fluctuations in the parent distribution will show up in our fluctuation measure as width in excess of the CLT expectation ($\Delta\sigma_{p_t} > 0$).

\section{Numerical Results}

\subsection{$<p_t>$ Fluctuations in STAR Central Events at $\sqrt{s_{NN}}=130$ GeV}

For the 15\% most central $\sqrt{s_{NN}}=130$ GeV STAR events we measure the $<p_t>$ fluctuation excess to be ${\Delta\sigma_{p_t}}_{\Sigma}=52.6 \pm 0.3$ MeV and ${\Delta\sigma_{p_t}}_{\Delta}=-6.6 \pm 0.6$ MeV (errors are statistical only) with $\overline{N}=735$, $\sigma_{\hat{p_t}}=359$ MeV, and $\hat{p_t}=535$ MeV.  This analysis includes only the best quality primary tracks (see section 5.3) from 183k events. 

\begin{figure}[ht]
\centering
\includegraphics[width=4in]{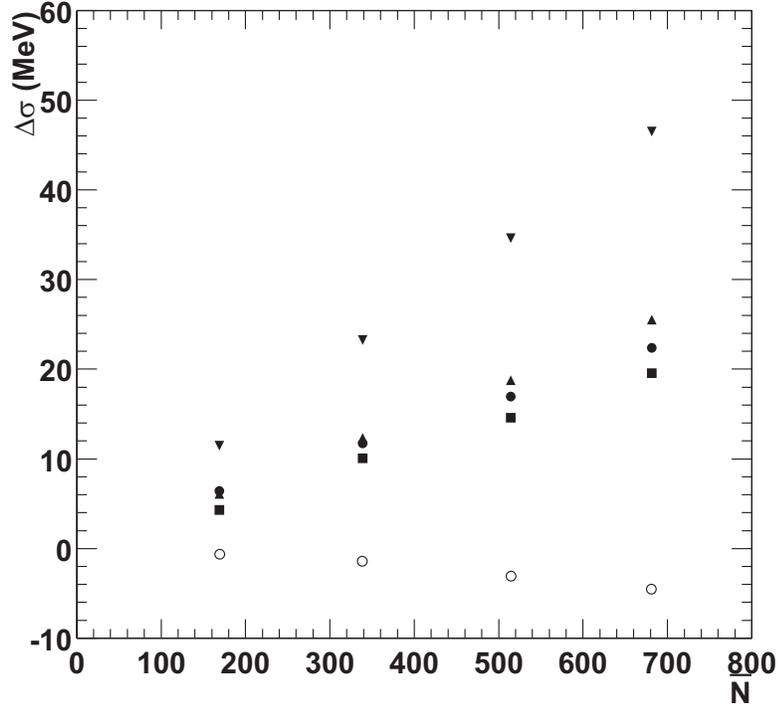}
\caption[Study of the effect on $\Delta\sigma$ of varying acceptance.]{Study of the effect on $\Delta\sigma$ of varying acceptance.  Tracks were randomly rejected in four separate analyses (75\%, 50\%, 25\%, and 0\%).  Circle, square and triangle markers are results from $\Delta\sigma_{+}$, $\Delta\sigma_{-}$, and $\Delta\sigma_{+-}$ respectively.  Inverse triangle (point down) and open circle markers show results for $\Delta\sigma_{\Sigma}$ and $\Delta\sigma_{\Delta}$.}
\label{Acceptance}
\end{figure}

This result is not the whole story, we still need to extrapolate to the full acceptance of the experiment.  To understand how to do this properly we randomly reject tracks to simulate decreasing acceptance and plot the change in the measured fluctuations with multiplicity.  For these data the acceptance variation is linear in both the sum and difference measures with slopes $m_{\Sigma}=0.0665$ and $m_{\Delta}=-0.00974$ (see figure~\ref{Acceptance}).

%
%

Since we have a measure for the full acceptance mean multiplicity \cite{STARSpectra} and this acceptance study shows a linear relationship between the measured fluctuations and this multiplicity, we can extrapolate the fluctuation results to the full detector acceptance: ${\Delta\sigma_{p_t}}_{\Sigma}=75 \pm 11$ MeV and ${\Delta\sigma_{p_t}}_{\Delta}=-9 \pm 1.4$ MeV, with $\overline{N}=1050$.  Errors on the extrapolated values reflect the full statistical and systematic error.  The dominant contribution is a $\pm15\%$ systematic error reflecting the uncertainty in the extrapolation procedure \cite{STARpt}.

\subsection{Centrality Dependent $<p_t>$ Fluctuations in STAR at $\sqrt{s_{NN}}=130$ GeV}

\begin{figure}[ht]
\centering
\includegraphics[width=5.4in]{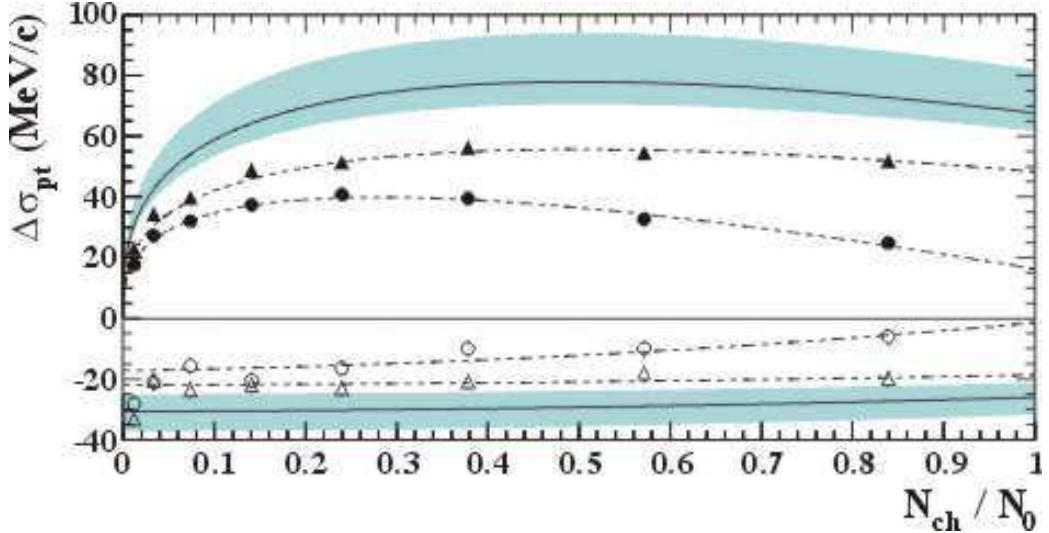}
\caption[$<p_t>$ fluctuation analysis results for 205k minimum-bias triggered events at $\sqrt{s_{NN}}=130$ GeV.]{$<p_t>$ fluctuation analysis results for 205k minimum-bias triggered events at $\sqrt{s_{NN}}=130$ GeV.  Results using quality cuts equivalent to those used on central events are shown with triangle markers.  An additional cut ($\chi^2 < 2$) was applied for results shown with circle markers.  Solid markers are $\Delta\sigma_{\Sigma}$ results, open markers are used for $\Delta\sigma_{\Delta}*3$.  The extrapolation to full acceptance is also shown with blue error bands determined by the dominant $\pm15\%$ systematic error.}
\label{mptCentrality}
\end{figure}

Using the minimum-bias trigger data centrality classes were defined by binning the total TPC track multiplicity.  The $N_{ch}$ distribution was divided into 8 centrality classes, each containing approximately 12.5\% of the events in the total minbias data sample.  Results are shown in figure~\ref{mptCentrality} for ${\Delta\sigma_{p_t}}_{\Sigma}$ and ${\Delta\sigma_{p_t}}_{\Delta}$ in these 8 centrality bins with extrapolation to full detector acceptance using the same scaling as for central data. 

%
%

\section{Systematic Error Sources}

In addition to filtering for data quality to insure robust results, great care was taken to account for sources of systematic error in the $<p_t>$ fluctuation analysis.

\subsection{Tracking Across the TPC Central Membrane}

The STAR TPC is physically divided into two separate active detector volumes by a mylar central membrane.  The TPC operates by drifting the charge deposited in the gas-filled active volume to the detectors at the endcaps.  This charge-drift occurs independently in the two TPC halves.  If the calibrations of the drift velocity of the TPC gas ($v_{drift}$) or the time at which the drift was initiated ($t_0$) are imperfect, tracks that travel across the membrane may not be properly reconstructed.  The resulting track discontinuities at the membrane can cause a significant error in the physical parameters of the reconstructed tracks.  This could cause an apparent increase in $p_t$ fluctuations.

We can insure that this effect does not significantly bias our results by checking the measured $p_t$ fluctuations of tracks in forward (positive) and backward (negative) $\eta$ separately.  If the primary vertex is in the forward part of the TPC then none of the forward directed tracks will cross the central membrane.  Similarly, for vertices in the backward part of the TPC no backward directed tracks will cross the membrane.  Thus, if we calculate the fluctuations of the tracks in the same $\eta$ region as the vertex separately from tracks in the opposite region, we can compare results for a population of tracks that have no membrane-induced discontinuities to a population that has a maximal number of membrane crossing tracks.

To insure the results of these analyses are comparable we must limit the accepted vertex positions to a small region near the central membrane; if the vertices are too far from the central membrane then the accepted multiplicities in the separate halves might differ enough that an acceptance correction would be required to make meaningful comparisons of the results.  For vertices within $25$ cm of the central membrane the same-side result is ${\Delta\sigma_{p_t}}_{\Sigma}=53.68$ MeV ($\overline{N}=378$) and the opposite-side result is ${\Delta\sigma_{p_t}}_{\Sigma}=53.66$ MeV ($\overline{N}=376$).  The result for tracks chosen independent of vertex position is ${\Delta\sigma_{p_t}}_{\Sigma}=53.67$ MeV ($\overline{N}=377$).  The differences here are certainly within our existing systematic error, so we can ignore central membrane effects with impunity.

\subsection{$\chi^2$ Cut Systematics}

In trying to insure track quality it is logical to remove tracks that do not effectively represent the path taken by a particle produced in the collision.  After the track reconstruction is performed on the TPC cluster points a goodness-of-fit is calculated ($\chi^2$) to measure how well this potential track fits the trajectory that was actually measured.  To insure only the highest track quality a strict cut of $\chi^2 < 2$ was placed on the track population.  

Unfortunately, application of this cut is problematic.  The inclusive reduced $\chi^2$ distribution is peaked at around 0.9, suggesting the error estimates for the cluster positions are too small.  Also, the $\chi^2$ distribution did not exhibit the expected $P_{\chi}$ behavior, suggesting that it does not properly measure of the goodness-of-fit.  Reconstruction experts admit \cite{Helen} that the calculated value for $\chi^2$ only loosely represents a measure of the goodness of fit of a track.  In addition to (or perhaps because of) the problems in the $\chi^2$ variable, there are a number of efficiency and acceptance problems in applying a goodness-of-fit cut.    

First of all, in very central events the tracking quality deteriorates relative to peripheral events due to the high density of clusters in the TPC.  This causes a significant broadening in the $\chi^2$ distribution for central events and a dynamic cut efficiency in centrality for the $\chi^2$ cut that favors peripheral and mid-peripheral events.  

Secondly, the cutting on reduced $\chi^2$ has inherent problems because of the broad range of degrees of freedom available to the various tracks.  Acceptable tracks can be made with as few a 5 clusters or as many as 45.  To remove bias from different numbers of fit points we have defined a bias-free form $\chi^2_{fix}=\sqrt{\nu}(\chi^2/\nu-1)$, similar to the reduced $\chi^2$.  This normalizes the widths of the reduced $\chi^2$ distributions so that a single cut rejects approximately the same percentage of tracks for varying $\nu$.  This could be achieved more precisely by changing the cut limits for different values of the number of degrees of freedom.  But this adds a level of complexity that makes characterization of the cut acceptance difficult.

Furthermore, the maximum number of clusters on a track is effected by the physical acceptance of the TPC with very large $|\eta|$ tracks having fewer possible points.  This makes the eta acceptance of the $\chi^2$ cut non-trivial and invites unintended consequences.  The acceptance dependence issue is at the heart of most of the most significant problems affecting the $\chi^2$ cut.  In addition to a complicated eta acceptance this cut also suffers from a complicated $p_t$ acceptance.  The $p_t$ acceptance has a direct effect on the measured $<p_t>$ and the measurement of the fluctuations of $<p_t>$.  Without a clear understanding of the reconstructed $\chi^2$ cut we are forced to abandon it; we can't trust the effect it will have on the $<p_t>$ distribution to preserve the physics of the collision.

\subsection{DCA Cut Systematics}

This analysis depends strongly on the exclusion of tracks that do not originate from the primary vertex.  The technical definition of a primary track in STAR is any track that passes within 3 cm of the primary vertex position, which has then been refit with the primary vertex as a track point.  This will certainly include some secondaries ({\em e.g.,} heavy resonance decays).  Ideally, by tightening our definition of a primary vertex track we decrease the total number of tracks accepted while increasing the percentage of true primaries.  Unfortunately, the DCA cut has the same problem as the $\chi^2$ cut; changing the DCA cut significantly changes our $p_t$ acceptance. 

Tracks that come directly from the primary vertex are less likely to experience a significant Coulomb scatter if they have very high $p_t$.  Thus, tightening the cut on primary vertex DCA has the unintended consequence of increasingly rejecting low $p_t$ tracks.  Even with this expected $p_t$ dependence we were surprised by the results of a DCA cut systematic study.
Varying the DCA upper limit from 3 cm (standard primary track definition) to 2 cm we find $\Delta\sigma_{p_t}=60$ MeV.  Going further and placing the DCA cut at 1 cm we find $\Delta\sigma_{p_t}=165$ MeV.  Clearly the correlation per particle is rising quite rapidly.  As we will see in the STAR results in the next chapter, taking the high or low $p_t$ populations alone increases the correlation per particle since the high-low cross-correlations contain a substantial anti-correlation component.  Thus, the sculpting of the $p_t$ distribution by both the DCA and $\chi^2$ cuts may cause abnormally high fluctuation results because of these cross-correlations.

\subsection{Elliptic Flow Punch-Through}

The measurement of significant elliptic flow effects at RHIC are well documented \cite{STARFlow} \cite{PHENIXFlow}.  We expect that there might be some effect of the elliptic flow signal on the measured non-statistical $<p_t>$ fluctuations.  To model this we sculpt the $\phi$ distribution to introduce a large, known azimuthal anisotropy.  This mimics a particle distribution that would arise from a very dramatic elliptic flow signal.  

The procedure we use is to create three different distributions with different levels of azimuthal anisotropy: 1) we discard half of the tracks in the event by removing 100\% of the tracks in back-to-back $\phi$ quadrants; it is as if the reaction plane of each event is aligned and there is no out-of-plane emission.  This inserts the largest possible elliptic flow signal into the data.  2) we discard 25\% of the tracks from two opposing quadrants and 75\% of the tracks from the others.  This models an intermediate flow signal (but still much stronger than we would reasonably expect from the data).  3) we discard half the tracks in the event independent of azimuth.  This inserts no additional flow signal beyond that present in the data.  

The results for the analysis of the simulated flow effects are as follows:  1) for the largest possible azimuthal anisotropy $\Delta\sigma_{\Sigma}=43.4$ MeV.  2) for the intermediate flow signal $\Delta\sigma_{\Sigma}=29.5$ MeV.  3) for no additional flow signal $\Delta\sigma_{\Sigma}=25.4$ MeV.  For all data sets $\overline{N} \sim 375$.  A realistic elliptic flow signal would be at least an order of magnitude smaller than the second case.  Thus, we conclude that we can safely ignore elliptic flow effects on the $\Delta\sigma_{p_t}$ measure.

\section{Graphical Fluctuation Measures}

Tannenbaum \cite{Tann} has taken the CLT reference further than a width comparison. He has shown that a Gamma distribution with parameters determined by the data (inclusive mean $p_t$, $N$) can act appropriately as a reference to the full $<p_t>$ distribution.  This provides, for the first time, a theoretical reference for this type of fluctuation analysis.  In the past the most popular graphical reference used has been a mixed-event reference.  The concept behind a mixed-event comparison is that the inclusive correlations in the data can be removed by constructing simulated events from random particles taken from different collision events.  Because these particles are taken from the same event population as the data analysis the relevant global parameters remain intact while the event-by-event correlations are removed.  

This works well in principle as an approximate CLT reference because we are, by construction, randomly sampling from the same parent as the data.  With the mixed-event $<p_t>$ distribution as a reference we expect the difference between data and reference to measure only the width contribution of the event-by-event fluctuations.  However, because the mixed-event reference is only a numerical approximation of the true CLT reference (the gamma distribution) there is no point in wasting the significant computational effort required for its construction.  There is significant emotional attachment within the community to the mixed-event reference that many find comfortable and familiar, but we reject it in favor of the more precise, and less wasteful CLT reference.  Why use a numerical approximation to a theoretical reference that can be trivially derived?

\section {Graphical Data Comparisons}

For the graphical data/reference comparison we choose to present the $<p_t>$ distribution in a format that is portable.  This allows for the comparison of results from different experiments with varying acceptance and energy.  Subtracting the inclusive mean $p_t$ and dividing by the width (second moment) of the distribution puts the $<p_t>$ distribution on a universal axis where the mean is at the origin, and deviations from the mean are measured in units of the width of the distribution.  This is compared to a reference gamma distribution whose parameters are determined by the data:
\begin{eqnarray}
g_{\overline{n}}(<p_t>)&=&\frac{\alpha_0}{\hat{p_t}}\frac{e^{-\alpha_0\overline{n}<p_t>/\hat{p_t}}}{(\alpha_0\overline{n}-1)!}(\alpha_0\overline{n}\frac{<p_t>}{\hat{p_t}})^{\alpha_0\overline{n}-1}. \\ \nonumber
\end{eqnarray}

%
\begin{figure}[ht]
\centering
\includegraphics[width=4in]{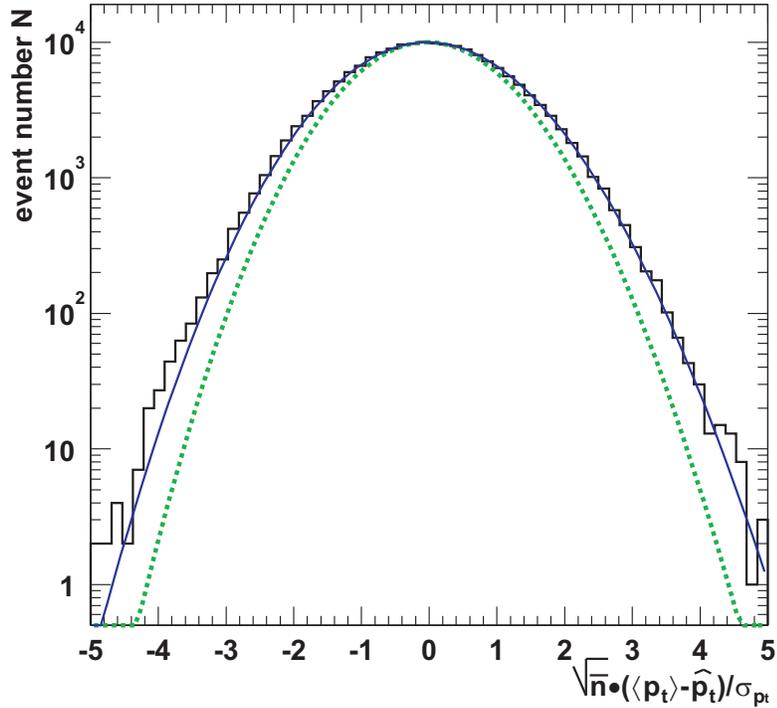}
\caption[The $<p_t>$ distribution for 183k STAR cental events at $\sqrt{s_{NN}}=130$ GeV.]{The $<p_t>$ distribution for 183k STAR cental events at $\sqrt{s_{NN}}=130$ GeV.  The data is presented in the aforementioned universal format along with two references.  The green dotted line is the reference gamma $g_{\overline{n}}(<p_t>)$.  The blue solid line is the same reference gamma scaled to include the non-statistical excess width measured in the numerical analysis (see section 5.6).}
\label{mPtvG}
\end{figure}
%

\begin{figure}[ht]
\centering
\includegraphics[width=4in]{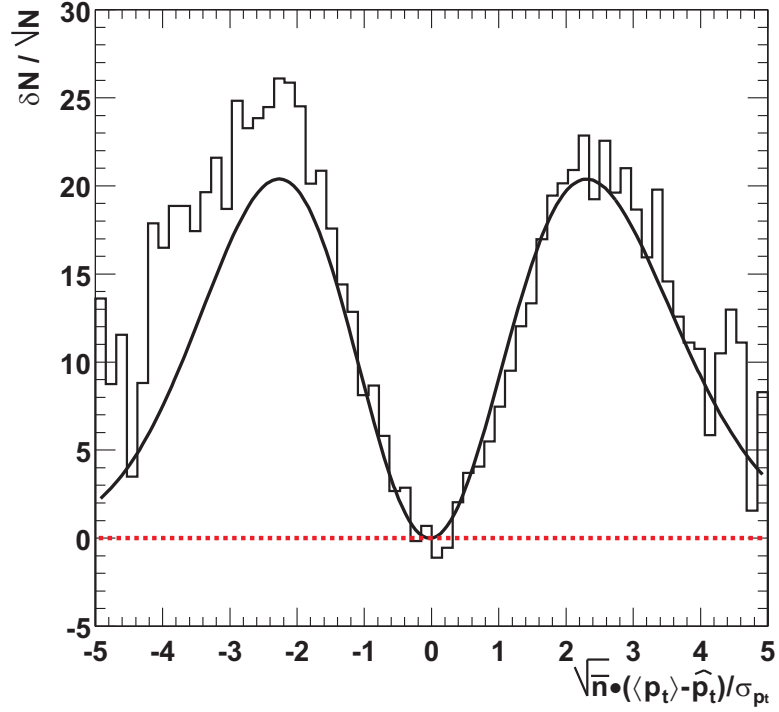}
\caption[The normalized difference between the STAR central $<p_t>$ data and a CLT reference.]{The normalized difference between the STAR central $<p_t>$ data and a CLT reference.  The solid black line shows the difference between the CLT gamma (red dashed line) and the excess width-scaled gamma.}
\label{mPtvGd}
\end{figure}

Figure~\ref{mPtvG} shows the data and reference $<p_t>$ distribution comparisons.  The data is in excellent agreement with the gamma reference that has been scaled to include the excess width measured in the numerical analysis.  The slight differences between the data and width-scaled gamma are highlighted in a difference measure plot (see figure~\ref{mPtvGd}).

%
%

Graphically we observe a significant fluctuation excess in close agreement with our numerical results in the $-2 < \sigma < 2$ range.  Farther out in the $<p_t>$ distribution the data is somewhat higher than the width-scaled gamma reference.  This is the region where the error is largest, nevertheless there does seem to be a significant trend with the data that is not addressed by the reference gamma.  However, keeping this result in context, these apparently anomalous events consist of a few hundred out of a sample of 183k events ($\sim$0.2\%).  Additionally, beyond the $5\sigma$ range of these plots there were 19 outlier events that were rejected from the analysis.  Most of these events had anomalous $<p_t>$ values due to known detector and reconstruction software problems.

Beyond clear confirmation of our numerical result the graphical comparison allows us to rule out the existence of a small class (between 1 and 10 percent) of anomalous events above statistical expectation at high $<p_t>$ values.  Early predictions \cite{EarlyMeanpt} suggested that QGP events might present themselves in such a direct way.

\section{Comparisons With Other Experiments}

\begin{figure}[ht]
\centering
\includegraphics[width=5in]{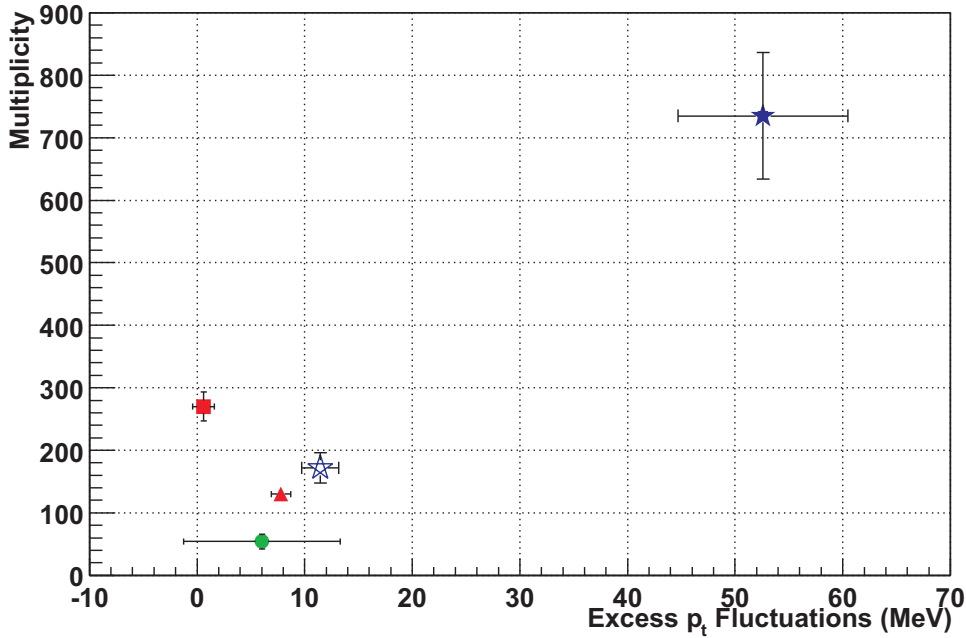}
\caption[$<p_t>$ fluctuation world data set for heavy-ion experiments.]{$<p_t>$ fluctuation world data set for heavy-ion experiments.  Red points are at SPS energies ($\sqrt{s_{NN}}=17$ GeV), green and blue at RHIC ($\sqrt{s_{NN}}=130$ GeV).  The square marker corresponds to NA49 data taken at forward rapidity ($4 < y_{\pi,cm} < 5.5$) \cite{NA49pt}.  The triangle marker represents the CERES/NA45 data taken at mid-rapidity ($2.1 < \eta_{lab} < 2.6$) \cite{CERESpt}.  The circle marks the PHENIX data taken at $|\eta| < 0.35$ in $\Delta\phi=58.5^{\circ}$ \cite{PHENIXpt}.  The filled star represents the STAR measurement in the widest acceptance $|\eta| < 1$ with full angle coverage.  The open star is the result from the STAR acceptance study in which 75\% of the tracks were randomly rejected.}
\label{world}
\end{figure}

Comparisons of non-statistical fluctuation measurements are tricky.  The measures of choice ($\Phi_{p_t}$ and $\Delta\sigma_{p_t}$) are approximately equal algebraically, but their explicit multiplicity dependence makes it impossible to directly compare values from different experiments.  Because each experiment applys a different set of quality cuts and operates with different angle and rapidity coverage there is no way to make direct comparisons.  However, it is important to survey what is known and has been measured to appreciate the context of the STAR result.  Thus, keeping in mind the caveat that these data represent significantly different measurements, we present the world data set for charge-independent non-statistical $<p_t>$ fluctuations (see figure~\ref{world}).

%
%

We have chosen to plot the fluctuation measures against multiplicity to highlight the variation in results with acceptance.  
The error bars on the fluctuation measures are statistical and systematic errors taken together, these are dominated by systematic errors for all experiments.  The errors presented on the multiplicity value are taken to be the width of the multiplicity distribution of events used in the different analyses.  There is no multiplicity error bar reported on the CERES result since the width of the multiplicity distribution is not reported in \cite{CERESpt}.

It is worth noting that the CERES and PHENIX results are consistent with the line determined by the STAR acceptance dependence study.  This is a provocative result, but it is by no means definitive.  If each experiment independently observed the same (random-rejection) acceptance dependence slope this might be meaningful.  Otherwise, it is just coincidence, which is likely considering the PHENIX error bar and acceptance.  The STAR measurement for the same multiplicity as PHENIX ($\overline{N}\sim 60$) is $\Delta\sigma_{\Sigma}=2$ MeV.  If we improve on this comparison by abandoning random-rejection in favor of cutting the STAR acceptance to match exactly that of the PHENIX measurement ($|\eta| < 0.35$ in $\Delta\phi=58.5^{\circ}$) the STAR result becomes $\Delta\sigma_{\Sigma}=8$ MeV \cite{QJLiu}.  Both results are within the broad PHENIX error bar, but the direct acceptance comparison is clearly superior.  This shows precisely how these results are highly dependent on not just the accepted multiplicity, but how those accepted tracks are chosen.  Thus, no direct comparison to the NA49 result is possible since it is measuring tracks in a completely different rapidity region from the other measurements.  It is entirely possible that there is a rapidity dependence of the non-statistical fluctuations that causes them to vanish outside of central rapidities, explaining the apparent disagreement between NA49 and the other experiments.

\section{Conclusions}

In conclusion, we have a striking result from this first fluctuation analysis in STAR.  We measure a 14\% $<p_t>$ fluctuation excess for the full acceptance corrected charge-independent analysis.  This is a provocative result that is not inconsistent with the previous measurements of non-statistical fluctuations, but also stands on its own as unique.  No other experiment has been able to measure fluctuations with so many primary particles.  As for interpretations of this result, the unique centrality dependence might be explained by suppression of initial state scattering effects in central collisions \cite{TATISS}. 

The charge-dependent results are consistent with slight global temperature fluctuations, and are compatible with the two-particle correlation results that will be discussed in the next chapter.  The comparison of the charge-dependent results to the NA49 measurement at SPS energies suggests that these fluctuations are slightly smaller at RHIC and deserve further attention.

%
%
%

\chapter{Two-Particle Correlation Analysis}

\section{Introduction}

One of the most basic analytical methods used in characterizing nuclear interactions is multiparticle correlation analysis.  It is present in Hanbury-Brown Twiss interferometry \cite{HBT}, intermittency \cite{Intermit}, $\Phi_{p_t}$ calculations \cite{PhiPt}, and a variety of other methods.  For the event-by-event multiparticle correlation analysis presented here we have chosen to focus entirely on two-particle correlations.

The $\Phi_{p_t}$ variable ($\Phi_{p_t} \equiv \sqrt{\frac{\overline{Z^2}}{\overline{N}}}-\sigma_{p_t}$) provides strong motivation for using two-particle correlations for event-by-event physics analysis.  To understand the relationship between fluctuations and correlations we must understand the components of $\Phi_{p_t}$, most importantly $\overline{Z^2}=\overline{N^2(<p_t>-\overline{p_t})^2}$.  This is the data piece of a data/central limit reference comparison.  We start our investigation of $\Phi_{p_t}$ by breaking $\overline{Z^2}$ into two parts: a central limit term from which interesting correlations deviate, and an excess fluctuation term that measures all of the correlations beyond those expected from the CLT.  
\begin{eqnarray}
\overline{Z^2}&=&\frac{1}{M}\sum_{e=1}^{M}[N^2_e(\frac{1}{N_e}\sum_{i=1}^{N_e}p_{t_i}-\overline{p_t})^2] \\ \nonumber
&=&\frac{1}{M}\sum_{e=1}^{M}[(\sum_{i=1}^{N_e}p_{t_i}-N_e\overline{p_t})^2] \\ \nonumber
&=&\frac{1}{M}\sum_{e=1}^{M}[(\sum_{i=1}^{N_e}p_{t_i})^2-2N_e\overline{p_t}(\sum_{i=1}^{N_e}p_{t_i})+N_e^2\overline{p_t}^2] \\ \nonumber
&=&\frac{1}{M}\sum_{e=1}^{M}[(\sum_{i,j=1}^{N_e}p_{t_i}p_{t_j})-2N_e^2<p_t>_e\overline{p_t}+N_e^2\overline{p_t}^2] \\ \nonumber
&=&\frac{1}{M}\sum_{e=1}^{M}[(\sum_{i\neq j}^{N_e}p_{t_i}p_{t_j})+(\sum_{i=1}^{N_e}p_{t_i}^2)-2N_e^2<p_t>_e\overline{p_t}+N_e^2\overline{p_t}^2] \\ \nonumber
&=&\overline{\sum_{i\neq j}^{N_e}p_{t_i}p_{t_j}}+\frac{1}{M}\sum_{e=1}^M(\sum_{i=1}^{N_e}p_{t_i}^2)-\overline{2N_e^2<p_t>_e\overline{p_t}}+\overline{N_e^2\overline{p_t}^2} \\ \nonumber
&=&\overline{\sum_{i\neq j}^{N_e}p_{t_i}p_{t_j}}+\frac{1}{M}{\cal P}^2_t-2\overline{p_t}\overline{N_e^2<p_t>_e}+\overline{N^2}\overline{p_t}^2 \\ \nonumber
&=&\overline{\sum_{i\neq j}^{N_e}p_{t_i}p_{t_j}}+\overline{N}\overline{p_t^2}-2\overline{p_t}\overline{N_e^2<p_t>_e}+\overline{N^2}\overline{p_t}^2
\end{eqnarray}
In this expression of $\overline{Z^2}$ we have used the fact that the total transverse momentum squared for all particles in all events can be written as ${\cal P}_t^2=M\overline{N}\overline{p_t^2}$.  With $\overline{Z^2}$ broken down in this form the path is clear, we must add and subtract the terms we need to get the requisite central limit term:
\begin{eqnarray}
\overline{Z^2}&=&\overline{Z^2}+(\overline{N}\overline{p_t}^2-\overline{N}\overline{p_t}^2)+(\overline{N^2}\overline{p_t}^2-\overline{N^2}\overline{p_t}^2) \\ \nonumber
&=&\overline{\sum_{i\neq j}^{N_e}p_{t_i}p_{t_j}}+\overline{N}\overline{p_t^2}-2\overline{p_t}\overline{N_e^2<p_t>_e}+\overline{N^2}\overline{p_t}^2+(\overline{N}\overline{p_t}^2-\overline{N}\overline{p_t}^2)+(\overline{N^2}\overline{p_t}^2-\overline{N^2}\overline{p_t}^2) \\ \nonumber
&=&[\overline{\sum_{i\neq j}^{N_e}p_{t_i}p_{t_j}}-\overline{N^2}\overline{p_t}^2+\overline{N}\overline{p_t}^2]-[2\overline{p_t}\overline{N_e^2<p_t>_e}-2\overline{N^2}\overline{p_t}^2]+[\overline{N}\overline{p_t^2}-\overline{N}\overline{p_t}^2] \\ \nonumber
&=&[\overline{\sum_{i\neq j}^{N_e}p_{t_i}p_{t_j}}-\overline{N_e(N_e-1)}\overline{p_t}^2]-2\overline{p_t}[\overline{N_e^2<p_t>_e}-\overline{N^2}\overline{p_t}]+\overline{N}\sigma_{p_t}^2. \\ \nonumber
\end{eqnarray}

Thus, if we define $A^2\equiv \overline{\sum_{i\neq j}^{N_e}p_{t_i}p_{t_j}}-\overline{N_e(N_e-1)}\overline{p_t}^2$ and $B^2\equiv 2\overline{p_t}[\overline{N_e^2<p_t>_e}-\overline{N^2}\overline{p_t}]$, we can write $\overline{Z^2}=A^2-B^2+\overline{N}\sigma_{p_t}^2$.  Substituting back into the definition of $\Phi_{p_t}$ we see why this formulation is useful:
\begin{eqnarray}
\Phi_{p_t}&=&\sqrt{\frac{\overline{Z^2}}{\overline{N}}}-\sigma_{p_t} \\ \nonumber
&=&\sqrt{\frac{A^2-B^2+\overline{N}\sigma_{p_t}^2}{\overline{N}}}-\sigma_{p_t} \\ \nonumber
&=&\sqrt{\frac{A^2-B^2}{\overline{N}}+\sigma_{p_t}^2}-\sigma_{p_t}. \\ \nonumber
\end{eqnarray}
When $A^2-B^2$ vanishes, then $\Phi_{p_t}$ also vanishes.  Thus, the meaning of $\Phi_{p_t}$ as a correlation measure hinges on the correlations measured by $A^2$ and $B^2$.

\section{The $A^2$ Term of $\Phi_{pt}$}

First, consider $A^2=\overline{\sum_{i\neq j}^{N_e}p_{t_i}p_{t_j}}-\overline{N_e(N_e-1)}\overline{p_t}^2$.  It's first term is the sum over the product of all transverse momentum pairs in an event (excluding self-pairs).  Thus, if we create a $p_t\otimes p_t$ space where each point in the space is defined by the ordinal pair ($p_{t_i},p_{t_j}$) we can calculate the first term in $A^2$ by integrating the product of the pair values over all pairs in this space.  Note that for a given event $e$ this space will be populated with $N_e(N_e-1)$ pairs.  This is simply the two-point correlation integral.  The second $A^2$ term can now be seen to be a reference term that is the integral over a space that is populated by $N_e(N_e-1)$ pairs each with value ($\overline{p_t},\overline{p_t}$).  Thus, $A^2$ is a measure of the two-particle correlations in the event sample beyond what you would expect to see from the simplest possible transverse momentum distribution.  This motivates us to look at the full two-particle transverse momentum correlation space.  We choose to look at the full correlation space instead of the reference subtracted integral to avoid the problems of $\Phi_{pt}$ and other central limit measures that summarize all correlations in a single number.  

While an integral measure such as $A^2$ may seem to be easily interpretable, there is some danger in using a single number to represent the correlation content of the data.  Significant correlations in one part of correlation space may conspire with equally significant anti-correlations in another region to add to zero, and thus present the erroneous picture that the data in question is trivially correlated.  The only way to unambiguously address the full correlation content of the data is to consider the full, unintegrated correlation space.

\section{The $B^2$ Term of $\Phi_{pt}$}

For completeness we must also examine the $B^2$ term.  It's meaning becomes clear if we rearrange it in terms of the total transverse momentum in an event ${P_t}_e$ and make the approximation $\overline{N^2}\approx\overline{N}^2$:
\begin{eqnarray}
B^2&\equiv&2\overline{p_t}[\overline{N_e^2<p_t>_e}-\overline{N^2}\overline{p_t}] \\ \nonumber
&=&2\overline{p_t}[\overline{N_e{P_t}_e}-\overline{N^2}\overline{p_t}] \\ \nonumber
&\approx&2\overline{p_t}[\overline{N_e{P_t}_e}-\overline{N}^2\overline{p_t}] \\ \nonumber
&\approx&2\overline{p_t}[\overline{N_e{P_t}_e}-\overline{N} \overline{P_t}] \\ \nonumber
&\approx&2\overline{p_t}\sigma^2_{N,P_t}.
\end{eqnarray}
Because $B^2\propto\sigma^2_{N,P_t}$, the covariance between the event multiplicity and the total transverse momentum of the event, we know that it is a measure of correlation between $N$ and $P_t$.  Of course, we expect to see a significant correlation between multiplicity and total transverse momentum from energy conservation in the collision.  A high-energy collision will produce more particles with a higher total $p_t$ than a low-energy collision.  Energy conservation dictates that $B^2$ must be significantly non-zero.  There is no such clear restriction on $A^2$, suggesting that in the absence of significant two-particle correlations $\Phi_{p_t}$ will be abnormally low due to energy conservation.  Studies done on the covariance of $A^2$ and $B^2$ suggest that there may be a substantial contribution to $A^2$ from energy conservation as well \cite{GRoland}, but no understanding of the physical mechanism for this has been reached.

\section{Flattening the $m_t \otimes m_t$ Space}  

In generating two-particle $p_t$ correlation spaces there are some technical tricks that we can use to make the task simpler.  The biggest problem we face here is the variation of statistical power with $p_t$.  The frequency of high $p_t$ particle production vanishes exponentially, so correlation measures in the high $p_t$ region will have large error bars.  We are saved from an exponentially increasing error by redefining the variable we use in the analysis.  The $\frac{1}{m_t}\frac{dN}{dm_t}$ distribution is truly exponential for particles emitted from a thermalized source \cite{ThermalMt} and can be flattened with a temperature hypothesis and a simple transformation.  Since $p_t$ and $m_t$ are closely related such a move lets us approach the same physics from a slightly different (and advantageous) perspective.  

The flattening transformation rebins the $m_t$ distribution with bins of exponentially increasing width designed so that all bins will have roughly equivalent occupancy if the thermalized collision has temperature $T$.  This provides uniform statistics, which allows us to make direct comparisons between different regions of correlation space.  Such a transformation can neatly map the $m_t$ distribution ($[0,\infty ) \rightarrow [0,1)$) in such a way as to include the full $m_t$ range while highlighting the soft physics region ($0.2 < m_t < 0.8$) in which we are most interested.

To flatten the transverse mass distribution based on a temperature hypothesis we follow the example of \cite{NumRep}.  We want the function that will take a pure exponential density distribution ($\rho (m_t) = e^{\frac{-m_t}{T}}$) on $\frac{1}{m_t} \frac{dN}{dm_t}$ and transform it to a uniform density distribution ($\rho(x) = 1$) on $\frac{dN}{dx}$.  Using a substitution for the unitless quantity $y\equiv \frac{m_t}{T}$ we can express the relationship between the two spaces as:
\begin{eqnarray}
\rho (y) dy&=&{\vert\frac{dx}{dy}\vert}dy=e^{-y} dy.
\end{eqnarray}
Solving for x will yield the flattening function, so we solve the above equation for dx and integrate:
\begin{eqnarray}
x&=&\int dx=\int^y_0 y'e^{-y'} dy' \\ \nonumber
&=&[-y'e^{-y'}]^y_0 +\int^y_0 e^{-y'} dy' \\ \nonumber
&=&-ye^{-y}-[e^{-y'}]^y_0=-(ye^{-y}+e^{-y}-1) \\ \nonumber
&=&1-(1+\frac{m_t}{T})e^{-\frac{m_t}{T}}.
\end{eqnarray}

Using this transformation we can map any physical $m_t$ value into the unit interval, $m_{\pi} < m_t < \inf \mapsto 1-(1+\frac{m_{\pi}}{T})e^{\frac{m_{\pi}}{T}} < x < 1$.  This is a little unwieldy and we would rather have x range from zero to 1, so we just shift x down by the offending factor and rescale.  This yields the desired mapping:
\begin{eqnarray}
x&=&1-[(1+\frac{m_t}{T})e^{-\frac{m_t}{T}}][(1+\frac{m_{\pi}}{T})e^{-\frac{m_{\pi}}{T}}]^{-1}.
\end{eqnarray}

%
\begin{figure}[ht]
\centering
\includegraphics[width=4in]{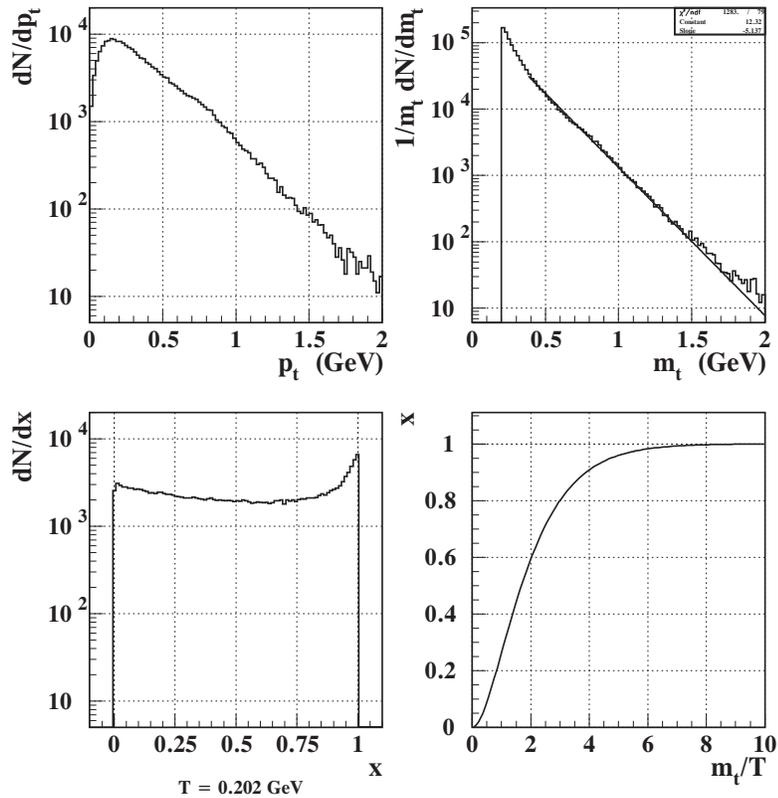}
\caption{Flattening transformation $m_t \rightarrow x$ applied to NA49 data with temperature hypothesis $T=202~MeV$.}
\label{mtTrans}
\end{figure}
%

Figure~\ref{mtTrans} shows this transformation applied to NA49 data.  A fit of the $\frac{1}{m_t}\frac{dN}{dm_t}$ vs. $m_t$ distribution gives us a good temperature hypothesis that is then incorporated into the flattening transformation.  Once the transformation is applied the distribution on $x$ is not perfectly flat because of the physical features in the $m_t$ distribution that deviate from thermal.  However, by selecting the proper reference any features present in the inclusive distribution can be removed from the problem.

\section{Sibling and Mixed Pair Spaces}

To extract event-scale physics from the full two-particle correlation space we will need to find a way of separating the
inclusive correlations in the space from the event-by-event correlation content.  To form a two-particle correlation space with both the inclusive and event-scale correlations we need only to form a space from pairs of particles taken within a single event.  This {\bf sibling pair} space is populated by all possible non-self-pairs for each event we are analyzing.  If we have $M$ events, each with $N$ particles, the sibling pair space will be populated by $M*N*(N-1)$ pairs.  To remove the inclusive correlation content from this space we must have an inclusive reference pair space for comparison.  The reference space we choose to use is a {\bf mixed pair} space.  This is a space made of pairs of particles where each particle in a pair comes from an event different from the other pair particle.  This is useful because the space of particle pairs from different events contains no event-by-event physics; we have designed the space to contain only correlations present in the inclusive distribution.  Thus, the mixed pair reference will contain any inclusive physics correlations and systematic effects that arise in the inclusive distribution (such as acceptance effects).  Now that we have an idea of how to form the data and reference correlation spaces we must consider carefully the details of generating these spaces and making comparisons between them.

\section{Forming Two-Point Correlation Spaces}

Single-point distributions from distinct events $a$ and $b$ can be used to generate pairs by taking the coordinates of two points from the single-point distributions as the x and y coordinates of a point in a two-dimensional pair space.  We can generate two distinct types of pairs this way: those using points from the same event (sibling pairs: $a\otimes a$, $b\otimes b$) and those using points from different events (mixed pairs: $a\otimes b$, $b\otimes a$).

\begin{figure}[ht]
\centering
\includegraphics[width=5in]{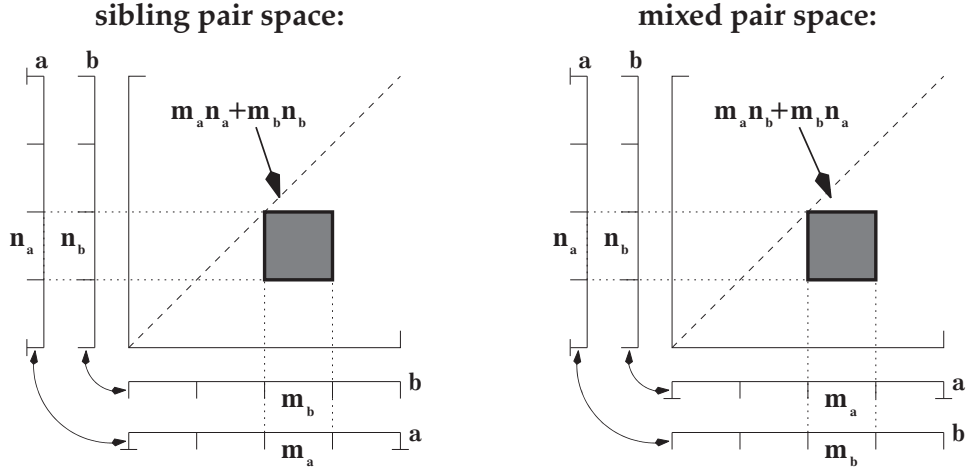}
\caption{A cartoon of the method used to construct two-point histograms from single-point distributions for events a and b.}
\label{evprod}
\end{figure}

When we have generated all the sibling and mixed point pairs for the two events we can bin the two-dimensional pair spaces and make comparisons between the 2d mixed and sibling pair histograms.  For off-diagonal bins in the sibling and mixed pair histograms we determine the bin contents to be ${\cal S}=m_{a}n_{a}+m_{b}n_{b}$ and ${\cal M}=m_{a}n_{b}+m_{b}n_{a}$ respectively, where we have defined single-point distribution bin contents as in figure~\ref{evprod} (following the example of \cite{Zajc}).

%
%

With this prescription for creating sibling and mixed pair distributions, a Monte Carlo study was carried out.  Approximately 40,000 single-point distributions of 160 points each were generated from uniform random distributions of points on the interval [0,1).  We binned the inclusive single-point distribution with a 25-bin histogram.  Using simple Poisson statistics we determined the root-mean-square error on the occupancy of a single bin in this histogram to be $2\cdot10^{-3}$.  From the single-point distributions of these Monte Carlo events we created sibling and mixed pair distributions as outlined in figure~\ref{evprod}.  These spaces were binned using 25x25 bin two-dimensional histograms.  If the pairs in these pair spaces were randomly distributed on the 2d space, the {\em rms} error on the occupancy of a single 2d histogram bin would be $0.8\cdot10^{-3}$.  This, however, is not what was observed in the simulation.  Instead, the error on the occupancy of a bin in the pair space exceeded $10^{-3}$.  This discrepancy in errors arises from correlations due to single-point statistical fluctuations appearing in the pair space as a ``statistical plaid", which has an amplitude comparable to the {\em rms} fluctuations in the single-point distribution ($2\cdot10^{-3}$).  The single-point errors are present in the 2d comparison measure because in this study we generated the sibling and mixed pair spaces using different events.  If we had used the same pool of events for generating the two spaces (as we will outline in detail in the next section) these errors would cancel in the comparison space and we would be left with an error appropriate to 2d Poisson statistics.

In the present example, if the mixed pair reference is properly formed then the projected marginal of the ratio distribution should have an {\em rms} amplitude of ${0.8\cdot10^{-3}}/{\sqrt{25}} \sim 0.16\cdot10^{-3}$ (uncorrelated 2d errors).  If not, this marginal error could be as large as the initial single-point error ($2\cdot10^{-3}$) illustrating a dramatic difference between correct and incorrect reference construction.

\section{Error Minimization in Two-Point Correlation Comparisons}

To understand how to remove the correlations in the pair spaces that arise from statistical fluctuations of the single-point or marginal distributions we focus on calculating contributions to the expected variance in the pair-space histogram bins.  We can make this task easier by rewriting the single-point histogram bin contents ($m_a$, $m_b$, $n_a$, $n_b$) to separate the mean behavior from fluctuations about the mean: $m_a\equiv m+\mu_a$, $m_b\equiv m+\mu_b$, $n_a\equiv n+\nu_a$, and $n_b\equiv n+\nu_b$.  Here we have defined $m$ and $n$ to be the mean single-point histogram bin contents relevant to the pair-space bin of interest ($m=\frac{1}{\cal N}\sum m_i$ where the sum is over all events ${\cal N}$, similarly for $n$).  Fluctuations about this mean for a particular event and bin being considered are then $\mu_a$, $\nu_a$, etc.  This formulation allows us to rewrite $\cal S$ and $\cal M$ as:
\begin{eqnarray}
{\cal S}&=&(m+\mu_a)(n+\nu_a)+(m+\mu_b)(n+\nu_b) \\ \nonumber
&=&2mn+m(\nu_a+\nu_b)+n(\mu_a+\mu_b)+\mu_a\nu_a+\mu_b\nu_b
\end{eqnarray}
\begin{eqnarray}
{\cal M}&=&(m+\mu_a)(n+\nu_b)+(m+\mu_b)(n+\nu_a) \\ \nonumber
&=&2mn+m(\nu_a+\nu_b)+n(\mu_a+\mu_b)+\mu_a\nu_b+\mu_b\nu_a,
\end{eqnarray}
which along with the mean values of ${\cal S}$ and ${\cal M}$, $\overline{\cal S}=2mn=\overline{\cal M}$, allows us to 
calculate the variances.
\begin{eqnarray}
\sigma_{\cal S}^2&=&<[{\cal S}-2mn]^2> \\ \nonumber
&=&<[m(\nu_a+\nu_b)+n(\mu_a+\mu_b)+\mu_a\nu_a+\mu_b\nu_b]^2> \\ \nonumber
&=&<[m\nu+n\mu]^2+2[m\nu+n\mu][\mu_a\nu_a+\mu_b\nu_b]+[\mu_a\nu_a+\mu_b\nu_b]^2> 
\end{eqnarray}
\begin{eqnarray}
\sigma_{\cal M}^2&=&<[{\cal M}-2mn]^2> \\ \nonumber
&=&<[m(\nu_a+\nu_b)+n(\mu_a+\mu_b)+\mu_a\nu_b+\mu_b\nu_a]^2> \\ \nonumber
&=&<[m\nu+n\mu]^2+2[m\nu+n\mu][\mu_a\nu_b+\mu_b\nu_a]+[\mu_a\nu_b+\mu_b\nu_a]^2> 
\end{eqnarray}
\begin{eqnarray}
\sigma_{\cal SM}^2&=&<[{\cal S}-2mn][{\cal M}-2mn]> \\ \nonumber
&=&<[m\nu+n\mu+\mu_a\nu_a+\mu_b\nu_b][m\nu+n\mu+\mu_a\nu_b+\mu_b\nu_a]> \\ \nonumber
&=&<[m\nu+n\mu]^2+[m\nu+n\mu][\mu_a\nu_a+\mu_a\nu_b+\mu_b\nu_a+\mu_b\nu_b] \\ \nonumber
&&+[\mu_a\nu_a+\mu_b\nu_b][\mu_a\nu_b+\mu_b\nu_a]> 
\end{eqnarray}

We have defined $\mu\equiv \mu_a+\mu_b$ and $\nu\equiv \nu_a+\nu_b$ for simplification.  If we now consider the difference histogram ${\cal D}={\cal S}-{\cal M}$ where $\cal S$ and $\cal M$ are created from the same 2 events, the variance of $\cal D$ is simply:
\begin{eqnarray}
\frac{\sigma_{\cal D}^2}{{\cal D}^2}&=&\frac{1}{\overline{\cal S}^2}(\sigma_{\cal S}^2+\sigma_{\cal M}^2-2\sigma_{\cal SM}^2)=\frac{\sigma_{\cal S}^2}{\overline{\cal S}^2}+\frac{\sigma_{\cal M}^2}{\overline{\cal M}^2}-\frac{2\sigma_{\cal SM}^2}{\overline{\cal S}\,\overline{\cal M}} \\ \nonumber
&=&\frac{1}{(2mn)^2}(<[\mu_a\nu_a+\mu_b\nu_b]^2+[\mu_a\nu_b+\mu_b\nu_a]^2 \\ \nonumber
&&-2[\mu_a\nu_a+\mu_b\nu_b][\mu_a\nu_b+\mu_b\nu_a]>) \\ \nonumber
&=&\frac{<[\mu_a\nu_a+\mu_b\nu_b-\mu_a\nu_b-\mu_b\nu_a]^2>}{(2mn)^2} \\ \nonumber
&=&\frac{<[(\mu_a-\mu_b)(\nu_a-\nu_b)]^2>}{(2mn)^2} 
\end{eqnarray}

If 1) the event multiplicity is approximately constant and 2) the parent distribution for the event ensemble doesn't change significantly from event to event then the deviations from the mean bin contents in the marginal distributions are exactly what we would expect from simple Poisson statistics.  Thus, $\mu_a\sim\sqrt{m}\sim\mu_b$ and $\nu_a\sim\sqrt{n}\sim\nu_b$.  Using these assumptions, if the sibling and mixed pairs are made from precisely the same event sample then the variance of the sibling-to-mixed difference is of order $(mn)^{-1}\sim \sqrt{\cal S}\sim \sqrt{\cal M}$.  Not surprisingly, we find this is also the variance of the sibling-to-mixed ratio (${\cal R}=\frac{\cal S}{\cal M}$):

\begin{eqnarray}
\frac{\sigma_{\cal R}^2}{{\cal R}^2}&=&\frac{\sigma_{\cal S}^2}{\overline{\cal S}^2}+\frac{\sigma_{\cal M}^2}{\overline{\cal M}^2}-\frac{2\sigma_{\cal SM}^2}{\overline{\cal S}\overline{\cal M}}=\frac{\sigma_{\cal S}^2+\sigma_{\cal M}^2-2\sigma_{\cal SM}^2}{(2mn)^2} \\ \nonumber
&=&\frac{<[(\mu_a-\mu_b)(\nu_a-\nu_b)]^2>}{(2mn)^2}
\end{eqnarray}

Now, consider sibling and mixed pairs created from different event populations ({\em e.g.,} sibling made from $a\otimes a$, $b\otimes b$; mixed made from $c\otimes d$, $d\otimes c$).  Here the correlations from single-point distribution fluctuations in the two pair spaces will differ, and there will be no cancelation of terms as before.  Thus, in this case, the leading term in the variance of the difference is: $(<[m\nu_{\cal S}+n\mu_{\cal S}]^2-[m\nu_{\cal M}+n\mu_{\cal M}]^2>)(2mn)^{-2}$, which is of order $\frac{m^2n+n^2m}{(2mn)^2}=\frac{m+n}{4mn}$.  Also, the covariant error between ${\cal S}$ and ${\cal M}$ vanishes when the pair spaces are made from different events, $\sigma_{\cal D}^2=\sigma_{\cal S}^2+\sigma_{\cal M}^2$, which allows us to confirm the order of the error on ${\cal D}$ in this case by using a simple error calculation:
\begin{eqnarray}
\frac{\sigma_{\cal M}^2}{{\overline{\cal M}}^2}\sim\frac{\sigma_{\cal S}^2}{{\overline{\cal S}}^2}&=&\frac{1}{(2mn)^2}\left(\left[\frac{\partial {\cal S}}{\partial m}\right]^2\sigma_m^2+\left[\frac{\partial {\cal S}}{\delta n}\right]^2\sigma_n^2\right) \\ \nonumber
&=&\frac{m^2n+n^2m}{(2mn)^2}=\frac{mn(m+n)}{4m^2n^2}=\frac{m+n}{4mn}
\end{eqnarray}

Thus, we can understand how the error on the comparison measures between the sibling and mixed pair histograms behaves for the two different cases. In the case of pair spaces generated from the same single-point data, the single-point statistical fluctuations in those data cancel in the difference or ratio, revealing the true two-point statistical fluctuations of a 2d distribution and the physics of interest.  In the case where the sibling and mixed pair spaces are generated from {\em different} data, the fluctuations in the 1d distributions do not cancel and so they dominate the error in the 2d space.

\section{Event Multiplicity Fluctuations}

In the previous section we assumed that the event multiplicity was constant.  In order to minimize the error this approximation is necessary because the fluctuations away from the mean bin contents for each event ($\mu_a$, $\nu_a$, etc.) include any variation in multiplicity in addition to counting statistics.  To study the effect of multiplicity variation on the {\em rms} error of ${\cal R}$ we have simulated 4 sets of $\sim$40k events each with varying levels of multiplicity fluctuation. 

%
\begin{figure}[ht]
\centering
\includegraphics[width=5in]{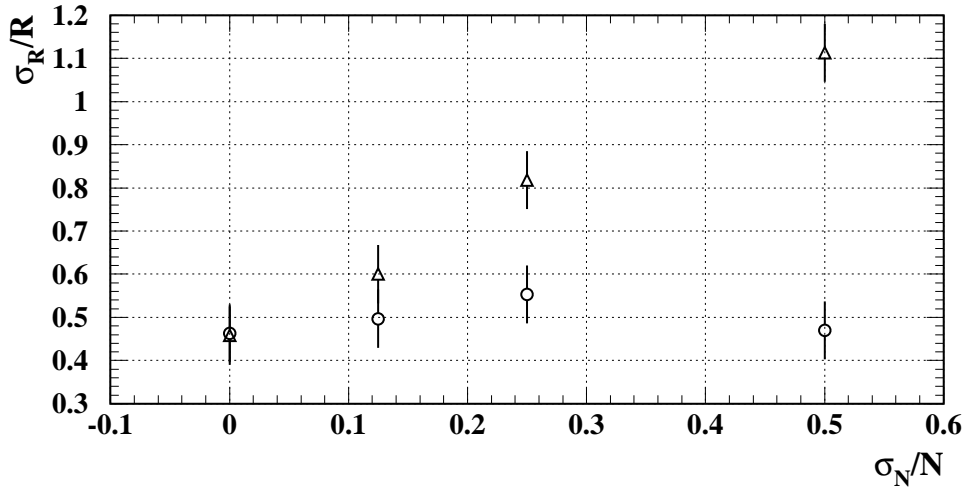}
\caption[Comparison of error for random and multiplicity-ordered mixing as a function of event multiplicity error.]{Comparison of error for random (triangles) and multiplicity-ordered (circles) mixing as a function of event multiplicity error.}
\label{poplo}
\end{figure}
%

These events were generated by constructing each event from a randomly generated uniform distribution, so they carry no correlations beyond statistical fluctuations.  By varying the degree of multiplicity variation we can observe the increasing error in the comparison measure (in this case the ratio) with increasing multiplicity fluctuations.  This increase in error arises from the single-point fluctuations that don't cancel completely when multiplicity fluctuations are present.  These correlations present themselves as a ``statistical plaid'' in the comparison measure histogram.

If the fixed-multiplicity stipulation is relaxed, the error in the ratio increases as shown in figure~\ref{poplo}.  However, additional correlations that give rise to this error can be removed by forming the proper reference using a more careful event mixing.  In the derivation of $\sigma_{\cal R}$ we only considered two events.  Thus, the approximation does not need to be valid for all events; only locally in an event space ({\em e.g.,} events sorted by multiplicity).  If we only mix events with similar multiplicity the approximation is valid for the error contribution from those mixed pairs.  If the approximation is valid for all mixed pairs in the space then we expect to see no increase in error or correlations.  Thus, by ordering events according to multiplicity and mixing only {\bf nearest-neighbor events} we construct a space that has the same error as a space built from randomly mixing events that all have the same multiplicity.

%
\begin{figure}[ht]
\centering
\includegraphics[width=5in ]{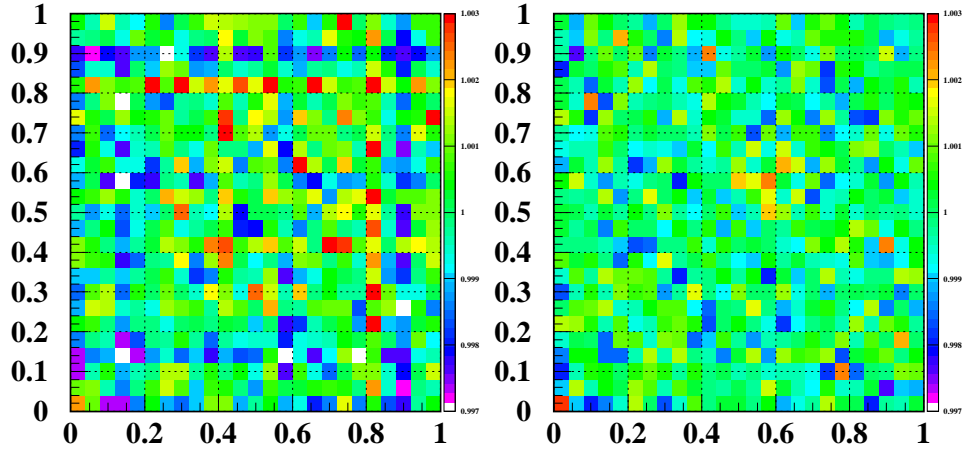}
\caption[Ratio of sibling to mixed pair histograms formed from simulated data with random and event-ordered mixing methods for $\frac{\sigma_N}{N}=0.5$.]{Ratio of sibling to mixed pair histograms formed from simulated data with random (left panel) and event-ordered (right panel) mixing methods for $\frac{\sigma_N}{N}=0.5$.  The vertical scale is $1 \pm 0.003$.}
\label{nfcomp}
\end{figure}
%

A similar approach can be taken to ease the restriction on the stability of the sampled parent distribution, assuming an event space can be defined in which events from the same or similar parent states are located close by, and events from different parent states are well separated.

\section{A Sample Analysis}

In the sample analysis presented here we considered two separate blocks (A and B) of 50k events and generated the sibling and mixed pair spaces separately for these two data blocks.  The sibling pairs were generated by taking all possible (non-self) pairs for each transverse momentum distribution.  The mixed pairs were generated by mixing multiplicity-ordered nearest-neighbor events (based only on an event index number: event $i$ is mixed with event $i+1$), which gives approximately twice as many mixed pairs as sibling pairs. 

%
\begin{figure}[ht]
\centering
\includegraphics[width=4.4in]{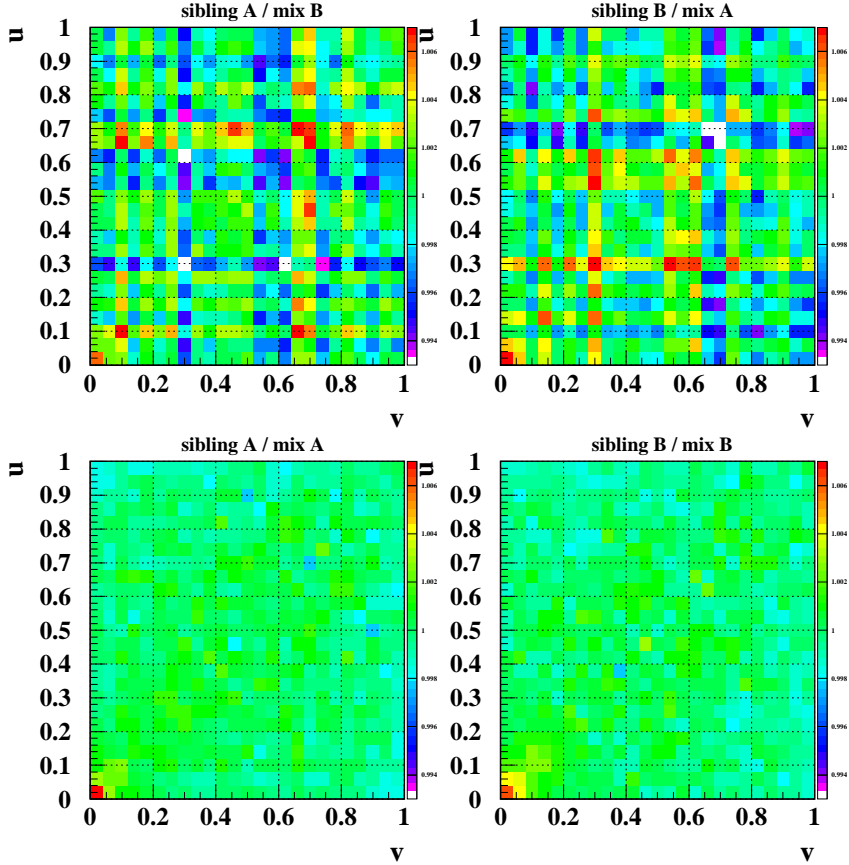}
\caption[Sibling/mixed pair histogram ratios illustrating the characteristic plaid pattern of an ill-formed reference.]{Sibling/mixed pair histogram ratios illustrating the characteristic plaid pattern of an ill-formed reference. 
The {\em top} two panels show the pair histogram ratio results for mixed pairs formed from events {\em different} from those used to generate the sibling pairs.  The {\em bottom} two panels show the pair histogram ratio results for mixed pairs formed from the {\em same} events as those used to generate the sibling pairs.  The vertical scale is $1 \pm .007$.}
\label{ABmake}
\end{figure}
%

When we look at the ratio of the sibling and mixed pair histograms generated from different data, the effect of the 1d statistical fluctuations from the transverse momentum distribution is obvious (the aforementioned ``statistical plaid''). These fluctuations obscure the physics content of the distribution, which can be seen clearly only when the sibling-to-mixed histogram ratio is taken with histograms generated from the same data, as shown in figure~\ref{ABmake}.

In forming a mixed pair reference from event pairs, sibling and mixed pairs must be generated using the same event population.  It is {\em essential} to use event pairs with matching global characteristics (in this case total multiplicity).  This conclusion follows from the recognition of statistical fluctuations as a form of correlation.  The single-point statistical fluctuations in marginal distributions can be successfully removed from the 2d pair distributions if a proper reference is formed, leaving only true two-point statistical fluctuations in the difference or ratio comparison distribution.  If events with significantly different multiplicity or distribution shape are combined to generate mixed pairs then the assumptions that allow us to cancel the single-point distribution statistical errors are invalid.

\section{NA49 $m_t\otimes m_t$ Correlation Analysis}

Using the aforementioned correlation space formation and comparison techniques, a two-particle correlation analysis was done with 100k central Pb-Pb events recorded in the NA49 detector \cite{Reid01}.  The track filtering cuts used in this analysis are similar to those used in the STAR fluctuation and two-particle correlation analyses.  The $\phi$ coverage is the same (full) but maintaining track quality in NA49 requires that we use tracks matched between the two main detector volumes (main and vertex TPCs), which means our rapidity coverage is somewhat forward ($4 < y_{\pi} < 5.5$) \cite{NA49}.  As in the STAR analysis particles in the region $0.01 < p_t < 2.0$ GeV were used and track splitting was avoided by using the standard ($\frac{n_{fit}}{n_{max}} < 0.5$) cut.  To remove secondaries the distance of closest approach of a track to the primary vertex was restricted to be within 3 sigma of the vertex position.  Events with abnormal multiplicities ($850 < N < 1450$ before cuts; $N < 100$ after) were also rejected to minimize contamination, particularly from non-central events.  The quality cuts used in this analysis have been chosen to be functionally identical to those used in the NA49 $<p_t>$ fluctuation analysis \cite{NA49pt}.  It was essential to preserve compatibility between these analyses because of the aforementioned connection between two-particle correlations and non-statistical fluctuations.  

\begin{figure}[ht]
\centering
\includegraphics[width=4.5in]{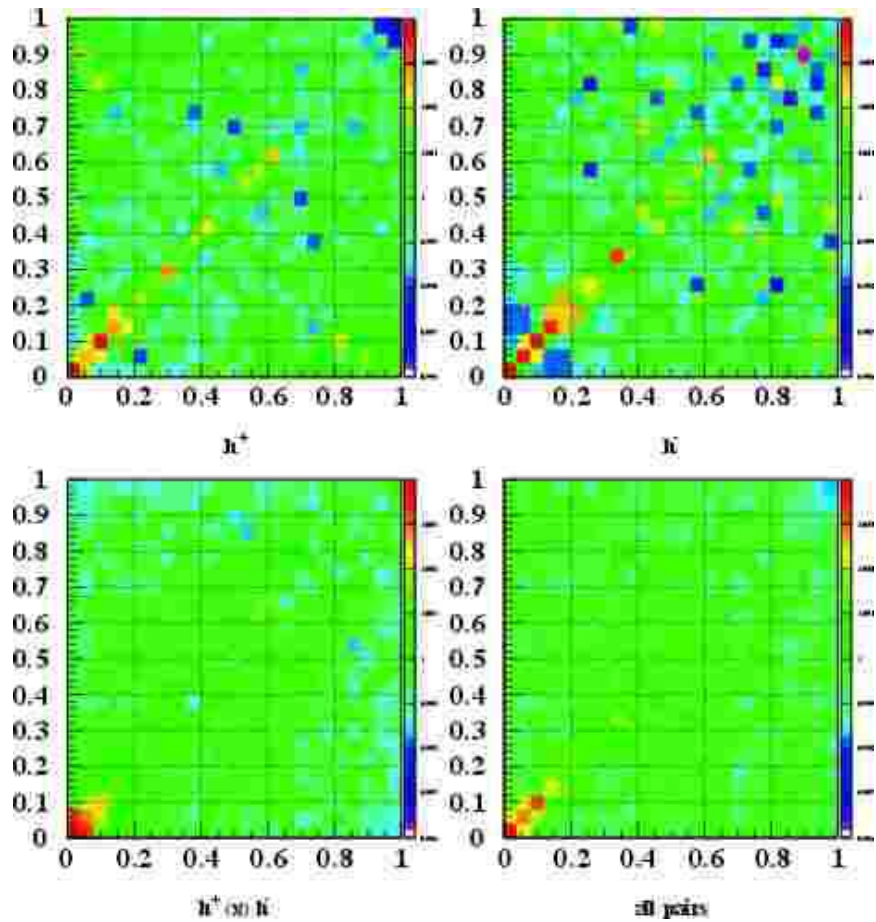}
\caption[NA49 two-point pair density ratio results for 100k central events.]{NA49 two-point pair density ratio results for 100k central events.  Ratio spaces shown are for pairs with the same charge ($[++]$ upper left, $[--]$ upper right), opposite charge ($[+-]$ lower left) and all pairs ($[cc]$ lower right).}
\label{NA49mtmt}
\end{figure}

After the cuts were applied we flattened the $m_t$ distribution as outlined in section 6.4.  Fitting the $m_t$ distribution yields a temperature of 202 MeV and flattens the distribution reasonably well (see figure~\ref{mtTrans}).  Event ordering was not necessary in this analysis because of the multiplicity restrictions placed on events in the cuts.  The results for the various charge-pair combinations shown in figure~\ref{NA49mtmt} bear this out since there is not an overwhelming statistical error as in the example shown in figure~\ref{ABmake}.

%
%

The most striking two-particle correlation features in the central NA49 data come at small scale and arise from well-understood physical mechanisms.  At low momentum in the same-sign spaces ($[++]$ and $[--]$) there is a clear Bose-Einstein signal that rises significantly above the mixed event reference.  We expect this from the substantial two-particle correlation effect of Bose statistics.  This has been well documented by HBT analysis and measured in NA49 \cite{NA49HBT}.  It is interesting to note that as the total momentum of the particle pairs increase, these Bose-Einstein correlations decrease.  In fact, for pairs with high total momentum there is a substantial anti-correlation in the positive charge same-sign pair space.  There is no corresponding (statistically significant) anti-correlation present in the negative charge pair space.  This could be an effect of the non-participant protons that sit at high $m_t$.

\begin{figure}[ht]
\centering
\includegraphics[width=4.5in]{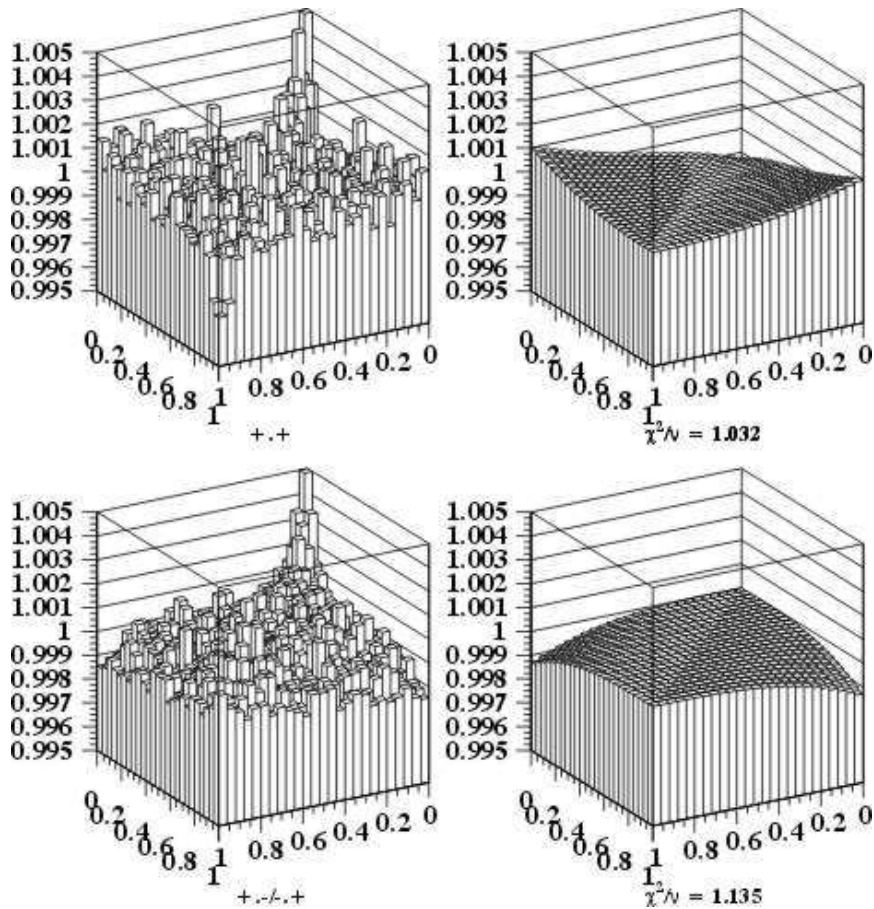}
\caption{NA49 two-particle correlation results for $[++]$ and $[+-]$ spaces compared to fits.}
\label{mtFit}
\end{figure}

In the opposite sign pair space ($[+-]$) there are no Bose-Einstein correlations (since we are looking at correlations between particles with different charge) but there is a substantial peak at very low momentum that comes from the Coulomb attraction of oppositely charged particles.  In addition to these small-scale correlation features there are substantial large-scale correlations.  This trend is made more clear by fitting the density ratio histograms.  Figure~\ref{mtFit} shows the positive charge same-sign and the opposite charge pair spaces in a perspective view (looking down the main diagonal from the far corner).  A multiparameter fit was done excluding the central corridor (particles with small $\Delta m_t$) to focus on the large scale features.  The fits show that there are significant correlations present in the high-low momentum same-sign pairs and significant anti-correlations present in the high-low momentum opposite-sign pairs.  

%
%

The low total momentum region of these fits tells an interesting physics story.  The fits show that there is a systematic deficiency of pairs at low total momentum in the same-sign pair space that is not present in the opposite-sign pair space.  This difference is due to the differing physical mechanisms that cause the small-scale features at low total momentum.  Because the density ratio spaces shown are normalized by the mixed pair background there is a conservation of pairs at work here.  Any feature in the space must come at the cost of another region to maintain the normalization.  Thus, the deficiency of pairs near the low total momentum region of the same-sign pair space should be expected.  Because Bose-Einstein correlations are relatively weak only pairs with small $\Delta m_t$ will be affected.  Pairs that make up the correlation feature must come from the local region of $m_t \otimes m_t$ space.  Even though we have removed the ridge of Bose-Einstein correlations by excluding the central corridor from the fit, we still see evidence of it in the low-momentum region of the same-sign $m_t \otimes m_t$ space.  The decreasing magnitude of the correlation ridge with increasing $m_t$ is consistent with the rapid decrease in the magnitude of the pair deficiency seen in the fit.  The Coulomb interaction, on the other hand, is quite strong and so particles from all over $m_t \otimes m_t$ space are swept into its low-momentum peak.  Thus, there is no analogous pair deficiency at low momentum in the opposite-sign pair space.

\section{STAR $m_t\otimes m_t$ Correlation Analysis}

\begin{figure}[ht]
\centering
\includegraphics[width=4.5in]{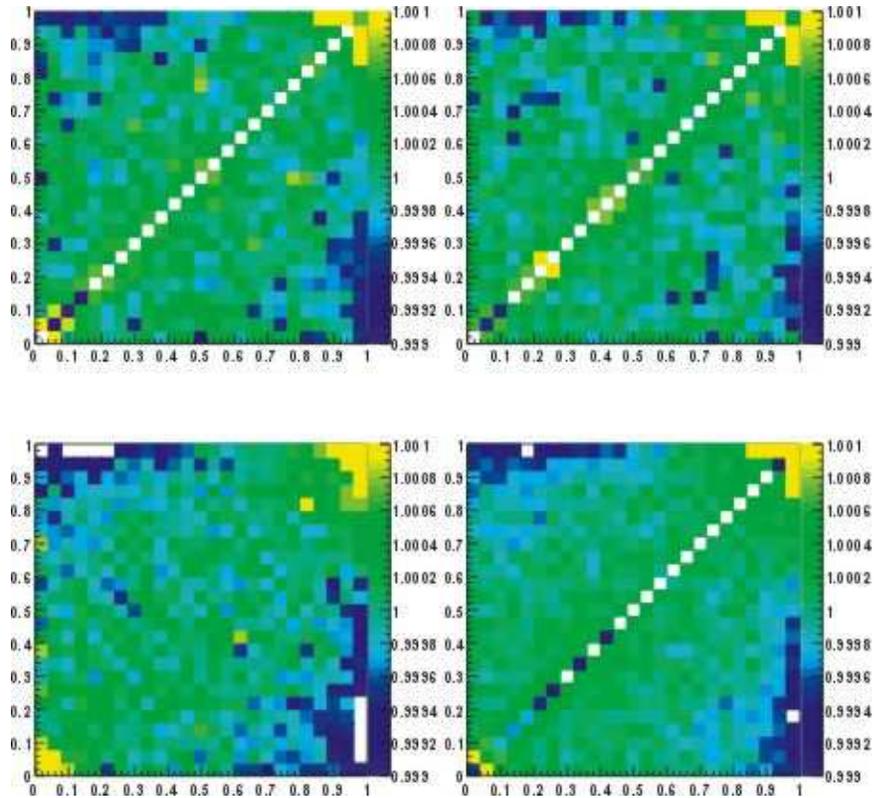}
\caption[STAR two-point pair density ratio results for 100k central events.]{STAR two-point pair density ratio results for 100k central (15\%) events.  Ratio spaces shown are for pairs with the same charge ($[++]$ upper left, $[--]$ upper right), opposite charge ($[+-]$ lower left) and all pairs ($[cc]$ lower right).}
\label{STARmtmt}
\end{figure}

The technique of event-ordering to minimize statistical error was essential in the STAR two-particle correlation analysis.  
The event multiplicities were much higher in STAR than in NA49 due to the higher energy collisions made at RHIC.  Thus, this data (100k of the 15\% most central STAR events at $\sqrt{s_{NN}}=130$ GeV) had a substantially larger spread in multiplicity than the NA49 data.  Additionally, in the STAR data the displacement of the primary vertex from the center ($\pm 75$ cm) biased the $\eta$ values of the tracks.  Thus, event ordering was necessary in both event multiplicity and vertex placement.  A coarse binning of the $N$ vs. $v_z$ space was made and only events within the same bin were used to generate mixed pairs following the error minimization techniques outlined in previous sections.  Two-point pair densities (sibling pairs/event-ordered mixed pairs) were calculated, results are shown in figure~\ref{STARmtmt}.

%
%

The STAR results show the same large scale saddle feature observed in the NA49 data.  This feature corresponds to a deficiency of pairs with high momentum difference in the sibling pairs when compared to the mixed pair reference.  The Bose-Einstein features at small scale have been intentionally excluded from this analysis by rejecting small $\Delta m_t$ pairs in the $[++]$ and $[--]$ spaces.  This is the source of the significant pair deficiency along the main diagonal in all but the $[+-]$ space.  There has been no systematic rejection of low momentum pairs enhanced by the Coulomb effect, and so there is still a small-scale feature present at low momentum in the $[+-]$ space.

There is one significant difference between the NA49 and STAR results.  The large scale feature highlighted by the fit shown in figure~\ref{mtFit} is different in the same-sign pair spaces.  The NA49 data has a correlation effect acting at large $\Delta m_t$ whereas the STAR data has an anti-correlation effect acting in the same region.  In the opposite-sign pair spaces both experiments have an anti-correlation effect acting in the large $\Delta m_t$ region.  

\section{Experimental Comparisons}

We can make comparisons between this analysis and the $<p_t>$ fluctuation analysis by decomposing the two-point ratios into a charge-dependent part and a charge-independent part.  To do this we combine the two-point ratio histograms as $([++] + [--])*([+-] + [-+])/4$ for the charge-independent part and $([++] + [--])/([+-] + [-+])$ for the charge-dependent part.  When this ratio histogram arithmetic is done we find results that are consistent with our fluctuation analysis.  

In NA49 the same-sign density histogram looks inverted compared to the opposite-sign histogram.  Thus, taking the product of the two yields a null results in the charge-independent formulation.  In the charge-dependent formulation this causes an enhancement of the correlation effect.  By comparison, the fluctuation results in NA49 central events found a charge-independent signal consistent with zero ($0.6 \pm 1$ MeV), and a significant (negative) excess fluctuation ($-8.5 \pm 1.5$ MeV).  

In the STAR central data the same- and opposite-sign histograms exhibit identical large-scale behavior.  Thus, when composing the charge-independent measure this correlation effect is enhanced, and when composing the charge-dependent measure the large-scale correlation feature will almost completely cancel itself out.  Again, the fluctuation results agree.  The charge-independent signal is substantial for central STAR data ($52.6 \pm 0.3$ MeV).  The charge-dependent signal is much smaller in magnitude ($-6.6 \pm 0.6$ MeV) but still significant.

\section{Physics Interpretations}

While it is gratifying that the two-particle analysis can confirm the results of the fluctuation analysis, that was not its main purpose.  Our main goal has been to identify the correlation features in the full two-particle correlation space and understand the physical mechanisms that give rise to them.  All of the small-scale features arose from well-understood physics (Bose-Einstein correlations, Coulomb attraction), but there is no easy explanation for the large-scale saddle feature that dominates these spaces.  

\begin{figure}[ht]
\centering
\includegraphics[width=4.5in]{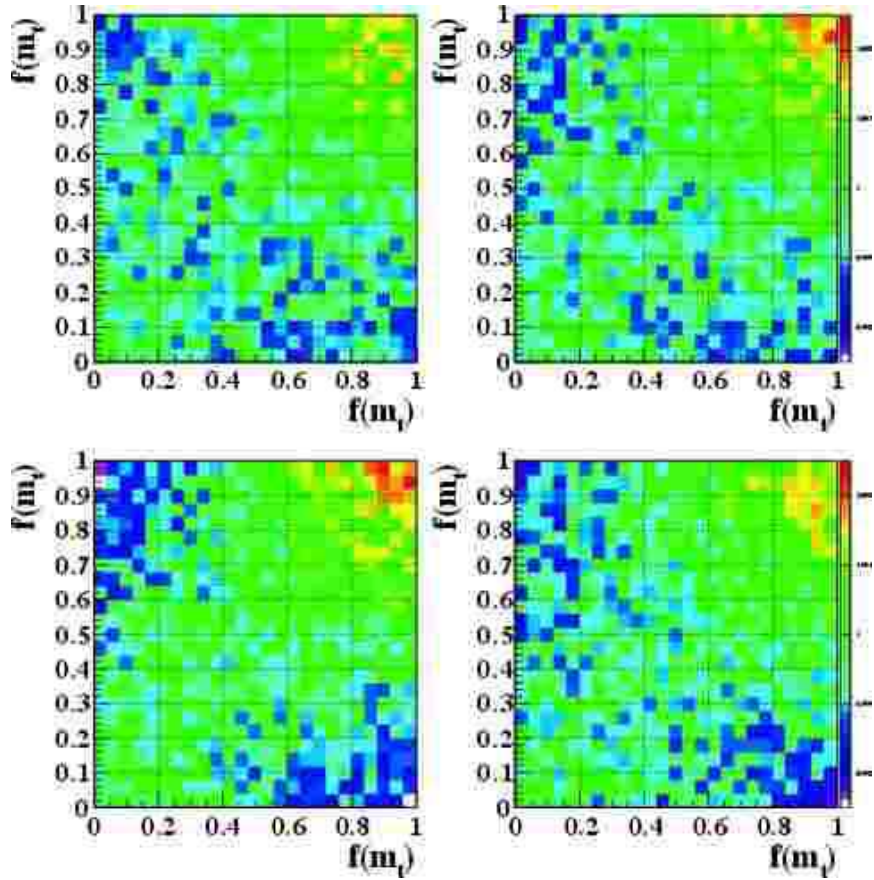}
\caption[A theoretical prediction of the effect of temperature fluctuations on pair-space density histograms.]{A theoretical prediction of the effect of temperature fluctuations on pair-space density histograms.  Histograms shown are for pairs with the same charge ($[++]$ upper left, $[--]$ upper right), opposite charge ($[+-]$ lower left) and all pairs ($[cc]$ lower right).}
\label{MCmtmt}
\end{figure}

Theoretical predictions suggest that event-by-event temperature fluctuations contribute to the variance of the event-wise mean $p_t$ \cite{CLT}.  Since an increase in the fluctuation measure must have some corresponding effect in the two-particle correlation space, a theoretical estimate of the effect of $T$ fluctuations on the pair density spaces was made.  These results are shown in figure~\ref{MCmtmt}.

%
%

Monte Carlo simulated data incorporating temperature fluctuations was also analyzed and served as a calibration of this effect \cite{Lanny}.  The STAR data, theory prediction, and Monte Carlo simulation are all in qualitative agreement.  All analyses show a substantial deficiency of sibling pairs in the large $\Delta m_t$ region (independent of charge combination) and an abundance of sibling pairs at high-$m_t$ near the main diagonal ($\Delta m_t \approx 0$).  The temperature fluctuations in the Monte Carlo study were at the 5\% level, and the order of magnitude of the resulting saddle feature was a factor of 10 larger than in the data.  Thus, we place an upper limit on temperature fluctuations in the STAR data at the 0.5\% level.  

Interpretation of the NA49 data is a more difficult problem.  The inconsistency between the temperature fluctuation results and the NA49 data in the same-sign pair space suggests that there is an additional physical mechanism at work.  The fact that this effect serves to cancel out the effect in the opposite-sign pair-space yielding a null result in the charge-independent formulation is suggestive.  Substantial additional work will need to be done to understand the charge-dependent results that are at the heart of this puzzle.

\section{Conclusions}

Starting from the algebra of the non-statistical fluctuation measures we found a deep connection between fluctuations and two-particle correlations.  Rather than try to summarize the rich structure of the two-particle correlation landscape in a single number or measure, we have chosen to approach the full two-particle correlation space directly.  In the process we have created a new method of event comparison using pair-space density ratios and a carefully constructed mixed pair reference.  

Substantial utility has been found in the connection between two-particle correlation spaces and non-statistical fluctuation measures.  By devising a method of highlighting the charge-(in)dependence of the correlation results we have found qualitative agreement between the fluctuation results of NA49 and STAR and their respective correlation contents.  While questions about the mechanism behind the charge-dependent results remain unanswered, theoretical predictions and simulations show qualitative agreement between the charge-independent results and temperature fluctuation models.  Calibration of the models with simulation allow us to place an upper limit on $T$ fluctuations in STAR at the 0.5\% level.

These methods have been quite productive in the hands of Aya Ishihara at the University of Texas in Austin.  She has provided significant insight into the nature of the emitting source by looking at correlations in axial phase space \cite{Aya} among other substantial contributions.  Increasingly these methods are being adopted in other groups within NA49 and STAR and the outlook is very positive for adoption by the broader community \cite{QM02Pratt}.  This has been the true success of this work,  providing the community with a set of self-consistent and useful tools to help it move toward an understanding of deconfined quark matter.

%
%
%

\chapter{Summary and Conclusions}

\section{Overview}

We have tackled the complex problems facing event-by-event physics in relativistic heavy-ion collisions with three distinct approaches.  We used a general approach in scaled correlation analysis, improved upon the $\Phi_{p_t}$ fluctuation measure finding striking results in the STAR data, and we extended the theory of fluctuation analysis to incorporate a connection to two-particle correlations.  The goal has been to provide significant direction to the burgeoning event-by-event physics community, and we have succeeded.

\section{Scaled Correlation Analysis Summary}

In the development of a model-explicit analysis system we generalized the partition used to calculate the entropy.  By pushing the entropy beyond the zero-scale limit and incorporating dithered binning technology we were able to formulate a novel method for measuring multiparticle correlations in a model-independent way.  Once the scale generalization of the entropy was made, scale-local measures of the correlation integral, information, dimension, dimension transport, and volume all logically followed.  We probed the behavior of these newly scale-localized topological measures with a randomly generated uniform distribution.  Comparing those results with a reference derived analytically from an ideal uniform distribution, we found that a cutoff factor had to be incorporated into the IUD reference to make it realistic and useful.  Understanding this cutoff and the related pseudoinformation provided insight into the relationship between the scaled entropy and factorial moments.

To apply the scaled correlation analysis system effectively to RHI collision data we needed to understand how to interpret the results that might come from the analysis.  We examined a variety of toy-model generated distributions to calibrate our expectations for scaled correlation analysis results.  We started by focusing on the most simplistic point distributions, one-dimensional randomly generated uniform distributions.  That example was extended by scaling the data relative to the embedding space, providing clear evidence that SCA results are unique for each correlation mechanism and independent of the scale of the embedding space.  We moved past one-dimensional distributions to 2d examples.  First, we confirmed the extrapolation of 1d results into 2d, then we looked at simple clustering simulation data to understand how a simplistic phase transition might appear in the SCA system.  

We looked at increasingly complex point distributions including a model of a space-filling-curve and one of the well-known strange attractors of the H\'enon map.  In the case of the H\'enon map attractor we discovered the rich and unique structure of its scaled-dimension, which cannot be adequately expressed in the zero-scale limit dimension of conventional topological approaches.  An extended example was analyzed using grey scale digital images of human faces to illustrate the utility of using an event-space approach with SCA results.   By calculating the dimension transport between each individual image and an ensemble average we were able to form an event space in which different images of a single subject's face could be distinguished from the faces of other subjects.  This same approach was taken in analyzing NA49 data and we found anomalous events at the 10\% level.  These were identified as pile-up events in which beam/gas interactions occured in addition to the primary beam/target interaction.  A pile-up rejector was added to the detector and in later runs these contaminating events were successfully removed.

\section{Fluctuation Analysis Summary}

Starting from the $\Phi_{p_t}$ measure and the work of Ga\'zdzicki and Mr\'owczy\'nski we have expanded significantly on the field of fluctuation analysis in heavy-ion collisions.  By defining a fluctuation measure based directly on the central limit theorem \cite{CLT} we have provided the community with an unbiased and interpretable measure.  Defining the measure is simply the beginning, the real work of this analysis is in maintaining data quality and insuring robust results.  By applying a strict set of event and track cuts we have minimized the contamination of the primary particle data sample from
particles originating from secondary interactions.  After these quality cuts are applied to STAR data a numerical fluctuation analysis found fluctuations above the statistical expectation at the 15\% level.  To insure our results were robust we did simulations and estimations of systematic error sources including analysis of tracking effects, cut systematics, and signal contamination from other sources.  A graphical data/reference comparison was made that confirmed our numerical results.  Comparisons with fluctuation results from other experiments revealed that this was a unique measurement because of the large multiplicity of our events and the statistical significance of our result.  Further comparisons need to be made, specifically the multiplicity scaling of the results in other experiments remains to be understood, but the preliminary results are promising.

\section{Two-Particle Correlation Analysis Summary}

The inherent relationship between the non-statistical fluctuations and two-particle correlations in an event motivated our development of a comprehensive two-particle correlation analysis.  Rather than look at an over-simplified correlation measure like the integral over the two-particle correlation space we looked at the full spectrum of two-particle correlations.  We first flattened the $m_t$ distribution with a temperature-dependent transformation that maps the whole transverse mass space into the unit interval ($[0,\infty) \rightarrow [0,1)$).  This mapping served to make the statistical error uniform over the whole space as well as highlight the soft physics region ($0.2 < p_t < 0.8$).  With the flattened $m_t$ space we formed all possible pairs between same and different events and built a sibling/mixed pair density ratio histogram.  This space presented a picture of the full scale spectrum of two-particle correlations in the data.  The STAR two-particle correlation analysis showed a rich structure consistent with the fluctuation analysis results.

\section{Conclusions and Future Directions}

This work represents a significant beginning for event-scale physics techniques in heavy-ion collisions.  Never before have the statistics at the event level been so favorable.  With the recent 200 GeV full-energy run at RHIC and the startup of the CERN LHC on the horizon, event multiplicity will continue to grow, increasing the opportunities for event-by-event analysis.  Clearly this is a critical time for event-by-event physics.  With that in mind, it is important to emphasize that a more coherent fluctuation analysis community is needed.  Right now these are the most exciting results coming out of the event-by-event physics programs.  Unfortunately, because of the acceptance dependence of the measures used, making coherent comparisons among the experiments is nearly impossible.  To make the results understandable and convincing to the broader physics community a method of standardizing the results and making comparisons must be defined by the relevant experts.  Also, there needs to be more theoretical support for this analysis.  While there is a strong community producing some great work, we are still lacking directly testable hypotheses from theorists.  The experimental results we have are quite provocative, but without solid predictions from theory they are of limited utility.

For our part, we must continue to refine and simplify the techniques we have been developing.  Finding digestible component pieces of the larger analysis picture that can be broken off and explored fully and independently is important.  It is useful and satisfying to completely specify and understand the intricate relationships between all of these analyses, but this is not enough.  To truly serve the physics community we must strive to find self-contained approaches to the specific problems at hand that can be applied and understood independently of the larger picture.  Much like the theory community, we too must serve the greater interest by providing tractable analysis tools and interpretable results.  

Through our efforts to understand the effect of various physical mechanisms on SCA analysis results we found that simulations are essential to this work.  Increasing the number of simulations done to provide context for the experimental results will be a key part of the future analysis efforts.  This is not to say that we need yet another event generator or overly complex physics simulation package.  Rather, it is much more convincing (and tractable) to approach the problem piecemeal by using simple toy simulation models to represent different physics effects.  It is not that the existing event generators are not useful, on the contrary they are essential.  However, we have enough of them already and they are often too complicated (and poorly understood) for the job of isolating the effect of a specific physics mechanism on the analysis results.

Event-by-event physics is the most exciting and rapidly growing subfield in heavy-ion physics.  There is no doubt that in the next 20 years it will prove to be a key part in the quest to understand quark matter.


%
%
%
%
\nocite{*}   
\bibliographystyle{plain}
\bibliography{the}

%
%
\appendix
\raggedbottom\sloppy

\vita{Jeff Reid was born to George and Sonja Reid on 10 November 1970 in Renton, Washington.  He attended public schools in Renton, Kent, and Sunnyside.  After dropping out of high-school to attend Johns Hopkins University, where (among other things) he learned to play pool, he transferred to Harvey Mudd College in 1989.  He graduated from HMC in 1993 with a B.S. in physics and a minor in film studies.  While in Claremont he was a DJ and music director for KSPC, the Pomona College radio station.  

After three unsuccessful attempts at gaining admission to the doctorate program at UW, he was finally admitted in 1996 after working with Tom Trainor at the nuclear physics laboratory as an hourly employee.  While studying for his doctorate he received an M.S. in physics from UW and became a valued member of the NA49 and STAR collaborations.  He gave numerous talks during this time, including a talk for NA49 at the 14th international quark matter conference in 1999, a talk for STAR at the 15th international quark matter conference in 2001, and an invited seminar at the CERN heavy-ion forum in 2001.  In addition to participating in the collaboration publications of NA49 and STAR, he published a paper on statistical fluctuations in two-point correlation analysis in {\em Nuclear Instruments and Methods} with Tom Trainor \cite{Reid02}.  With the completion of this dissertation in December of 2002, he has fulfilled the requirements for his doctorate.

Jeff has also participated in several sound-art installations and performances over the last 10 years, including opening for MSBR at Thee Last Supper Club in Seattle.  He currently resides in Renton, Washington.}

\end{document}